\documentclass[fleqn,usenatbib]{mnras}
\usepackage{graphicx}	% Including figure files
\usepackage{amsmath}	% Advanced maths commands
\usepackage{amssymb}	% Extra maths symbols
\usepackage{newtxtext,newtxmath}
\usepackage[justification=centering]{caption}
\usepackage{adjustbox}
\usepackage{longtable}
\usepackage[flushleft]{threeparttable}
\usepackage{hhline}
\usepackage{comment}
\usepackage{rotating}
\usepackage{array}
\usepackage{hyperref} 
\usepackage[T1]{fontenc}

\DeclareMathOperator{\EX}{\mathbb{E}}% expected value

\title{\Large \bf Paving the Way for Euclid and JWST via Optimal Selection of High-$z$ Quasars}

\author[0000-0002-2579-4789]{Riccardo Nanni,$^{1,2}$\thanks{E-mail: riccardonanni@ucsb.edu} 
Joseph F. Hennawi,$^{1,2}$
Feige Wang,$^{3}$
Jinyi Yang,$^{3}$
\newauthor
Jan-Torge Schindler,$^{1,4}$
Xiaohui Fan$^{3}$
\\
$^{1}$Leiden Observatory, Leiden University, PO Box 9513, 2300 RA Leiden, The Netherlands \\
$^{2}$Department of Physics, University of California, Santa Barbara, CA 93106-9530, USA \\ 
$^{3}$Steward Observatory, University of Arizona, 933 North Cherry Avenue, Tucson, AZ 85721, USA \\
$^{4}$Max Planck Institut f\"ur Astronomie, K\"onigstuhl 17, D-69117, Heidelberg, Germany
}

\date{Accepted XXX. Received YYY; in original form ZZZ}

\pubyear{2020}

\begin{document}
\label{firstpage}
\pagerange{\pageref{firstpage}--\pageref{lastpage}}
\maketitle

\begin{abstract}
We introduce a probabilistic approach to select $6\le z \le8$ quasar candidates for spectroscopic follow-up, which is based on density estimation in the high-dimensional space inhabited by the optical and near-infrared photometry. Density distributions are modeled as Gaussian mixtures with principled accounting of errors using the extreme deconvolution (XD) technique, generalizing an approach 
successfully used to select lower redshift ($z\lesssim 3$) quasars. 
We train the probability density of contaminants on 733,694 7-d flux measurements from the 
1076 deg$^2$ overlapping area from the DECaLS ($z$), VIKING ($YJHK_s$), and unWISE ($W1W2$) imaging surveys, after requiring they dropout of DECaLS $g$ and $r$, whereas the
distribution of high-$z$ quasars are trained on synthetic model photometry. Extensive
simulations based on these density distributions and current estimates of the quasar luminosity function indicate that this method achieves a completeness of $\ge75\%$ and an efficiency of $\ge15\%$ for
selecting quasars at $6 < z < 8$ with $J_{\rm AB} < 21.5$. Among the classified sources are 8 known $6<z<7$ quasars, of which 2/8 are selected  suggesting a completeness $\simeq 25\%$, whereas
classifying the 6 known ($J_{\rm AB} < 21.5$) quasars at $z > 7$ from the entire sky, we select 5/6 or a completeness of $\simeq 80\%$. The failure to select the majority of $6<z<7$ quasars arises 
because our model of quasar SEDs underestimates the scatter in the distribution of fluxes. This new optimal approach to quasar selection paves the way for efficient spectroscopic
follow-up of Euclid quasar candidates with ground based telescopes and JWST. 
\end{abstract}


\section{Introduction} \label{sec:intro}
Luminous high-redshift quasars (QSOs) are one of the best probes of the primordial Universe at the end of the dark ages.
Their spectra provide important information regarding the properties of the intergalactic medium (IGM) during
%% JFH later statges --> during 
%% Euclid will find quasars up to z ~ 9, so I would not restrict to late stages
%% RN done
the epoch of reionization (EoR). In fact, deep spectroscopy of $z>6$ QSOs showed that the IGM is significantly neutral at $z \ge 7$ \citep[e.g.,][]{Banados18,Davies18,Wang20,Yang20b}, but highly ionized at $z \le 6$ \citep[e.g.,][]{Mcgreer11,Mcgreer15,Yang20a}. 
%% JFH I would add a citation to the dark pixel gaps paper by McGreer et al. here to "highly ionized"
%% RN done

In addition, the engines of the most distant QSOs, the super massive black holes (SMBHs), are crucial for understanding the formation mechanisms of
the first generation of black hole seeds \citep[see][, for a recent review]{Inayoshi20}, for a recent review). Their existence up to $z = 7.6$ \citep[e.g.,][]{Wang21}, and hence formation in less than 1 Gyr, poses the most stringent constraints on the masses of black hole seeds. In fact, making the standard assumptions about Eddington-limited accretion, current BH masses in the highest-$z$ quasars appear to rule out the expected $\sim100$ $M_{\odot}$ seeds from Pop III remnants, and instead require more massive seeds \citep[e.g.,][]{Volonteri10,Volonteri12}.
%% JFH2 This could be more clearly stated. Reading this, one does not
%% get the basic idea namely. 
%% 1) Making the stnadard assumptions about accretion
%% current BH masses in the highest-z quasars appear to rule out the expected
%% ~100 Msun seeds frmo Pop III remnants and instead require more massive seeds. %% 2) This list of scenarios solves this problem in these ways... 
%% RN done
Thus, the following theoretical scenarios have been proposed: the growth from massive black hole seeds ($10^{4-6}\,  M_{\odot}$) through the direct collapse of a primordial cloud \citep[e.g.,][]{Habouzit16,Schauer17,Dayal19}, lower-mass seeds ($10^{2-3}\,  M_{\odot}$ which are the remnants of PopIII stars) with Eddington limited or even super-Eddington accretion and very rapid growth \citep[e.g.,][]{MadauRees01,TanakaHaiman09,Inayoshi16}, or presence of radiatively inefficient accretion modes \citep[e.g.,][]{Trakhtenbrot17,Davies19}.

As of today, more than 200 quasars have been discovered at redshift $z \ge 6$ \citep[e.g.,][]{Fan01,Wu15,Jiang16,Banados16,Matsuoka16,Wang17,Reed17,Yang19,Matsuoka19a}
%; see \citealt{Bosman20} for an up-to-date list), 
%% JFH The citation to Bosman's updated list is messed up here. I think you want to add that in brackets somehow. 
%% RN done
thanks to the advent of wide-field multi-band optical and NIR imaging surveys such as: the Sloan
Digital Sky Survey \citep[SDSS; e.g.,][]{Fan01}, the Canada-France-Hawaii Telescope Legacy Survey \citep[CFHTLS; e.g.,][]{Willott09}, the Panoramic Survey Telescope and Rapid Response System 1 \citep[Pan-STARRS1; e.g.,][]{Banados16}, the United Kingdom Infrared Telescope Infrared Deep Sky Survey \citep[UKIDSS; e.g.,][]{Mortlock11}, the VISTA Kilo-degree Infrared Galaxy survey \citep[VIKING; e.g.,][]{Venemans13}, the VLT Survey Telescope ATLAS \citep[VST-ALTAS; e.g.,][]{Carnall15}, the Dark Energy Survey \citep[DES; e.g.,][]{Reed15}, the DESI Legacy Imaging Surveys \citep[DELS; e.g.,][]{Wang17}, the UKIRT Hemisphere Survey \citep[UHS; e.g.,][]{Wang19}, and the Hyper Suprime-Cam survey \citep[HSC; e.g.,][]{Matsuoka16}).

%% JFH I recommend a pargraph break here. 
%% RN done
At the highest redshifts, 
%% JFH Among the highest-redshift QSOs --> At the highest redshifts, 
%% RN done
there are only eight quasars known at $z \ge 7$ \citep{Mortlock11,Banados18,Wang18,Yang19,Yang20a,Matsuoka19a,Matsuoka19b,Wang21} with two of them at $z = 7.5$ \citep{Banados18,Yang20b}, and the most distant one at $z=7.6$ \citep{Wang21}.
This sample of $z \ge 7$ QSOs is still very limited -- owing to the opacity of the 
intervening high-$z$ IGM, distant quasars are brightest redward of their Ly$\alpha$ emission line which is redshifted to NIR wavelengths at $z \ge 7$,
%% JFH I would write something like "This sample of $z \ge 7$ QSOs is still very limited -- owing to the opacity of the 
%% intervening high-$z$ IGM, distant quasars are brightest redward of the Ly$\alpha$ emission which is redshifted to NIR wavelengths at z > 7"
%% RN done
making both imaging and spectroscopic observations more challenging. Furthermore, the expected number density of $z \ge 7$ quasars is low ($10^{-3}$ deg$^{-2}$ at $J=21$; \citealt{Wang19}), while the contaminants, mostly Galactic cool dwarfs and early type galaxies, are far
more numerous ($\approx 20$ deg$^{-2}$ at $J=21$). 
%% JFH Would be best to be quantitative here, even it is only approximate. How much more numerous. You have a plot showing
%% this, you could even cite it. 
%% RN done
As a result, quasar target selection in this redshift range is largely inefficient (efficiency $\sim 1\%$; \citealt{Banados18}; \citealt{Wang21}), and
thus requires large amounts of telescope for spectroscopic confirmation, 
%leading to the increase of the time requested with large telescopes for spectroscopic confirmation, and consequently
making it extremely challenging to find more bright $z > 7$ quasars with
existing datasets.

On the other hand, the advent of the next generation photometric and spectroscopic telescopes, such as Euclid or the James Webb Space Telescope (JWST), should prove to be a watershed moment 
%% JFH landmark --> watershed moment (or something like that). Landmark is more like a place, you mean a point in time
%% for which watershed is more appropriate. 
%% RN done
in high-redshift quasar studies (\citealt{Euclid19}).
In fact, Euclid's wide field IR imaging should enable the discovery of $\sim 100$ quasars with $7.0 < z < 7.5$, and $\sim$ 25 beyond the current record of $z = 7.6$, including $\sim$ 8 beyond $z =8.0$ \citep{Euclid19}, and JWST will deliver exquisite spectra of them. Ground based telescopes will play an essential role in discovering the brighter Euclid quasars, whereas fainter $J_{\rm AB} > 21.5$ ones will likely require JWST. Although current selection methods based on simple color-cuts were able to discover most of the $z>7$ known QSOs (\citealt{Banados18}; \citealt{Yang20b}; \citealt{Wang21}), their $\sim1\%$ efficiency is far too low to make confirmation of the on average fainter Euclid QSOs feasible,
as this would require excessive amounts of ground based and JWST
observations. 
%integration times from ground based telescopes or expansive JWST observations,
%% JFH I would explain why here. Basically Euclid will find fainter quasars, because since the number density 
%% is decreasing you need to work fainter to find them. Bright ones will be do-able from the 
%% ground, but will be fainter than what is found to date. Fainter ones would require JWST. Both of these points
%% long integration times from the ground or expensive JWST confirmation demand much higher efficiencies. 
%% RN done
It is thus clear that more efficient selection methods are required. 
%% JFH enhanced request --> demands
%% RN done

So far, two different statistical methods 
%% JFH You need a statement here like color-cuts have been used as well, before going intot he statistical stuff. 
%% Indeed nearly all z > 7 quasar discoveries were made by color-cuts. 
% RN done
for selecting high-$z$ QSOs have been proposed. The first one is based on the Bayesian model comparison (BMC) technique laid out by \citet{Mortlock12}, while the second uses a simpler minimum-$\chi^2$ model fitting
method to the quasars' spectral energy distribution (SED) (\citealt{Reed17}). These methods are based on improved population models for the key contaminants: MLT dwarf types, and compact early-type galaxies, and they both require model colours for each population. The BMC method additionally requires a model for the surface density of each source as a function of apparent magnitude. Although these methods have been successfully used in the past to select high-$z$ QSOs (\citealt{Mortlock11}; \citeyear{Mortlock12}; \citealt{Reed17}), including in the VIKING survey (\citealt{Barnett21}), they mostly rely on constructing a contaminant model of the entire sky in the color-range in question to very faint magnitudes.
The efficacy and feasibility of this approach has not yet been demonstrated,
and it appears extremely challenging given our currently poor knowledge about the different kind of contaminants. Another quasar method that
has been employed uses the random forests machine learning
algorithm in conjunction with color-cuts for quasar selection and photometric redshift estimation (\citealt{Schindler17}; \citeyear{Schindler18}; \citeyear{Schindler19}; \citealt{Wenzl21}). While this method has been demonstrated
to successfuly select quasars at lower-$z$, its primary drawback is
that it cannot properly account for photometric uncertainties.

%% JFH I would be a bit more critical and state that they really rely on constructing a contaminant model of the entire sky %% in the color-range in question to very faint magnitudes, and maybe even something like the "feasibility of which has
%% not been demonstrated" or something like that. I guess it is in the details of how much you trust the Barnett paper. 
%% I mean they compare to COSMOS which is ~ 1 deg^2. 
%% RN done

%% JFH This paragraph is very weak and is the main justification for your approach. Not sure you say all of this below,
%% but you need to at least touch on it. The main merits of XD are 
%% 1) fully Bayesian and optimal, i.e. density estimation is the optimal approach (akin to Mortlock approach if models
%% contaminant model is perfectly known), 2) contaminant model is fully empirical and requires making no assumptions, could 
%% be generalized to depend on sky position, 3) full accounting of errors, i.e. noiseless distributions are inferred
%% which are then convolved with a given targets uncertaintites. 

In this paper, we describe our probabilistic high-$z$ quasar selection technique,
%% JFH2 It is awkward to define an acroynym before you actually spell out what
%% it is. Move this definition downstream. 
%% RN done
which uses density estimation in flux space to compute
the probability of being a high-$z$ quasar for each candidate. For density estimation, we use the extreme deconvolution method (XD; \citealt{Bovy11a}; \citeyear{Bovy11}),
which generalizes the familiar machine learning approach of describing
a probability density with  a mixture of Gaussians to the case of
heteroscedastic noise. 
%% JFH2 This is not the correct XD refrence. The original method goes
%% back to Bovy, Roweis, Hogg. Please make sure you cite that as well. 
%% RN done
XD enables one to deconvolve errors for noisy training data to construct
the true underlying noiseless probability density, and then reconvolution
of the associated noise to evaluate the probability at new arbitrary test
locations.  In the context of high-$z$ quasar selection, the main
merits of this approach are: 1) it is fully Bayesian and thus optimal, that
is density estimation constitutes the optimal approach to estimate a classification probability\footnote{Similar to Mortlock's approach if the contaminant and quasar models were perfectly known.},
%% JFH2 Shall we make this parenthetical a footnote instead?
%% RN done
 2) the contaminant model is fully empirical and requires making no assumptions,
%% JFH2 Let's take out "it could be generalized to depend on sky  position."
%% This is more of a detail and can be added to the summary. 
%% RN done
3) it fully accounts for errors in a principled fashion, i.e. noiseless distributions are inferred via deconvolution and then reconvolved with the given target uncertainties. 
%% JFH Need to add a better description of what XD is. Mention ML. Mixture of Gaussian density estimation approach 
%% but enabling deconvolution of errors for noisy training data, and reconvolution for evaluating probability density
%% at new locations using associated noise. 
% RN done
In the end, the target selection/classification problem becomes the task of training good number-density models for both the target population and the contaminant population to maximize the efficiency and completeness of the survey. 
We applied our target selection technique (hereafter XDHZQSO) to a set of possible high-$z$ candidates that are selected with the use of optical, NIR and MIR surveys, and construct our XDHZQSO quasar targeting catalog. This catalog will be used for future spectroscopic follow-up to confirm new high-$z$ QSOs in the NIR ground based survey area,
%% JFH2 Maybe you can say ground based and thus generalize to what you are
%% doing with UKIDDS
%% RN done
while this technique provides a better method for classifying and prioritize high-$z$ QSOs candidates in the near future, especially with the advent of \textit{Euclid} in 2022.
%% JFH2 can we mention EUCLID maybe here?
%% RN done

This paper is structured as follows. We present the XDHZQSO method in \S \ref{sec:prob_class}. In \S \ref{sec:train_data} we discuss the data used to train our probabilistic classifier, and in \S \ref{sec:dens_model}
%% JFH2 This next sentence is so ambiguous that it basically has no content.
%% Can you be more specific about what you do in section 4?
%% RN done
we describe the construction of the XDHZQSO models from the training data and its application to classify our candidates. In \S \ref{sec:qso_selection} we provide a detailed description of the analysis of source completeness and efficiency.
%% JFH2 We use completeness and efficiency later. Try to use consistent
%% terminology throughout. 
%% RN done
In \S \ref{sec:known_qsos} we show the results of our code in classifying both the known high-$z$ QSOs in the VIKING survey area, and the known $z>7$ QSOs on the entire sky. In \S \ref{sec:discussion} we discuss the limitations of our selection technique, compare it to other methods, and describe various extensions to the basic method described in this paper. We conclude in \S \ref{sec:conclusions}. Throughout the paper, we adopt a flat cosmological model with $H_0 = 68.5$
%% JFH2 Units should not be in math mode!
%% RN done
km s$^{-1}$ Mpc$^{-1}$ (\citealt{Betoule14}), $\Omega_{\rm M} = 0.3$, and $\Omega_{\Lambda} = 0.7$. All the magnitudes are given in the AB system, while the uncertainties of our reported measurements are at 1-$\sigma$ confidence level.

\section{Probabilistic classification method} \label{sec:prob_class}
The use of probabilistic methods for target selection is essentially 
%% JFH substantially --> essentially 
%% RN done
a classification problem in which objects are classified into one of a discrete set of classes, 
%% JFH samples --> classes
%% RN done
based on their measured physical attributes.
These classes can be modeled using a set of objects with class assignments available on which we can train the classification algorithm. Although this is a classical problem in data analysis/machine learning, the physical attributes of astronomical targets are rarely measured without substantial and heteroscedastic measurement uncertainties, and often there is also the problem of missing data. Knowing that, classification algorithms for astronomical target selection have to deal with these complications 
by naturally degrading the probability of an object being in a certain class if the measurement uncertainties imply that the object overlaps several classes.

Consider an object $O$ with ``true'' attributes $\{F_i \}$ that we wish to classify into class $A$ or class $B$. In our specific case, we would like to classify an object $O$ into classes ``high-$z$ QSO'' or ``contaminant'' based on its physical noiseless $\{F_i\}$ and noisy $\{f_i\}$ attributes (e.g., fluxes, magnitudes, colors, or relative fluxes), and the associated uncertainties $\{df_i\}$. This can be expressed using the Bayes’ theorem to relate the probability that object $O$ belongs to class $A$ to the density in attribute space:
\begin{equation}\label{eq:bayes}
        P(O\in A|\{F_i\})=\frac{p(\{F_i\}|O\in A)P(O\in A)}
        {p(\{F_i\})},
\end{equation}
where
\begin{equation}\label{eq:norm}
        p(\{F_i\})=p(\{F_i\}|O\in A)P(O\in A)+p(\{F_i\}|O\in B)P(O\in B),
\end{equation}
since $A \cup B$ contains all of the possibilities. In Eq. \ref{eq:bayes} we distinguish between discrete probabilities $P$ and continuous probabilities $p$. The $p(\{F_i\}|O\in A)$ factor in the numerator of the right-hand side of Eq. \ref{eq:bayes} is the density in attribute space evaluated at the targets’s attributes $\{F_i\}$, while $P(O\in A)$ is the total number of $A$ objects in a prior probability. 
%% JFH unbiased sample sounds weird. Perhaps more appropriate to calling this the prior probability.
%% RN done
The denominator $p(\{F_i\})$ is a normalization factor, and expresses the total probability that the object $O$ belongs to either class $A$ or class $B$. It is easy to see that this
probability is a true probability since it always lies between zero and one,
and the sum of the probabilities for the two classes is equal to one.

Measurement uncertainties are handled in this framework through marginalization over the ``true'' properties $\{F_i\}$ given the observed ones $\{f_i\}$ and the measurement-uncertainty distribution $p(\{f_i \}|\{ F_i \})$:
\begin{equation}\label{eq:errors}
        p(\{f_i\}|O\in A)=\int d\{F_i\}p(\{F_i\}|O\in A)p(\{f_i \}|\{ F_i \}).
\end{equation}
%% JFH2 Let's talk about this. I think there may be a math error here!
%% RN done
We take $p(\{f_i \}|\{ F_i \})$ to be Gaussian, which is
an extremely good approximation for flux measurements. As a result
XD provides a simple mechanism to 1) infer the the true underlying ``noise deconvolved distribution'' $P(O\in A|\{F_i\})$, as well as 2) performs the convolution integral in Eq. \ref{eq:errors}. Since the model is a mixture of Gaussians
and the errors are Gaussian, the normally complex operations of
deconvolution/convolution reduce to trivial algebraic operations.

%% JFH I would add a note here saying XD allows provides a simple mechanism to 1) infer the 
%% the true underlying "noise deconvolved distribution" P(O\in A|\{F_i\})$ as well as 2) perform the convolution integral
%% in this equation.  Both are trivial for mixture of Gaussians models and Gaussian errors since deconvolutions/convolutions
%% become algebraic operations. 
%% RN done

%In the end, the target selection/classification problem becomes the task of training %good number-density models for both the target population and the contaminants %population to maximize the efficiency and completeness of the survey. 
%% JFH Consider making this stronger statement in the intro. 
%% RN done
Compared to other probabilistic selection methods, the great advantage of our approach is that the poorly understood contaminants are modeled fully empirically, rather than relying on physical models (e.g., \citealt{Mortlock12}; \citealt{Barnett21}), and the contaminant classes are all grouped into a single all-inclusive contaminant class. In this way, the density models for the contaminant class can be simply trained using real data from the entire sky. This method was already applied in the past to select SDSS QSOs (\citealt{Bovy11}; \citealt{Bovy12}), 
%% JFH can you also cite the other Bovy papers that use XD. There is the photo-z one, and there is also a paper by 
%% Myers, Bovy. I'm not on that one, but it is led by Myers postdoc and adds WISE data. All of those should be cited here. 
%% RN done
and was shown to be effective even in the challenging redshift range $2.5 \le z \le 3$
where the stellar contamination is significant. 
%% JFH Move these references to the random forest approach to either the intro or comparison to previous work. It is
%% a distrction here, and I don't think the approach is that similar. Random forest is a decision tree classifer. 
%% RN done
%Similarly, another classification method based on random forest machine learning algorithms was used on SDSS and WISE photometry for high-$z$ quasar classification and photometric redshift estimation (\citealt{Schindler17}; \citeyear{Schindler18}; \citeyear{Schindler19}), however it was applied in combination with color-cuts for target selection. 

%% JFH Add a Table to this section indicating the surveys used, the filters they cover, 
%% total area of VIKING or something that they cover, as well as the x\sigma magnitude limits in each filter. 
%% RN done
\section{Training data} \label{sec:train_data}
To construct probability density models we trained on either real or simulated photometry, depending on whether we are considering ``contaminants'' or ``quasars''.  Contaminants were trained on  $1076$ deg$^2$ of overlapping
imaging from VIKING ($YJHK_s$), DECaLS ($grz$), and unWISE ($W1W2$)\footnote{To compute the area covered by the sources in our sample we used the $healpy$ Python package, based on the Hierarchical Equal Area isoLatitude Pixelization (HEALPix). We used $healpy$ to subdivide a spherical surface in 200 pixels, in which each pixel covers the same surface area as every other pixel, and summed the areas of the pixels that includes one or more sources from the VIKING survey area.}. In Table \ref{tab:surveys}
%% JFH2 Should you just omit the area column in Table 1. It seems silly
%% given that only the overlap matters. 
%% RN done
we summarize the properties of the three surveys we used for our selection. The quasar models were trained on synthetic photometry from the \citet{McGreer13}  ``simqso'' simulator \footnote{\url{https://github.com/imcgreer/simqso/}}. This section describes the data used to train these density classification models.
\begin{table}
  \centering
  \captionsetup{justification=centering, labelsep = newline}
      \caption[]{Survey properties}
      \begin{adjustbox}{center, max width=\textwidth}
         \begin{tabular}{l c c}
            \hline
            \hline \rule[0.7mm]{0mm}{3.5mm}
            Survey &  Filters & 5$\sigma$ depth\\
            \hhline{~~~}
            \hline \rule[0.7mm]{0mm}{3.5mm}
            VIKING & $ZYJHK_s$ & 23.1, 22.3, 22.1, 21.5, 21.2 \\
            \rule[0.7mm]{0mm}{3.5mm}
            DECaLS & $grz$ & 23.95, 23.54, 22.50\\
            \rule[0.7mm]{0mm}{3.5mm}
            unWISE & $W1W2$ & 20.72, 19.97\\[2pt]
            \hline 
         \end{tabular}
        \end{adjustbox}
         \label{tab:surveys}
\end{table}
%% JFH Somewhere in the manuscript put in a sentence on how you computed the area of VIKING (using Healpix) 
%% and quote the exact number. 
%% RN done

\subsection{Contaminant Data}\label{sec:train_star}
The contaminant training set is generated using photometry from deep optical, and near- and mid-IR imaging surveys.

\begin{figure*}
 \begin{center}
 \includegraphics[height=21cm, width=15cm,keepaspectratio]{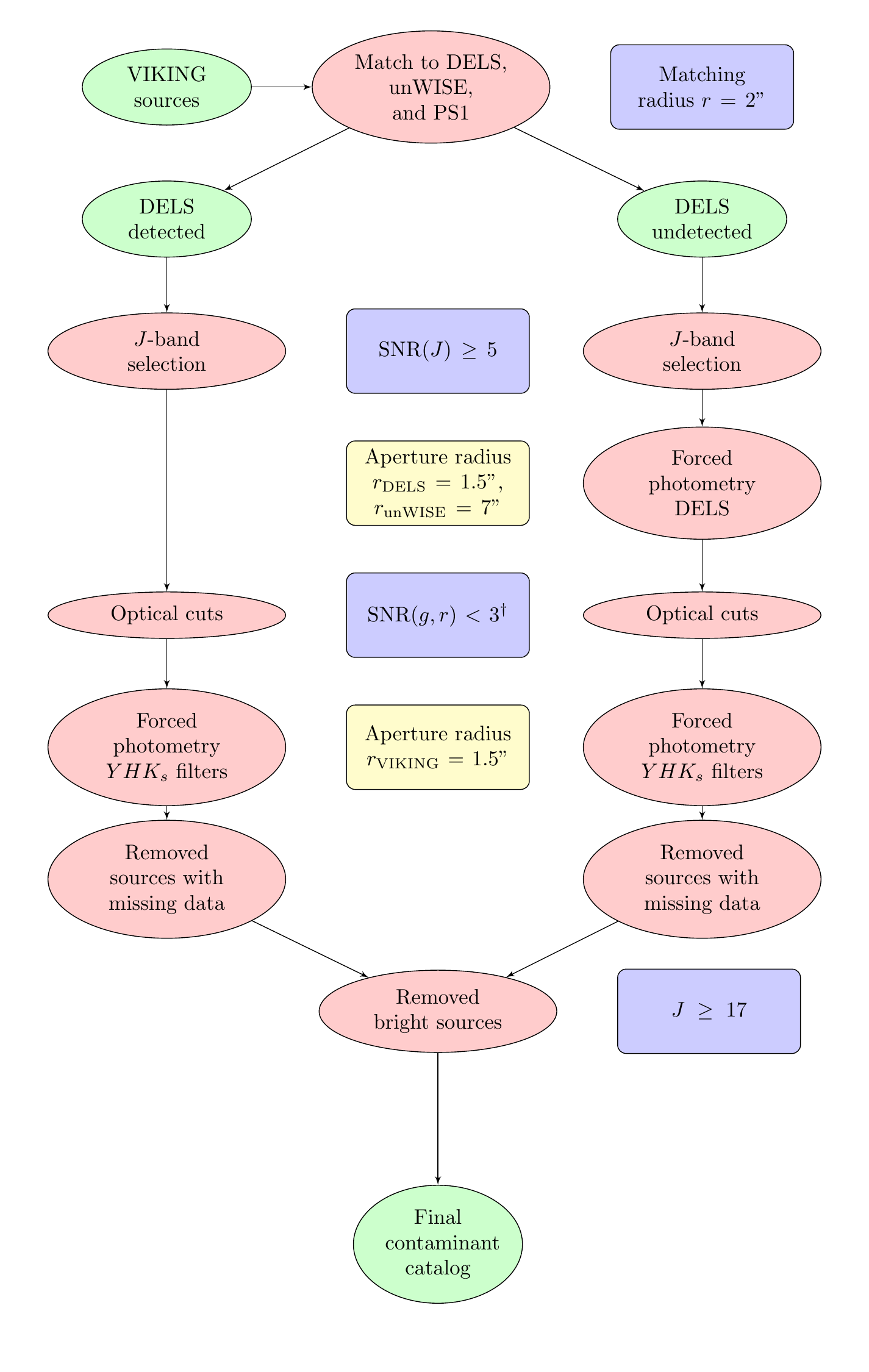}
 \caption{General steps (red ellipses) performed to construct the contaminant training sample. The blue boxes represent the conditions that the sources must satisfy to make it to the next step, while yellow boxes provide more information about some specific steps. After the first step (the match with other surveys), sources are divided into two sub-catalogs depending on their DELS counterpart: sources with a DELS detected counterpart (DELS detected), and sources with no detected counterpart but with DELS observations (DELS undetected). Sources with neither DELS counterpart nor observations are simply removed. \\$\dagger$ At this step we also removed sources with ${\rm SNR}(g_{\rm PS1},r_{\rm PS1})\ge3$, or ${\rm SNR}(i_{\rm PS1})\ge5$ and $i-z<2$, when these data are available.}
 \label{fig:flowchart}
 \end{center}
\end{figure*}

At NIR wavelengths, we used $Y$, $J$, $H$, and $K_s$ bands coming from VIKING DR4.
%% JFH2. You need to spell out the survey acronyms once in one location and then
%% use them consistently throughout. So it is weird that you redefine VIKING
%% here. 
%% RN done
The VIKING data were obtained from the VISTA Science Archive\footnote{\url{http://horus.roe.ac.uk/vsa/}}.
For optical bands, we mainly used data from the DESI Legacy Imaging Surveys (DELS)\footnote{\url{https://www.legacysurvey.org/}}, which combines three different imaging surveys: the Dark Energy Camera Legacy Survey (DECaLS), the Beijing-Arizona Sky Survey (BASS; e.g., \citealt{Zou19}), and the Mayall $z-$band Legacy Survey (MzLS). These three surveys jointly image $\sim$14,000 deg$^2$ of the extragalactic sky visible from the northern hemisphere in three optical bands ($g$, $r$, and $z$). The sky coverage is approximately bounded by $-18^{\circ} < \delta < +84^{\circ}$ in celestial coordinates, and $|b| > 18^{\circ}$ in Galactic coordinates, and it overlaps with most ($\approx80\%$) of the VIKING survey footprint. An overview of the DELS surveys can be found in \citet{Dey19}. 
When available, we also included Pan-STARRS (PS1) photometric data in our selection, which provides $3\pi$ sky coverage ($\approx70\%$ overlap with the VIKING footprint) in five different filters: $g_{PS1}$, $r_{PS1}$, $i_{PS1}$, $z_{PS1}$, and $y_{PS1}$. 
As described below, these data were used to further refine our training catalog. 
In the MIR, we used the $W1$ and $W2$ bands coming from the unWISE release \citep{Schlafly19}, that comes from the coaddition of all publicly available $3-5 \mu m$ WISE imaging \citep{Wright10}, including that from the ongoing NEOWISE \citep{Mainzer11} post-cryogenic phase mission. The steps used to construct
our catalog are illustrated schematically in Fig. \ref{fig:flowchart},
which we describe in detail in the following.

%% JFH2 Nice schematic. If possible remove the math-mode from the subscripts
%% for things like r_DELS and r_unWISE
%% RN done
To construct our contaminant training sample, we cross-matched the VIKING catalog with the DELS, PS1, and unWISE ones, using a radius $2''$. As we are interested in finding $6\le z\le 8$ QSOs, we used the $J$-band as the ``detection band''.
%% JFH2 I omitted to compute the relative fluxes as it is confusing here. 
%to compute the relative fluxes of the sources.
%% RN done
%% JFH This is the first time you mention relative fluxes. I would just say as the "detection band". 
%% RN done
In fact, at the very high-redshift ($z>7$) the Ly$\alpha$ 
%% JFH You are writing Ly_\alpha with a subscript which is not standard. Everyone writes Ly$\alpha$. I would change
%% throughout. 
%% RN done
drop falls in the $Y$-band, preventing the detection of very high-$z$ QSOs, while the VIKING $J$-band reaches a depth of 22.1 (at 5$\sigma$). We then selected all the sources with $J$-band signal-to-noise ratio ${\rm SNR}(J)\ge5$.
Since $z\ge6$ QSOs drop in the bluest optical filters, we further required our objects to have 
%% JFH SNR -- {\rm SNR}, should not be in math mode. 
%% RN done
${\rm SNR}(g,r)<3$\footnote{Sources detected in DELS have alreasy forced photometry for the DECaLS-$grz$ and the unWISE-$W1W2$ filters.}, and, when available, ${\rm SNR}(g_{\rm PS1},r_{\rm PS1})<3$. We also removed objects with ${\rm SNR}(i_{\rm PS1})\ge5$ and $i-z<2$, when these data were available. For sources covered by the DELS footprint but with no counterpart detected in the survey within $2"$, 
we performed forced photometry on the DECaLS and unWISE
%% JFH2 In a footnote or in the text remind the reader that the DELS
%% catalog has forced photometry already for all DELS detected
%% sources!
%% RN done
images with an aperture radius $1.5'',7''$, respectively, and removed all sources with ${\rm SNR}(g,r)\ge3$. For the surviving sources, we also performed forced photometry on the VIKING images ($YHK_s$ filters), using an aperture radius $1.5''$. Sources that have missing data in at least one of the requested filters (VIKING-$YHK_S$, DECaLS-$grz$, unWISE-$W1W2$) are removed from our catalog at this stage. Finally, we visually inspected bright sources ($J<17$), and found they were often artifacts or bright stars, so you decided to exclude them.
%% JFH2 The role of the visual inspection is unclear here. Makes it sound like
%% you only removed some. I think what you want to say is that you visually
%% inspected bright sources and found they were often artifacts or something
%% so you decided to exclude them. 
%% RN done
%% JFH I have no idea what " with fluxes that are five times below our survey depth in one or more bands.". 
%% Why would you remove faint sources? This does not make any sense. Avoid 
%% writing throw-away non-quantitative statements like this. Quote the survey depths. Are you using 3\sigma depths or
%% 5-sigma depths? Your reader does not know. Imaging that someone needs to reproduce your work, and so you need to state
%% all the cuts that you performed with numbers otherwise it is not reproducible. 
%% RN done
The resulting final ``contaminant'' training catalog contains 733,694 sources, while the number of sources that survived each filtering step are presented in Table \ref{tab:summary}. Among the final sources, we
%% JFH2 You might also remark that all of the known z > 6 quasars in the
%% footprint are in your list, i.e. your cuts are not losing quasars. Is
%% this true?
%% RN this is not true as in the footprint there are known QSOs that we don't select because either they are too faint SNR(J)<5 or they have missing data in one or more bands.
identified eight known $6\le z \le 7$ QSOs, indicating that the contamination of the contaminant training set with high-$z$ quasars is small. Therefore, we did not remove these known QSOs from the training set. 
%% JFH What does VIKING cross-matched mean? Did you cross match
%% before applying the SNR(J) > 5 cut? If so maybe just call the first line
%% in the Table "VIKING catalog" or something
\begin{table}
  \centering
  \captionsetup{justification=centering, labelsep = newline}
      \caption[]{Selection criterion on the ``contaminant'' training catalog}
      \begin{adjustbox}{center, max width=\textwidth}
         \begin{tabular}{l c c}
            \hline
            \hline \rule[0.7mm]{0mm}{3.5mm}
            Data sample & DELS detected & DELS undetected\\
            \hhline{~~~}
            \hline \rule[0.7mm]{0mm}{3.5mm}
            VIKING cross-matched & 55,540,784 & 21,575,809 \\
            \rule[0.7mm]{0mm}{3.5mm}
            ${\rm SNR}(J)\ge5$ & 35,061,078 & 2,538,387\\
            \rule[0.7mm]{0mm}{3.5mm}
            ${\rm SNR}(g,r)<3^{\dagger}$ & 408,974 & 397,960\\
            \rule[0.7mm]{0mm}{3.5mm}
            Sources with data in all bands & 399,898 & 335,521\\[2pt]
            \hline \rule[0.7mm]{0mm}{3.5mm}
            Total sources ($J\ge17$) & \multicolumn{2}{c}{733,694}\\[2pt]
            \hline 
         \end{tabular}
        \end{adjustbox}
        \begin{tablenotes}
      \small
      \item $\dagger$ At this step we also removed all the sources with ${\rm SNR}(g_{\rm PS1},r_{\rm PS1})\ge3$, or ${\rm SNR}(i_{\rm PS1})\ge5$ and $i-z<2$, when these data are available.
    \end{tablenotes}
         \label{tab:summary}
\end{table}

\subsection{Quasar Data}\label{sec:train_qso}
%% JFH Is the McGreer code quasar model magnitude dependent? You never state if it is or not. Also, it is rather 
%% unclear if you classifying the quasars in magnitude bins similar to the contaminants or how that was done. Generally
%% it feels like their are lots of omissions of information regarding how the quasars were modeled. 
We used a sample of 440,000 $6\le z\le 8$ QSOs simulated from the ``simqso'' code from \citet{McGreer13}, using the updated version described in \citet{Yang16}. The simqso code was used to generate a grid with a uniform distribution in redshift over the range $6\le z\le8$, and in magnitude over the range $17\le J\le 22.5$.
Assuming that the QSO spectral energy distributions (SEDs) do not evolve with redshift (\citealt{Khun01}; \citealt{Yip04}; \citealt{Jiang06}; \citealt{Banados18}), the quasar spectrum is modeled as a power-law continuum with a break at 1200 \AA. For redder wavelength coverage, we added four 
%% JFH Did you add these new breaks, or were they already in the code used by the references you cite?
%% RN Some breaks were added after discussing with JT
breaks at 2850, 3645, 6800, and 30000 \AA. The slope ($\alpha_{\lambda}$) from 1200 to 2850 \AA$\:$ follows a Gaussian distribution with mean $\mu(\alpha_{1200}) = - 0.5$ and dispersion $\sigma(\alpha_{1200}) = 0.3$; the range from 2850 to 3645 \AA$\:$ has a slope drawn from a Gaussian distribution with $\mu(\alpha_{2850}) = - 0.6$ and $\sigma(\alpha_{2850}) = 0.3$; from 3645 to 6800 we adopted a
Gaussian with $\mu(\alpha_{3645}) = 0.0$ and $\sigma(\alpha_{3645}) = 0.3$; finally, from 6800 to 30000, we used $\mu(\alpha_{6800}) = 0.3$ and $\sigma(\alpha_{6800}) = 0.3$. The parameters of emission lines are derived from the composite quasar spectrum from
%% JFH from the composite quasar spectrum from Glikman et al. 
%% RN done
\citep{Glikman06}, and the lines are added to the continuum as Gaussian profiles, where the Gaussian parameters (wavelength, equivalent width, and full with half maximum) are drawn from Gaussian distributions. These distributions recover
trends
in the mean and scatter of the line parameters as a function
of continuum luminosity, e.g., the Baldwin effect (\citealt{Baldwin77}), and blueshifted lines (\citealt{Gaskell82}; \citealt{Richards11}). The simulator also models absorption from
%% JFH The simulator also models absorption from ....
%% RN done
from neutral hydrogen absorption in Ly$\alpha$ forests based on the work of \citet{Wor_Pro11}.
The final noiseless photometry of simulated QSOs is derived from the model spectra by integrating them against the respective filter curves. 
%% JFH from the model spectra by integrating them against the respective filter curves. 
%% RN done
%To avoid singular inverse variances for the effectively noiseless model data, we added a tiny error (0.01) to the simulated noiseless relative fluxes, using a gaussian distribution, and used for consistency this small value of the error as the input error on the pohtometry in the XD code.  
%% JFH fitting issues is vague. Just say something like to avoid singular inverse variances for the effectively 
%% noiseless model data, you added a tiny error and used for consistency used this small value of the error as the
%% input error on the pohtometry in the XD code. 
%% RN done
%in the limit where there is no error during the XD deconvolution stage, we added a tiny error (0.01) to the simulated noiseless relative fluxes, using a gaussian distribution.

%% JFH you need to 1) explain why are modeling fluxes whereas the convention is colors, 2) explain why you choose to 
%% use relative fluxes (following Bovy). This would be a good moment to explain then the error covariance issue and 
%% provide that formula. The main problem is that you are quoting relative fluxes, but you never explicitly tell the reader
%% how that is defined or why you do that. 
%% RN move the following section in the next paragraph

\section{XDHZQSO density model} \label{sec:dens_model}
To estimate the density of contaminants and quasars in flux space (the $p(\{f_i\}|O\in A)$ factor from Eq. \ref{eq:bayes}), we used the XDGMM\footnote{\url{https://github.com/tholoien/XDGMM}} implementation of extreme deconvolution from
\citet{Holoien17}. XDGMM is a \textit{python} package that utilizes the
scikit-learn API (\citealt{scikit-learn}; \citealt{sklearn_api}) for 
%% JFH2 add a citation for sklearn
%% RN done
Gaussian mixture modeling. It performs density estimation of noisy, heterogenous, and incomplete data and uses the XD algorithm\footnote{\url{https://github.com/jobovy/extreme-deconvolution}} (\citealt{Bovy11}) for fitting, sampling, and
determining the probability density at new locations.
%% JFH2 Can we omit this next sentence. I don't think it is required. 
%% RN done
%In principle, XD can handle also missing data by assigning large errors to the missing properties, but we do not need it since we performed forced photometry on non-detected sources and removed those with missing data.
%%JFH I have no idea what this last sentence means. "The XD feature" is not defined anywhere. I think you may be
%% referrng to missing data but am confused. 
%% RN done
As described by \citet{Bovy11}, XD models the underlying, deconvolved, distribution as a sum of $N$ Gaussian distributions, where $N$ is a model complexity parameter that needs to be set using an external objective. It assumes that the flux uncertainties are known, as is in our case, and consists of a fast and robust algorithm to estimate the best-fit parameters of the Gaussian mixture. In \S \ref{sec:model_construction} we follow the approach used by \citet{Bovy11} to
%% JFH2 What are you describing? This sentence is awkwardly constructed? Is
%% the description your text, or the description the model?
%% RN done
construct the flux density model of the two classes.

Finally, to compute the probability of an object belonging to a certain class, we need to estimate the number counts of both quasars and contaminants (the $P(O\in A)$ factor from Eq. \ref{eq:bayes}): i.e., these are the prior factors of our Bayesian approach. For the contaminants, we compute this factor empirically from the number counts ($J$-band magnitude distribution of contaminants), while for the quasars we derived them from the high-$z$ QSO luminosity function. However, to derive the true number counts for the QSOs, which includes the survey incompleteness at the faint end, we used the empirical data to compute the incompleteness for the VIKING survey, and apply it to the QSO number counts. In \S \ref{sec:priors} we provide details about the computation of
these prior factors.

%% JFH2 COnsider adding more subsections to this section. There is a
%% lot going on in this section.
%% RN done

\subsection{The binning approach}\label{sec:binning}

%% JFH2 What is flux density? It is never defined. I think you mean
%% probability density. Which term in equation 1 are you referring to
%% here as the density? It would be help the reader if you used
%% consitent terminoology and also refrred to the math you wrote down
%% in previous sections. Provide that reference below.
%% RN done
The full model consists of fitting the probability density (the $p(\{F_i\}|O\in A)$ factor from Eq. \ref{eq:bayes}) in a number of bins in $J$-band magnitude for the two classes of objects. We opted to bin in $J$-band
because the probability density
%% JFH2 flux density -- probability density 
%% RN done
of quasars will have a dominant power-law shape corresponding to the number counts as a function of apparent magnitude, whereas the color distribution is much flatter. While the latter can be represented well by mixtures of Gaussian distributions, the power-law behavior cannot without using large numbers of Gaussians.
Thus the slow variation of the color distributions with magnitude is captured
by our model, since we use narrow bins in $J$-band magnitude.

\begin{figure}
 \begin{center}
 \includegraphics[height=7cm, width=9cm]{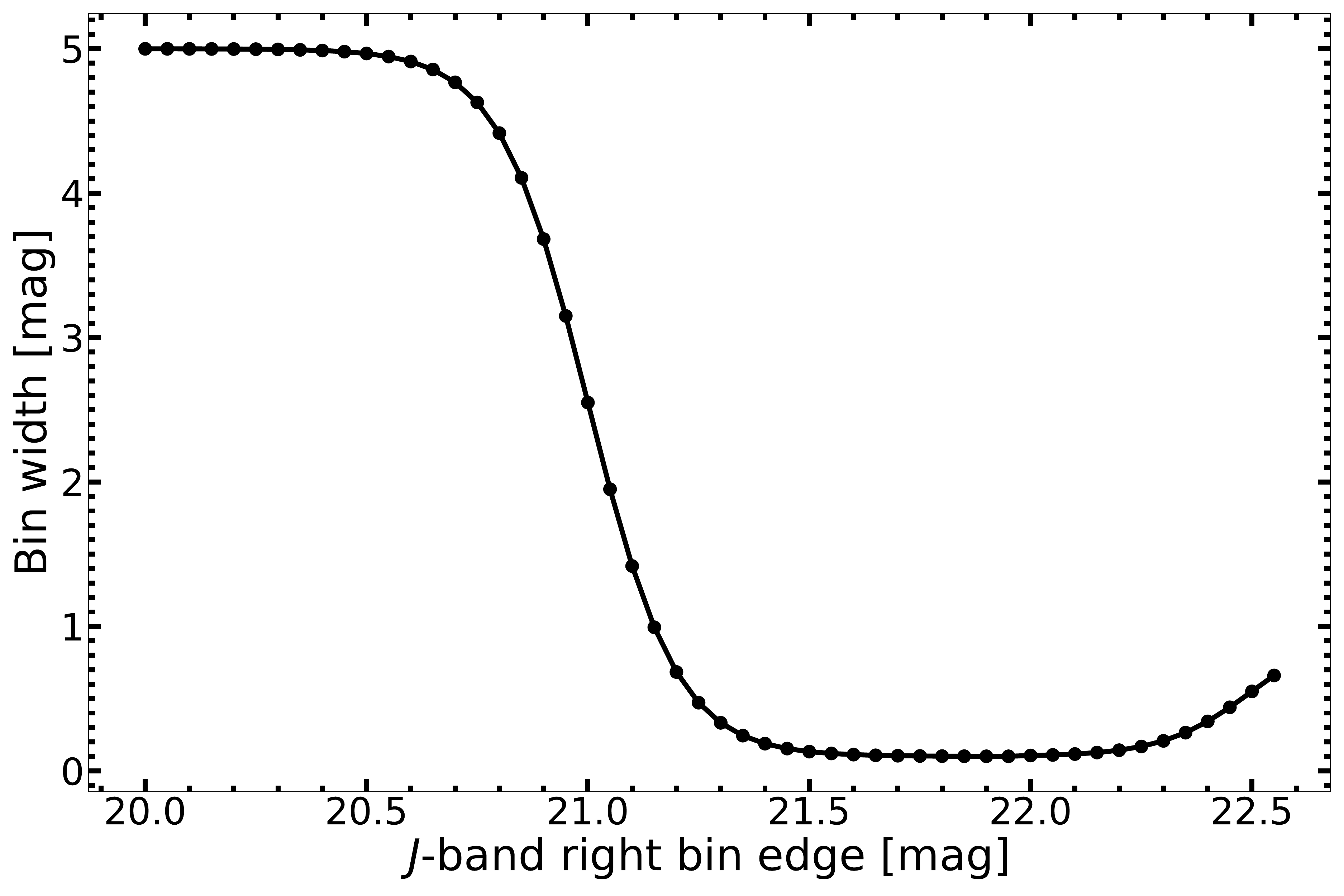}
 \caption{Double sigmoid function that displays the right edges and the width of the bins used to train the contaminant model.}
 \label{fig:sigmoid}
 \end{center}
\end{figure}

The full contaminant model consists of 50 overlapping bins where the right edges are uniformly distributed in the range $J=20-22.5$ with a step of 0.05 mag, while the width is given by a broken sigmoid function:
%%JFH2 Remove math mode from subscripts please!
%% RN done
\begin{equation}
    \begin{aligned}
    w &= bw + (bs_1 - bw) \frac{1}{1+e^{\frac{J_{\rm bre}-m_{\rm th_1}}{\Delta m}}} \quad {\rm for} \quad J_{\rm bre}\le 22,\\
    w &=bw + (bs_2 - bw) \frac{1}{1+e^{\frac{J_{\rm bre}-m_{\rm th_2}}{\Delta m}}} \quad {\rm for} \quad J_{\rm bre}> 22
    \end{aligned}
\end{equation}
where, $J_{\rm bre}$ is the $J$-band bin right edge, $bw=0.1$ represents the
minimum bin width
%% JFH2 minimum what?
%% RN done
and $bs_1=5$, $bs_2=1$ represent the maximum bin widths in the two $J$-band ranges, $m_{th_1}=21$, $m_{th_2}=22$, and $\Delta_m=0.1$.
The broken sigmoid for the contaminants is shown in Fig. \ref{fig:sigmoid}. The use of a variable bin width is driven by the need of having a model that is as continuous as possible, as the XD fits can jump between local maximums. In fact, this procedure guarantees that many ($>20\%$) of the objects in the bins overlap for adjacent bins, and thus the model varies smoothly. Furthermore, the use of a broken sigmoid guarantees that both at the bright and faint ends, where fewer objects are present, the bins are large enough to contain a sufficient number of sources. In fact, we have $>2000$ training objects in each bin to build the contaminant models.

%% JFH2 This seems confusing. Your data goes to J = 17, but your quasar
%% model cuts off at 19? This needs some explaining. 
%% RN, there was a type, done
As for the quasar model, we used 11 uniform spaced bins with a width 0.5 mag in the range $J=17-22.5$, and we further divided the quasar class into three subclasses corresponding to ``low-redshift'' ($6\le z\le 6.5$), ``medium-redshift'' ($6.5\le z\le 7$), and ``high-redshift'' ($7\le z\le 8$) quasars, constructing a QSO model for each bin. We opted to divide the QSO into these three redshift bins, instead of working with a broad $6\le z\le 8$ bin, for the following reasons: 
\begin{enumerate}
    \item As shown in \S \ref{completeness}, the efficiency and completeness of our selection method strongly depends on the $z$-bin in question owing to the changing overlap between quasars and contaminants.
    \item While the $6\le z\le 7$ range has been largely investigated in the past, few objects have been found at $7\le z\le 8$, making it the highest priority range that we are interested in investigating.
    \item Spectroscopic wavelength coverage is different for different
      instruments, with the dividing line between optical and near-IR
      spectrographs typically occurring around  $\approx10,000$ \AA$\,$
      ($z_{Ly\alpha}=7.2$). Thus, not all the $6\le z\le 8$ QSOs can simply be
      confirmed with a single instrument, and multiple instruments could
      be required to confirm candidates over such a broad redshift range.
      Hence, the redshift bins we adopted also facilitates in efficiently
      conducting follow-up observations.
\end{enumerate}
However, in the future we plan to introduce the redshift as one of
the modelled quantities as done by \citet{Bovy12}, so that one would no
longer needs to construct models in different redshift bins, as this approach
also provides photometric redshifts, which can be used to
select candidates over any desired redshift range. 

\subsection{Construction of the model}\label{sec:model_construction}
The XD code fits for all the $J$-band magnitude bins for a given class are initialized using the best-fit parameters for the previous bin, so to guarantee the continuity of the mode.
%% JFH2 Add a sentenece again saying that this guarantees continuity of the
%% the model. 
%% RN done
The starting bin (the one that is not initialized) is the closest to $J=21$, where we know we always have a quite large sample of objects ($>10^4$) for the training. Hereafter, we describe the model in a single bin first for a single example class, using the contaminant class as the example.
%% JFH Provide some more motivation for this variable bin width. The point is that we want the model to be as continuous
%% as possible, whereas the EM algorithm inherent to the XD fits can jump between local maxima. So this procedure
%% guarantees lots of the objects in the bins overlap for adjacent bins and thus model varies smoothly. You might also 
%% explain that the sigmoid is because at the bright end the bins needs to be larger to give sufficient numbers of sources. 
%% RN done

%% JFH Indicate units of bin width on the y-axis. 
%% RN done

In a single bin in $J$-band magnitude, we separate the absolute flux from the flux relative to the $J$-band in the likelihood in Eq. \ref{eq:bayes} as follows:
\begin{equation}\label{eq:prob_dens}
  \begin{aligned}
    %% JFH2 Please don't write text in mathmode. cont should
    %% be {\rm cont}. Please look over the text and see to it these changess
    %% are made everywhere. 
    %% RN done
    p(\{f_i\}|O\in``{\rm cont.}")=p(\{f_i/f_J\}|f_J,O\in``{\rm cont.}") \\
    \times p(f_J|O\in ``{\rm cont.}"),
\end{aligned}
\end{equation}
%% JFH2 Previously you used f_i for the fluxes now you switch to
%% f_x. Please try to use consistent notation everywhere. Either make
%% equation 1 f_x, or make this equation here and all that follow f_i
%% RN done
where $\{f_i\}$ are the $z$, $Y$, $H$, $K$, $W1$, $W2$ fluxes, $\{f_i /f_J\}$ are the fluxes relative to $J$-band, and $f_J$ is the $J$-band flux.
We model the two factors of the right hand side of Eq. \ref{eq:prob_dens} separately.

\begin{figure*}
 \begin{center}
 \includegraphics[height=17cm, width=17cm]
 {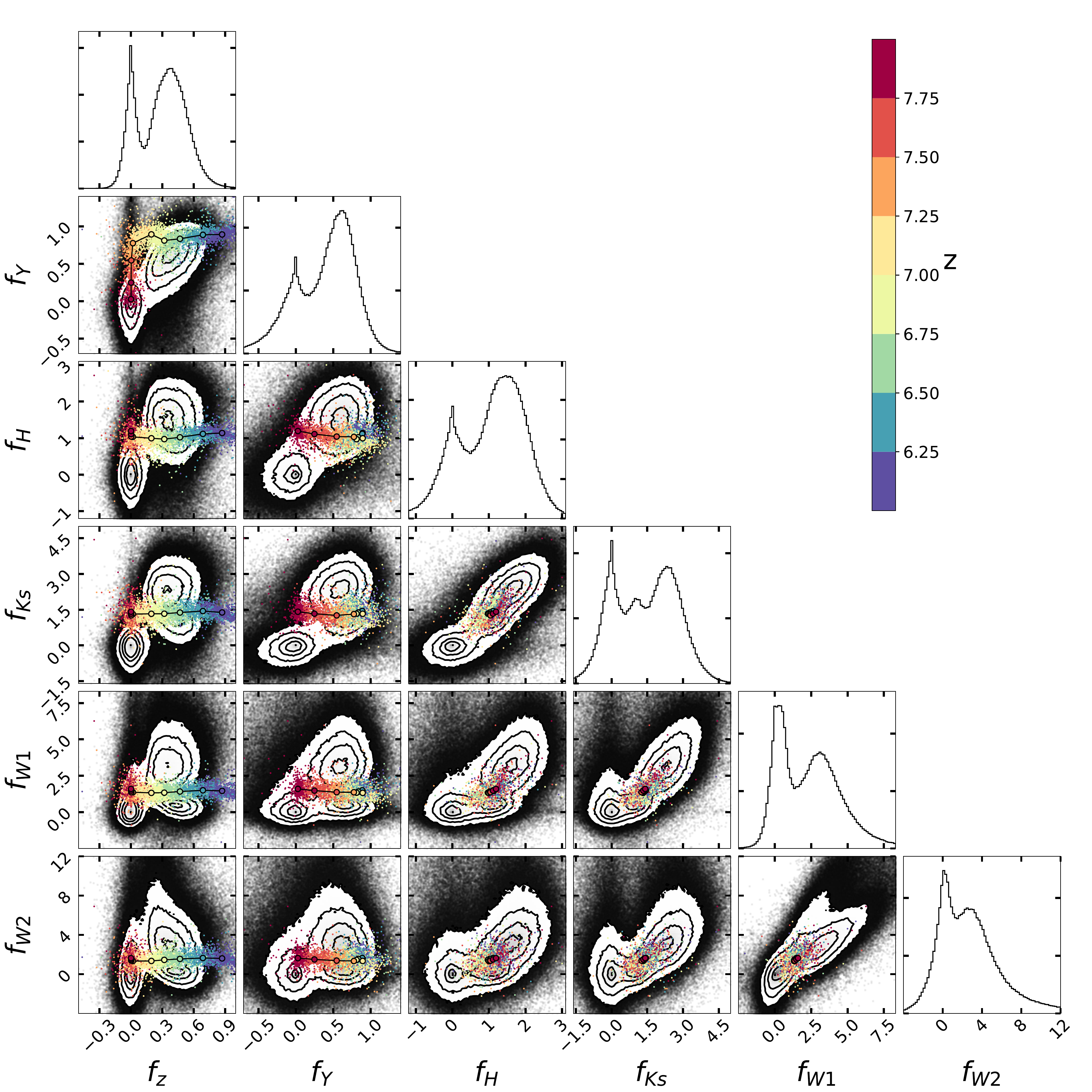}
%% JFH Calling this a "J-band relative flux plot" is a bit confusing. Call it a relative flux plot, and somewhere in 
%% in the paper define them (see my comments in the text). 
%% RN done
 \caption{Relative-flux plots for both the contaminant (black contours and points) and a sub-sample (5000) of high-$z$ QSO training data (coloured points). The color-bar shows the redshift of the simulated QSOs, while the labelled quantities are relative fluxes (i.e., fluxes in different bands divided by the $J$-band flux). For display purposes, we added to the simulated noiseless quasars the real errors coming from our contaminant catalog as explained in Appendix \ref{appendixa}, while the black line and colored filled circles represent the color-redshift relation predicted using our simulated QSOs. Although we do not know the real nature of our contaminants, we expect that most of them are cool brown dwarves and early type galaxies.}
 \label{fig:color_color}
 \end{center}
\end{figure*}

%% JFH Consider moving this earlier, since the reader needs to have read this to understand Figure 1, which arises
%% earlier in the text than this section. 
We modeled the $p(\{f_i/f_J\}|f_J,O\in``{\rm cont.}")$ 
%% JFH refrain from saying "first" and "second" factor. Just plug in the equation in math mode. It is more clear, 
%% especially since you don't provide eqn. references. 
%% RN done
factor using XD in narrow bins in $J$-band magnitude.
%% JFH I think the discussion of the bins belongs here. Not sure why you postpone to below. 
%% RN done
We use relative fluxes rather than colors since the observational uncertainties are closer to Gaussian for relative fluxes than they are for colors. Also, for sources where the flux measurement can be negative the magnitudes are badly behaved, while relative fluxes remain well behaved in this case. To evaluate the XD probabilities during training, we always convolved the underlying model with the object's relative-flux
uncertainties assuming that they are Gaussian distributed, such that the convolution of the Gaussian mixture with the Gaussian
uncertainty results in another Gaussian mixture.
Although the
ratio of two noisy Gaussian deviates is not itself Gaussian distributed,
Gaussianity is a good approximation provided that the $J$-band flux errors are
small. The validity of this approximiation is discussed further in Appendix \ref{appendixb}.  Note also that since all other fluxes are divided by the $J$-band
flux, the resulting uncertainties are covariant, and we provide the functional
form of this covariance matrix in  \ref{appendixb}.
To train for the QSO models, since the simulated quasar fluxes are noiseless, we simply need to fit their flux densities without deconvolving to derive the underlying deconvolved quasar model.
However, to avoid singular inverse variances for the effectively noiseless model data, we added a tiny error (0.01) to the simulated noiseless relative fluxes
drawn from a Gaussian distribution, and used for consistency this small value of the error as the input error on the photometry in the XD code. In Fig. \ref{fig:color_color} we show the relative-flux relative-flux diagrams of our training data: the contaminants are displayed using black contours, while a sub-sample (5000) of simulated $6\le z \le 8$ QSOs are shown as coloured points. For display purposes, we added to the displayed quasars real errors drawn from
a noise model based on our contaminant catalog which is described in Appendix \ref{appendixa}.

%% JFH2 Usually appendices are in the order in which they are referenced in the
%% text. 
%% RN done

%% JFH I would add a paragraph or a few sentences explaining the basic features of the flux-flux plots somewhere in 
%% in the text. Here is a reasonable place.  The main point is that readers have not really looked at these before so you  %% need to give them some intuition for what dropouts look like etc. In other words describe the basic features of the 
%% contaminants if you can (might need to look at WISE papers for this, but is there a stellar locus? galaxy locus) and 
%% then describe the properties of the quasar tracks. 

%% JFH Change the limits on the color-bar to dislay z =6 and z =8. this can be done by making the limits i.e. 5.99 and 8.01. %% It is also weird to me that the tick mark on your color-bar is not centered within the color. Perhaps that is something
%% you can adjust. 

%% JFH2 We desperately need subsections in this mega-section. 

We modeled the six-dimensional relative fluxes $\{f_i /f_J\}$, using 20 Gaussian components. The number 20 was chosen after performing XD fits with 10, 15, 20, and 25 components. While fits with less than 20 components overly smoothed the observed distribution, models with more than 20 components used the extra components to fit extremely low significance features in the observed distribution. The same number of components was also adopted by \citet{Bovy11}. Similarly, we also used 20 Gaussian components to fit for the quasar models. 

\begin{figure*}
  %% JFH2 How are the 1d histograms normalized? Maybe you should choose
  %% all subsampls to integrate to unity in these panels. What you
  %% currently have looks weird. 
  %% RN done
 \begin{center}
   \includegraphics[height=17cm, width=17cm]{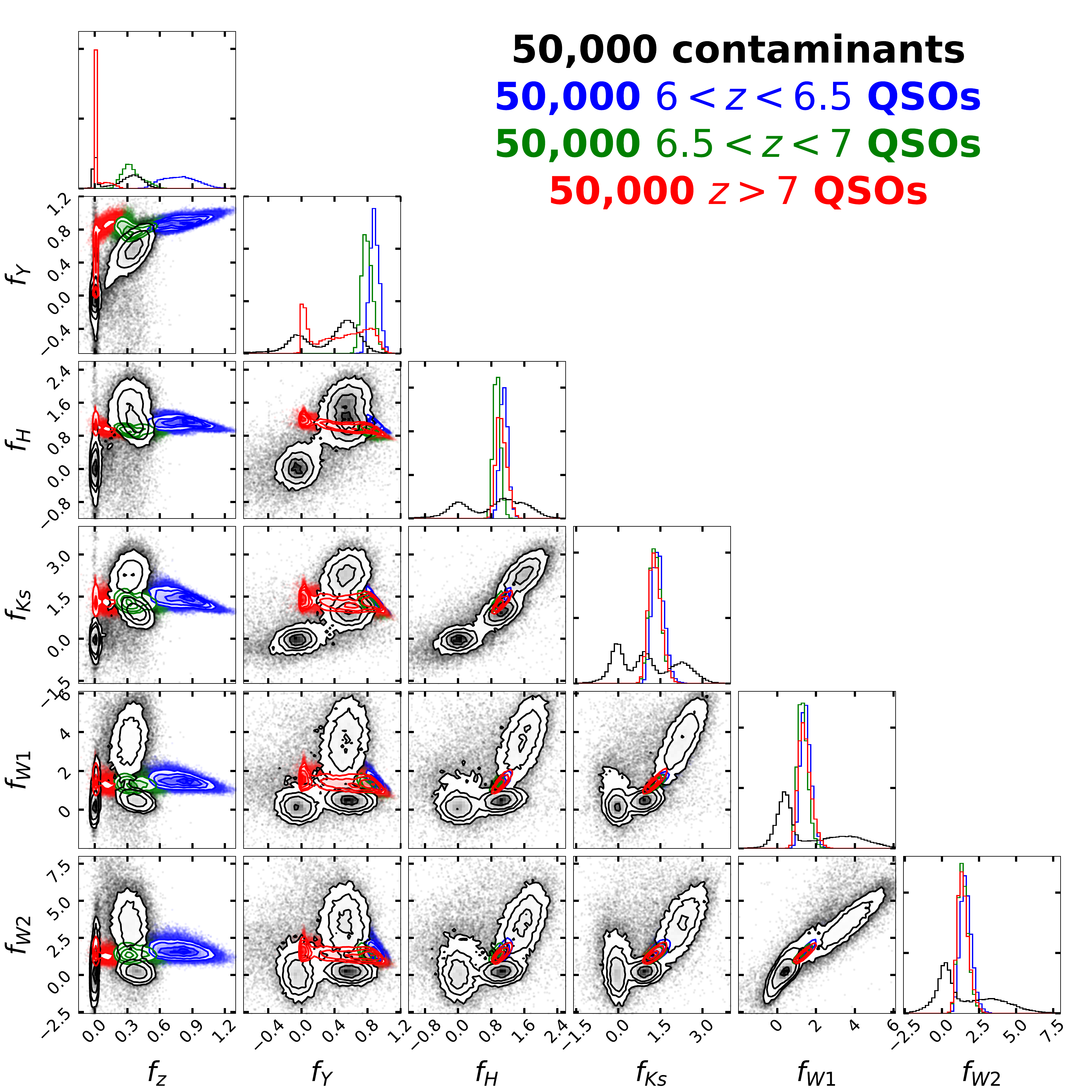}
   %% JFH2 Replace all occurences of flux-flux with relative-flux relative-flux
   %% RN done
   
 \caption{Relative-flux relative-flux contours for the $J=20.67-21.2$ deconvolved contaminant model (black), and for the deconvolved $J=20.5-21.0$ $6\le z\le 6.5$ (blue), $6.5\le z\le 7$ (green), and $7\le z \le8$ (red) QSO models. The labelled quantities are relative fluxes (i.e., fluxes in different bands divided by the $J$-band flux). To generate the displayed samples, we sampled 50,000 sources from the contaminant model, and 50,000 objects from each of the three QSO models.}
 \label{fig:model_comparison}
 \end{center}
\end{figure*}

\begin{figure*}
 \begin{center}
 \includegraphics[height=17cm, width=17cm]
 {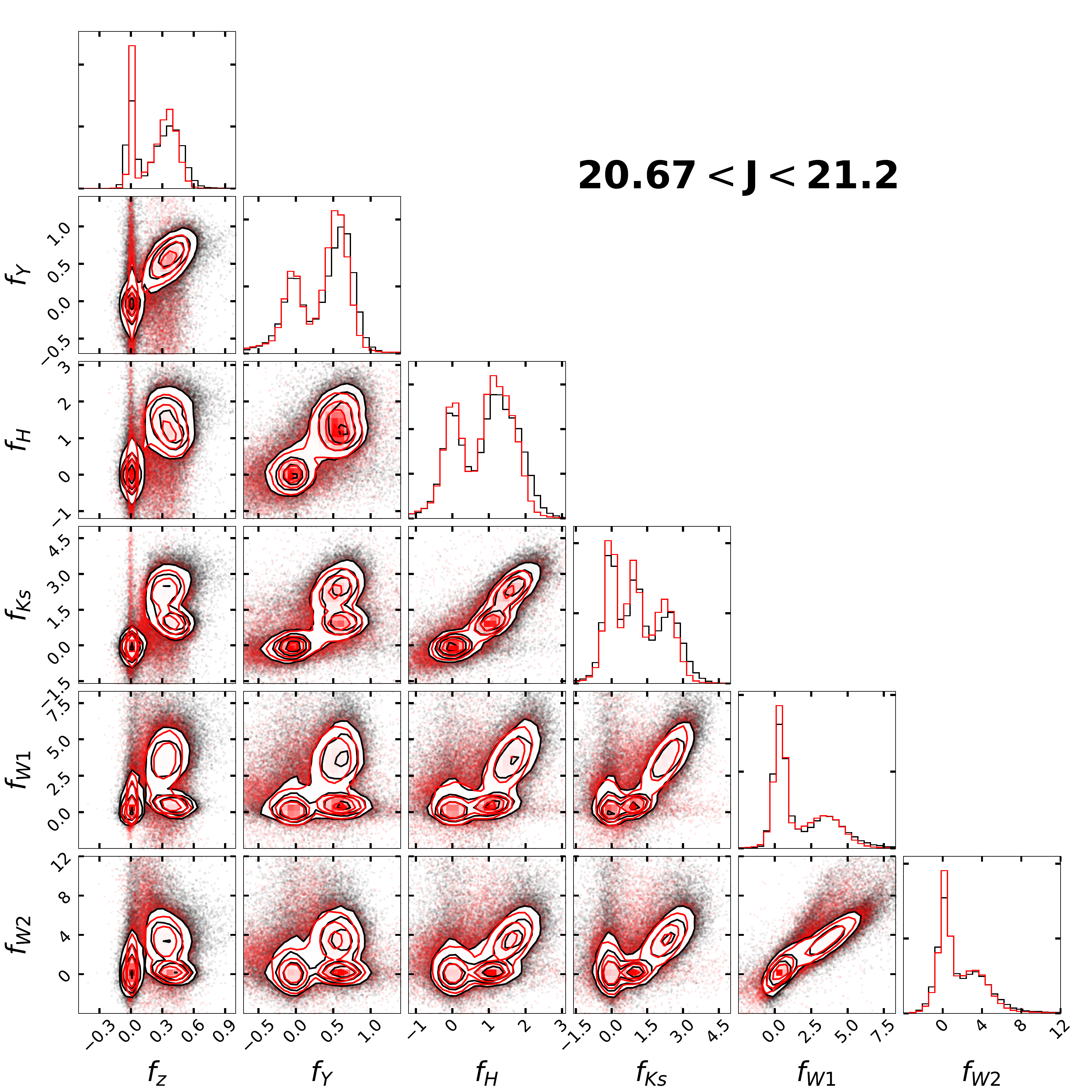}
 \caption{Relative-flux relative-flux contours for the deconvolved $20.67< J< 21.2$ contaminant model (red), compared to the real data distribution (black). The labelled quantities are relative fluxes (i.e., fluxes in different bands divided by the $J$-band flux). Overall, the red contours are tighter compared to the black ones showing the efficacy of XDHZQSO in deconvolving the nosy distributions.}
 \label{fig:deconvolved}
 \end{center}
\end{figure*}

\begin{figure*}
 \begin{center}
 \includegraphics[height=17cm, width=17cm]
 {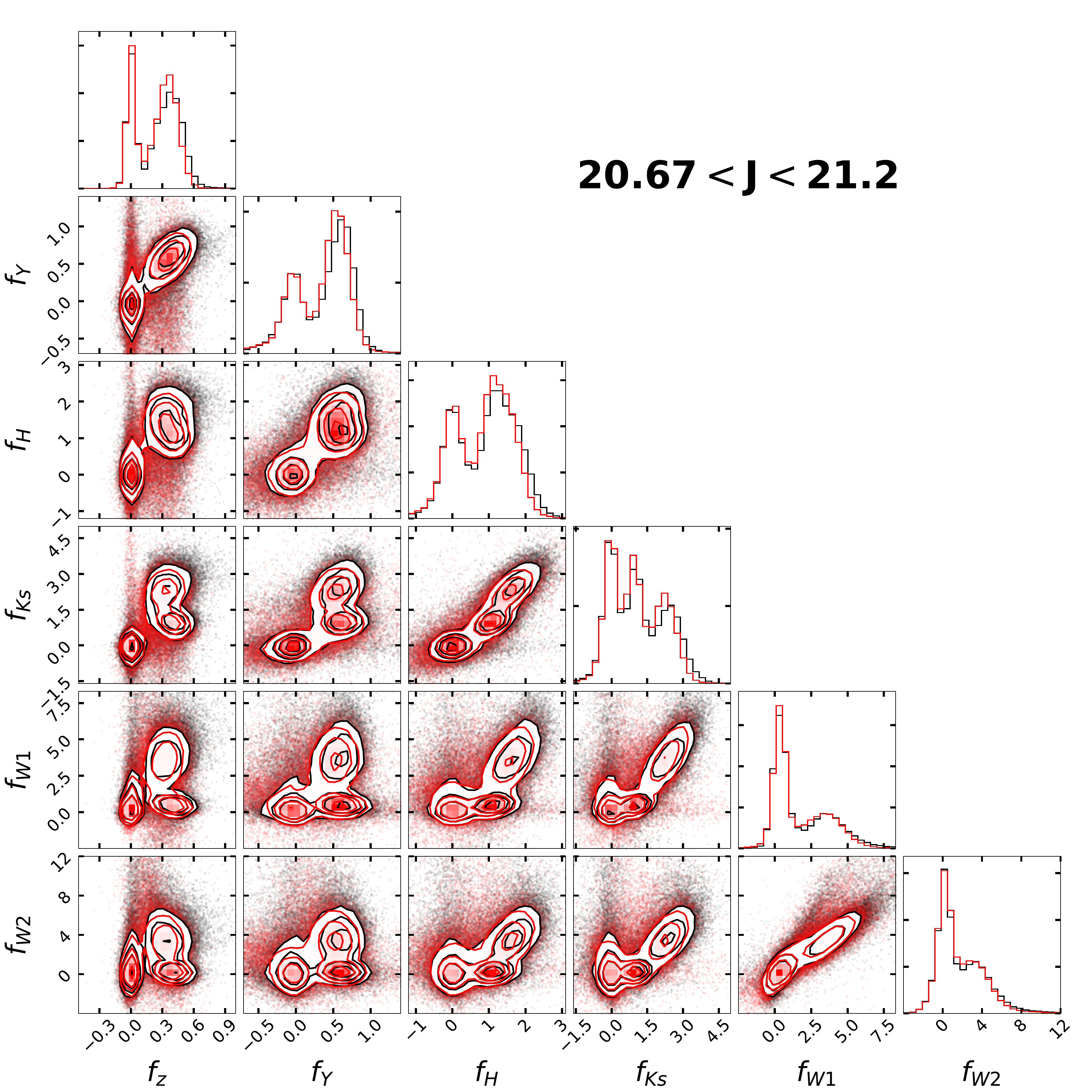}
 \caption{Relative-flux relative-flux contours for the noise added deconvolved $20.67< J< 21.2$ contaminant model shown in Fig. \ref{fig:deconvolved} (red), compared to the real data distribution (black). Errors have been added as explained in Appendix \ref{appendixa}. The labelled quantities are relative fluxes (i.e., fluxes in different bands divided by the $J$-band flux). It is apparent that, after re-adding the errors to the deconvolved quantities, we obtain a distribution that is consistent with the $20.67< J< 21.2$ real data.}
 \label{fig:convolved}
 \end{center}
\end{figure*}

To provide a visual example of the model generated by the XDHZQSO code, 
%% JFH Not sure why you call it the XD deconvoultion results. Deconvolution is part of the story, but you are illustrating
%% the quality of the models. 
%% RN done
we display in Fig. \ref{fig:model_comparison} the $20.67< J< 21.2$ deconvolved contaminant model (black contours) compared to the $20.5< J< 21.0$ QSO models in the three redshift bins: $6\le z \le 6.5$ (blue), $6.5 \le z \le 7$ (green), and $7 \le z \le 8$ (red). To generate the displayed samples, we drew 50,000 sources from the deconvolved contaminant model, and 50,000 objects from the three redshift-bins deconvolved QSO models.
%% JFH2 Are you plotting samples from the quasar model or the real simulated
%% quasars? Please clarify
%% RN done
It is apparent that the large overlap between the contaminant and the $6.5 \le z \le 7$ and $7 \le z \le 8$ QSO contours will greatly lower the efficiency in selecting QSO candidates in these two redshift ranges, as better explained in \S \ref{completeness}. To asses the quality of our contaminant deconvolved models, we sampled the deconvolved models in each $J$-band bin\footnote{For each bin we sampled a number of sources equal to the number of real VIKING sources from that bin.}, re-added the errors to the deconvolved fluxes following our noise modeling procedure described in Appendix \ref{appendixa}, and compared the relative-flux distribution of the reconvolved sample with the original real noisy data. In Fig. \ref{fig:deconvolved} we compare a simulated set of samples (red contours) from the deconvolved $20.67< J< 21.2$ contaminant model with the real data distribution (black), while in Fig. \ref{fig:convolved} we compare the same simulated sample after adding the errors, following Appendix \ref{appendixa} (red), with the real contaminant distribution (black). It is apparent that, after re-adding the errors to the deconvolved quantities, we obtain a distribution that is identical with the $20.67< J< 21.2$ real data.
%% JFH2 Figure 5 and Figure 6 just don't look that different. You can
%% hardly see the impact of deconvolution. Is that because the errors
%% are small? Is there another bin with larger errors (but without the
%% problem at really large errors) that makes this more clear?
%% RN unfortunately not, done
\subsection{Computation of the priors}\label{sec:priors}
\begin{figure*}
 \centering
 \includegraphics[height=6.5cm, width=8.5cm]{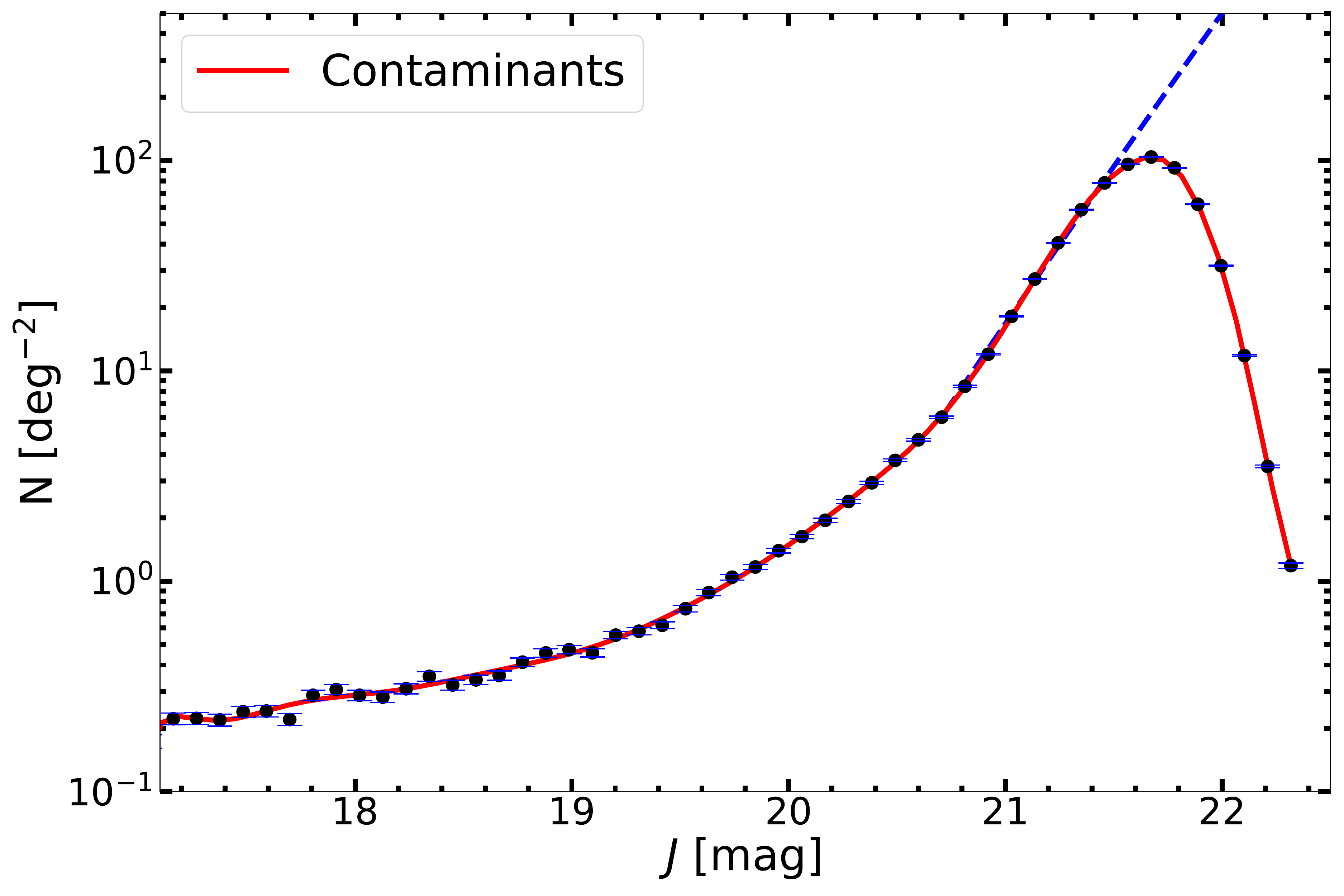}
 \includegraphics[height=6.5cm, width=8.5cm]{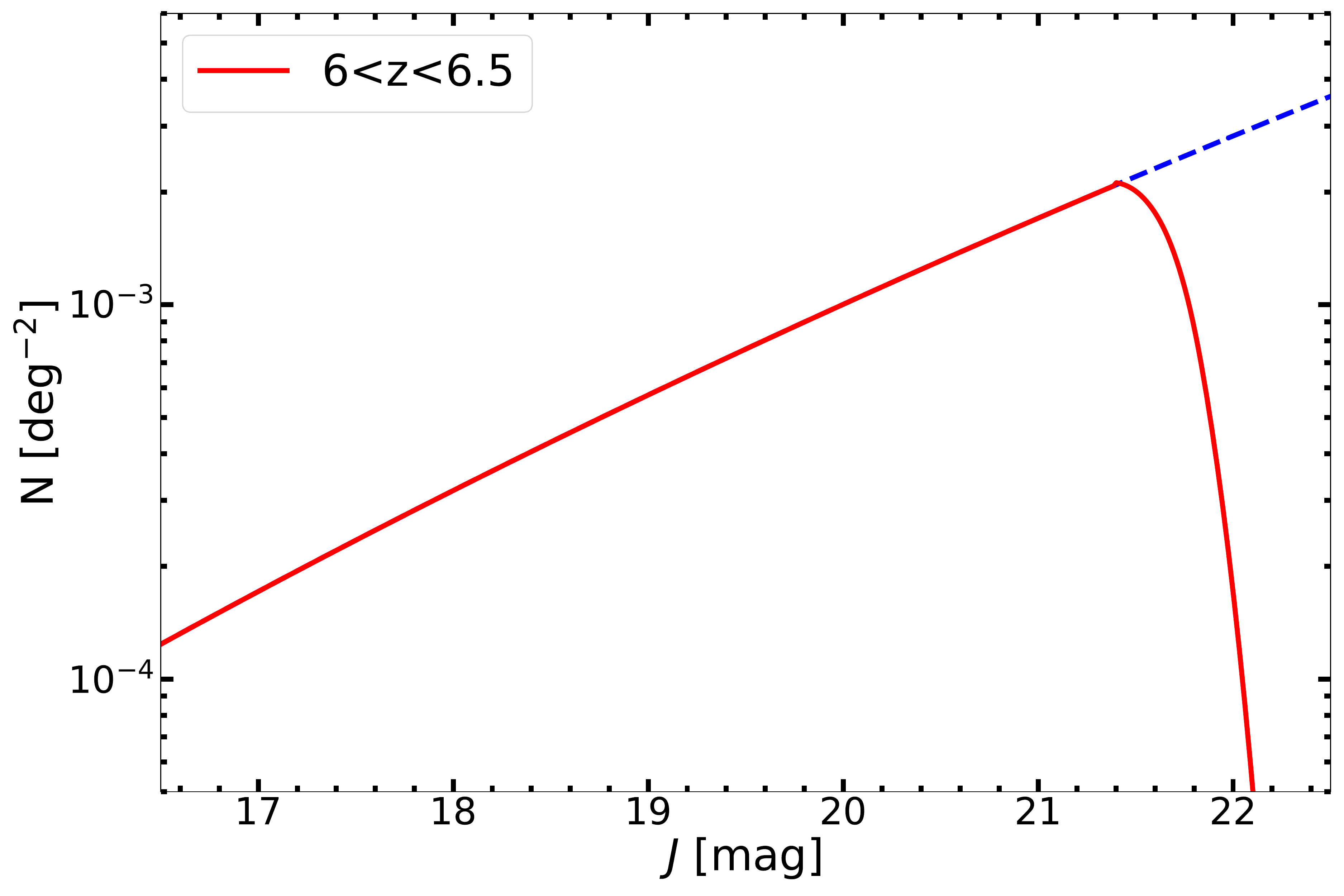}
 \includegraphics[height=6.5cm, width=8.5cm]{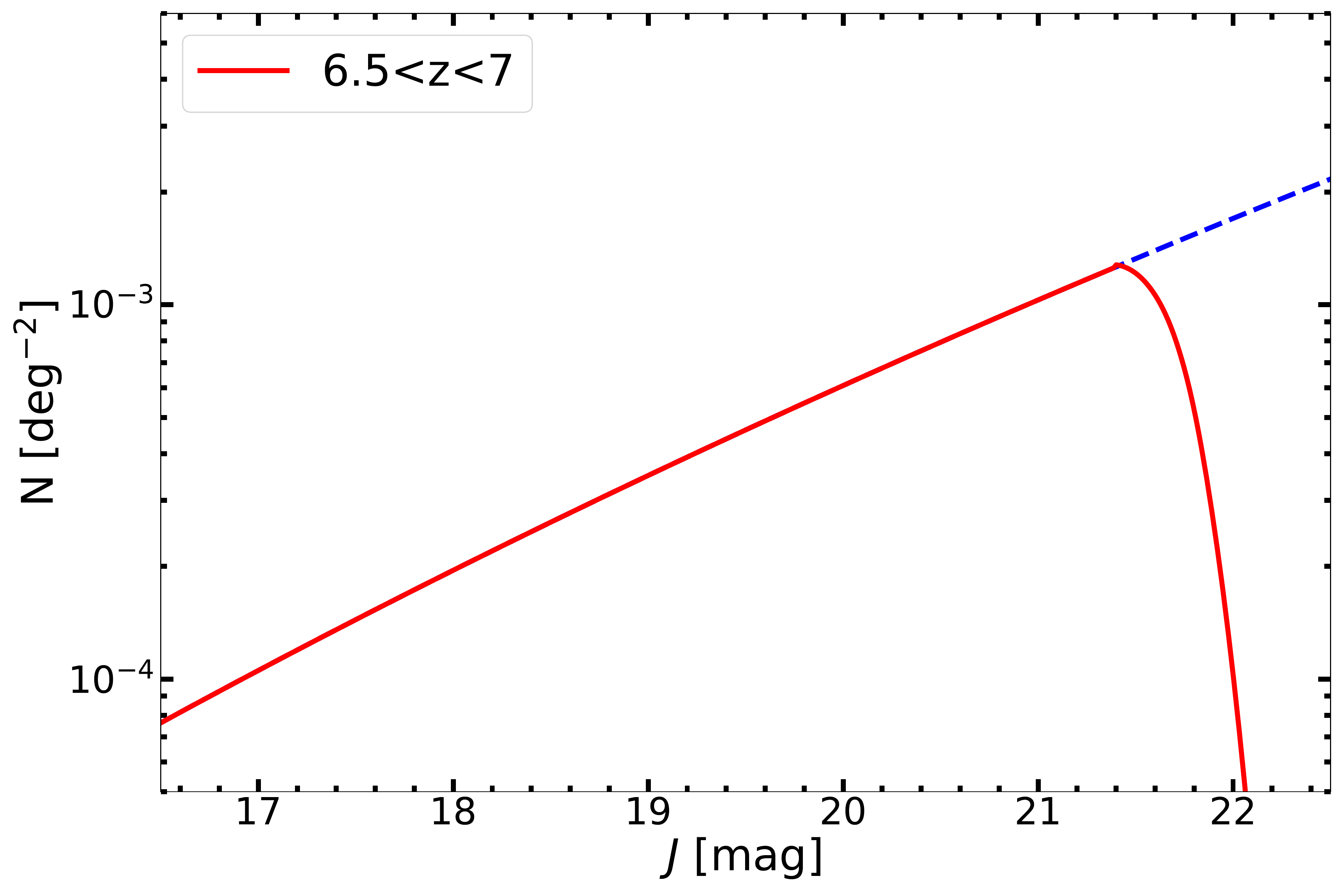}
 \includegraphics[height=6.5cm, width=8.5cm]{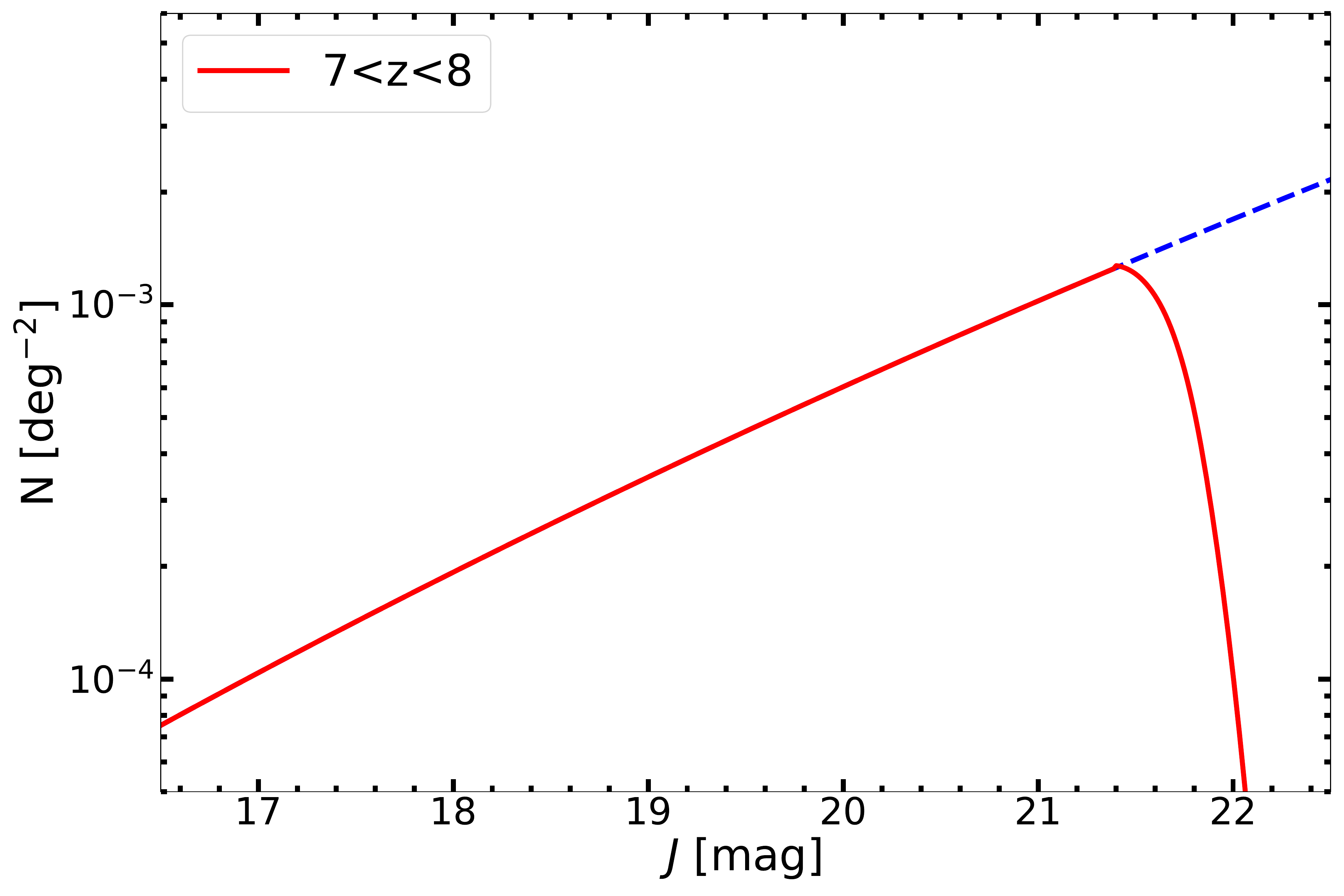}
 \caption{Number counts $p(f_x|O\in ``class'')$ priors for the contaminant (top left panel), and the $6\le z \le 6.5$ (top right panel),  $6.5\le z \le 7$ (bottom left panel), $7\le z \le 8$ (bottom right panel) QSO classes as a function of the $J$-band magnitude. In the top left panel, the black points are the real contaminant data from the VIKING survey, while we used a 40-order polynomial to perform a robust fit to the range $J\le 21.4$, and at $J> 21.4$ we used a cubic spline to interpolate the histogram, namely to capture the drop-off due to catalog incompleteness (red line). To model the effect of the incompleteness on the real data distribution, we fit a power-law to the range $20.7 \le J \le 21.4$, and extrapolated it to $J > 21.4$ (blue dashed line). The ratio between the value given by the power-law and the cubic spline interpolated number counts gives us the incompleteness correction term to apply to our QSO number counts at $J>21.4$. The 1$\sigma$ Poissonian errors are shown as short blue lines. The other three QSO panel show the the $z\sim6.7$ quasar LF from \citet{Wang19}, after the inclusion of the incompleteness (red line). The extrapolation of the LF at $J>21.4$ without the incompleteness correction is shown as a blue dashed line.}
 \label{fig:prior_dist}
\end{figure*}

The second factor of Eq. \ref{eq:prob_dens}, $p(f_J|O\in ``{\rm cont.}")$
%% JFH2 Cont should not be in math mode
%% RN done
represents the number counts of contaminants (or quasars) as a function of apparent magnitude, and is always expressed in units of ${\rm deg}^{-2}$. 
%% JFH provide equation references to avoid ambiguity. 
%% RN done
%% JFH It is unclear what "combining with the number factor means."
%% RN rephrased
%% JFH this should be ${\rm deg}^2$. Change throughout. 
%% RN done
For the contaminant class, we modeled the number counts directly using the number counts of the training data, by fitting the histogram of $J$-band magnitude number counts per square degree. 
%% JFH I'm confused, your results look too smooth to be interpolation? 
%% JFH You are confusing interpolation and fitting in a very confusing way. 
%% RN I know, it is just because we have the same verb in Italian (interpolate) to indicate both interpolation and fit. Btw, I corrected it
We used a
%% JFH2 40-order really? Seems extreme, but let's not change anything now. 
%% RN done
40-order polynomial to perform a robust fit to the range $J\le 21.4$, while
%\begin{equation}
%\begin{split}
%    & f(J) = c J^{\alpha_1} \quad for \quad J<J_{th1},\\
%    & f(J) = c J_{th1}^{\alpha_{1}-\alpha_{2}}J^{\alpha_{2}} \quad for \quad J_{th1}\le J<J_{th2},\\
%    & f(J) = c J_{th1}^{\alpha_1-\alpha_2} J_{th2}^{\alpha_2-\alpha_3} J^{\alpha_{3}} \quad for \quad J_{th2}\le J \le 21.4.
%\end{split}
%\end{equation}
%where $J_{th1}=20.6$, $J_{th2}=18.5$ are manually fixed, while we derived: $log(c)=-100.3$ for amplitude, and $\alpha_1=76.7$, $\alpha_2=44.9$, $\alpha_3=20.2$ for the slopes in the three $J$-mag ranges.
%% JFH I don't know what a triple power-law fit is. Write out the equation. Quote the values you obtained for the 
%% power laws from the fit. Explain how you did the fit, i.e. least squares, weighted/unweighted?
%% RN done
at $J> 21.4$ we used a cubic spline to
%% JFH Explain why you switched to spline, i.e. namely to capture the dropoff due to catalog incompleteness. State
%% if you used cubic spline
%% RN done
interpolate the histogram, namely to capture the drop-off due to catalog incompleteness. In order to model the effect of the incompleteness on the real data distribution, we fit a power-law to the range $20.7 \le J \le 21.4$:
\begin{equation}
  %% JFH2 For should not be in math mode!
  %% RN done
   f(J) = c J^{\alpha} \quad {\rm for} \quad 20.7\le J\le21.4
\end{equation}
where $log(c)=-95.3$ and $\alpha=73.0$, and extrapolated this power law fit
to $J > 21.4$. The ratio between the value given by the power-law and the cubic spline interpolated number counts gives us the incompleteness correction term to apply to our QSO number counts at $J>21.4$. 
We show in Fig. \ref{fig:prior_dist} (top right panel) the $p(f_J|O\in ``cont.")$ factor.

%% JFH2 Be more specific in the title sentence
%% that you are talking about the prior
%% RN done
For the ``quasar'' class, 
we used a model for the $z\sim6.7$ quasar luminosity function (LF) from \citet{Wang19}
%% JFH I think you should just put the equation here rather than in 5.5. It is not good writing style to forward cite
%% things and equations. Better to cite things you previoiusly discussed. 
%% RN done
to compute the number density of quasars as a function of the apparent $J$-band magnitude, in the three redshift 
%% JFH This is your first mention of these redshift bins. Whereas in your plots in Fig.2 they already showed up. You 
%% probably want to mention that you use three bins somewhere or here. 
%% RN done
bins ($6\le z\le 6.5$, $6.5\le z\le 7$, and $7\le z\le 8$). 
This LF is characterized by a double power-law:
\begin{equation}\label{eq:LF}
    \Phi(M_{1450},z)= \frac{\Phi^{*}(z)}{10^{0.4(\alpha+1)(M_{1450}-M^{*})}+10^{0.4(\beta+1)(M_{1450}-M^{*})}}
\end{equation}
%% JFH M is not defined anywhere. This should be M_{1450}
%% RN done
where $M_{1450}$ is the absolute magnitude at 1450 \AA, $\alpha$ and $\beta$ are the faint-end and bright-end slopes, respectively, $M^*$ is the characteristic magnitude, and $\Phi^*(z) = \Phi^*(z = 6) \times 10^{k(z-6)}$ is the normalization, where $k = -0.72$ as measured by \citet{Jiang16} for $5 < z < 6$ QSOs.
We fixed the four parameters to the $z\sim6.7$ LF measured by \citet{Wang19}: $\alpha=-1.9$, $\beta=-2.54$, $M^*=-25.2$, and
%% JFH2 log shold be \log in latex. And is this log baed 10 or log base e?
%% RN done
$\log_{10}(\Phi^*)=-8.5$.
To express the LF as a function of $J$-band apparent magnitude we convert the $M_{1450}$ to $J$-band magnitude using the distance module and the $k$-correction from \citet{Richards06}:
\begin{equation}\label{eq:kcorr}
  %% JFH2 math mode for the log --> \log. Is this log 10 or loge? One is ln,
  %% the other is \log_10
  %% RN done
  k{\rm-corr} = -2.5(1+\alpha_{\nu})\log_{10}(1+z) - 2.5*\alpha_{\nu}\log_{10}\left(\frac{145[\rm nm]}{1254[\rm nm]}\right)
  %% JFH2 no math mode for nm
  %% RN done
\end{equation}
where $\alpha_{\nu}=-0.5$.
Then, we multiplied in  the survey incompleteness at $J>21.4$ that we computed from the contaminant distribution.
We show in Fig. \ref{fig:prior_dist} the
%% JFH2 Fix the math mode for quasar, it should be normal text. 
%% RN done
$p(f_J|O\in ``6\le z\le 6.5 \; {\rm quasar}")$ factor (top right panel), the $p(f_J|O\in ``6.5\le z\le 7 \; {\rm quasar}")$ factor (bottom left panel), and the $p(f_J|O\in ``7\le z\le 8 \; {\rm quasar}")$ factor (bottom right panel).

%%JFH2 So I don't know what the points represent on the quasar plots and you
%% don't say in your caption. Given that you compute this all by integrating
%% the LF, I don't think the points with error bars have any value
%% per-se, but I'm not sure what you are plotting. It might be simpler
%% to just show all three LF curves alone on one plot, however it might
%% be messy with the incompleteness on there. 
%% RN done

%% JFH Is there a reason not to just show all three on this plot?
%% RN done

\section{high-$z$ QSO selection} \label{sec:qso_selection}

In this section we present the XDHZQSO source classification for all the objects
%% JFH2 selected by our initial cuts described in section XX
%% RN done
selected by our initial cuts described in \S \ref{sec:train_star}. 
This catalog was also used to train the contaminant model as described in \S \ref{sec:dens_model}, since we argued that the fraction of high-$z$ QSOs contained in this catalog
is negligible. Using the models of quasar and contaminant deconvolved relative fluxes, we computed the probability that every object is a high-$z$ QSO or a contaminant using Eq. \ref{eq:prob_dens}.
Specifically, we used the models from the previous section as follows. For an object with
$J$-band magnitude $J$, we first found the bin whose midpoint is the closest to this magnitude. Then, we used this bin to evaluate the relative-flux density $p(\{f_x/f_J\}|f_J,O\in``cont.")$ 
%% JFH put in the equation here for p
%% RN done
for this object’s relative fluxes by convolving the underlying 20 Gaussian mixture model with the object’s uncertainties. This uncertainty convolution is simply adding the object’s uncertainty covariance to the intrinsic model covariance for each component.
%% JFH maybe say covariance to be more precise 
%% RN done

Finally, we evaluated the number density as a function of the object's apparent  magnitude in $J$-band, using the interpolated relations described in \S \ref{sec:model_construction}. We did this for each of the classes (contaminant and the three quasar classes) and compute the probabilities using Eq. \ref{eq:bayes}.
%% JFH What do you mean by "normalize the probabilities". This sounds like your doing something extra, which you are not??
%%RN done
%% JFH put in the equation into the sentences above that you evaluate in this way. Provide eqn numbers etc. Overall the
%% referencing of equations is everywhere confusing and ambiguous. 
%% RN done

\begin{figure*}
 \centering
 \includegraphics[height=7.5cm, width=9cm]{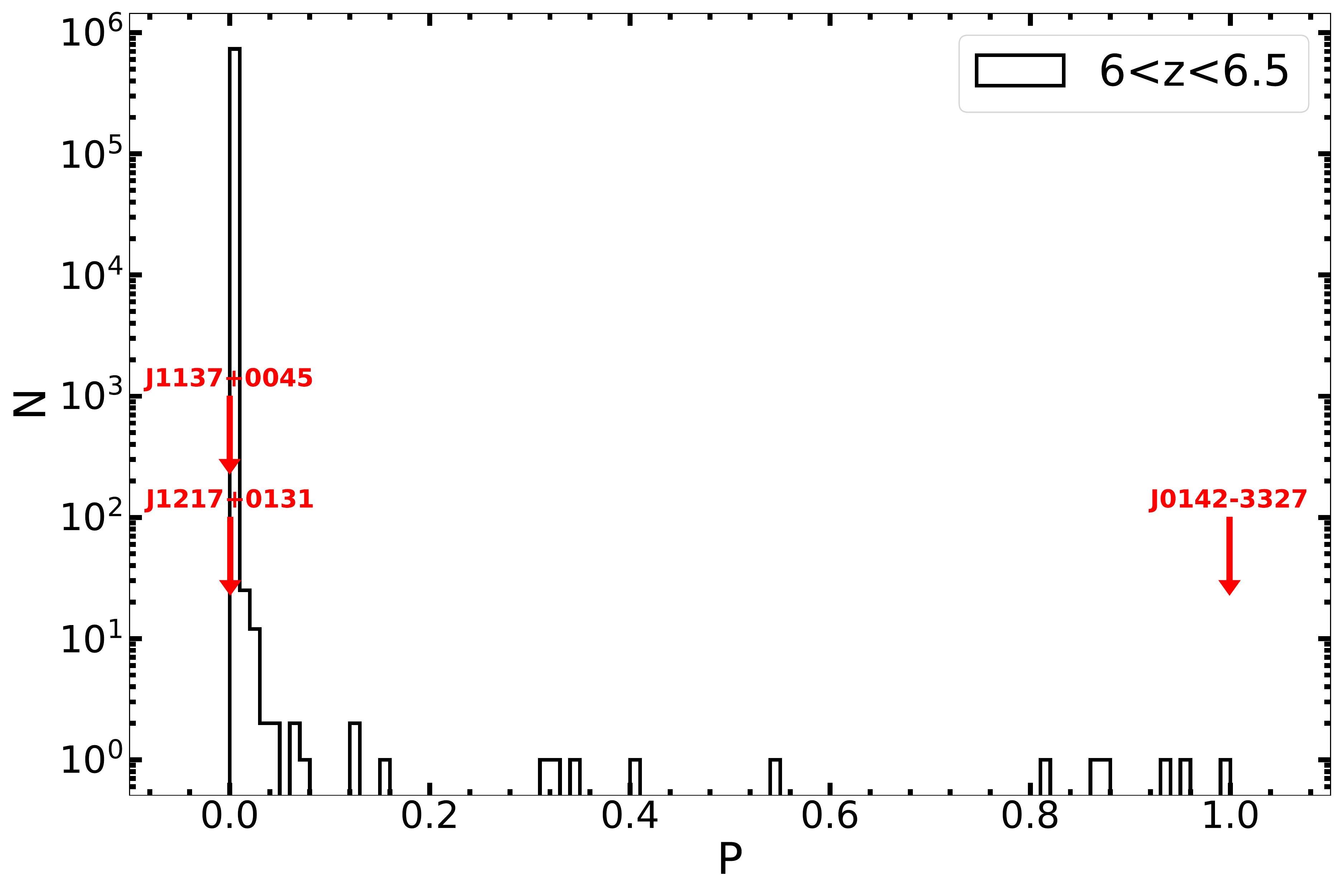}
 \includegraphics[height=7.5cm, width=9cm]{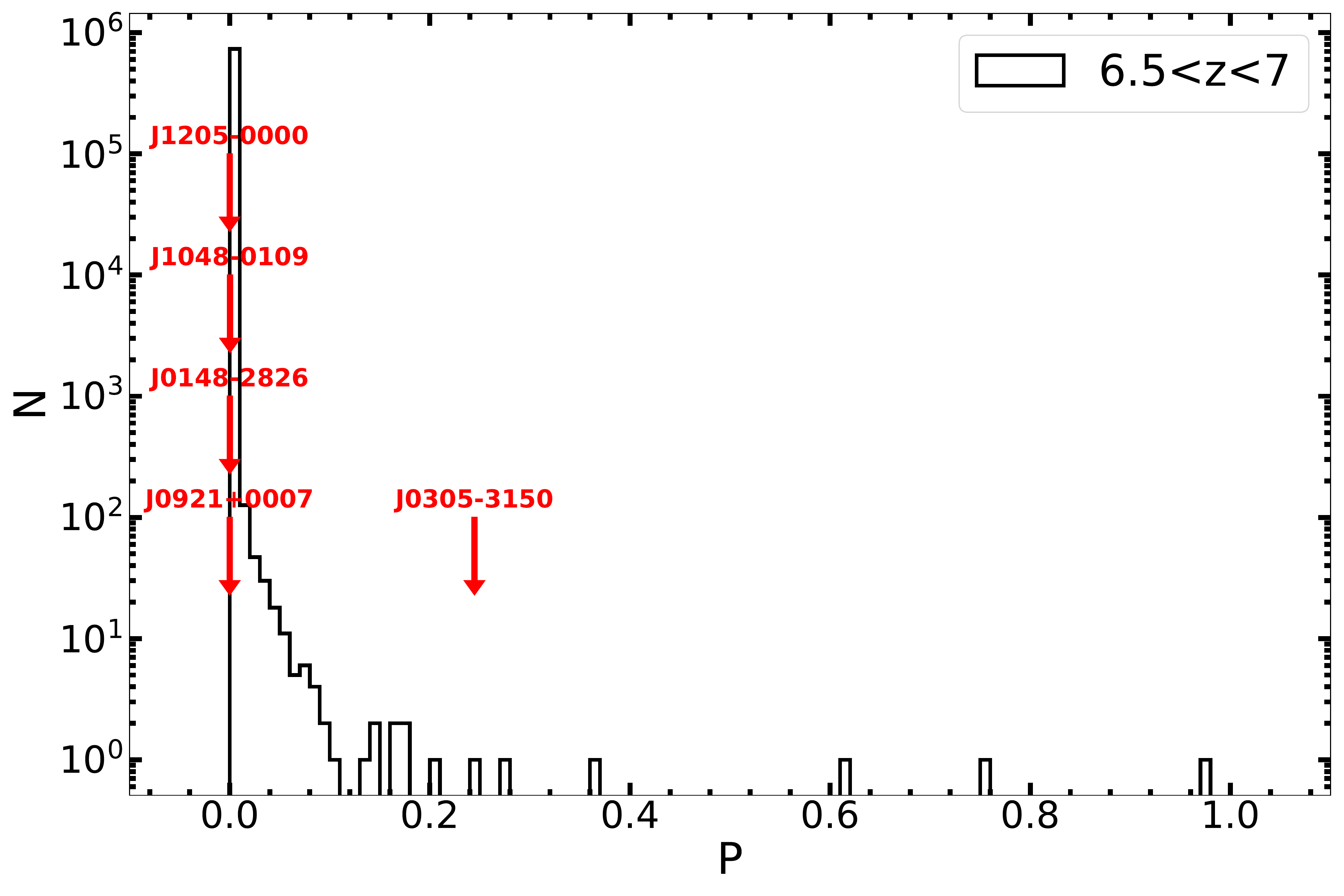}
 \includegraphics[height=7.5cm, width=9cm]{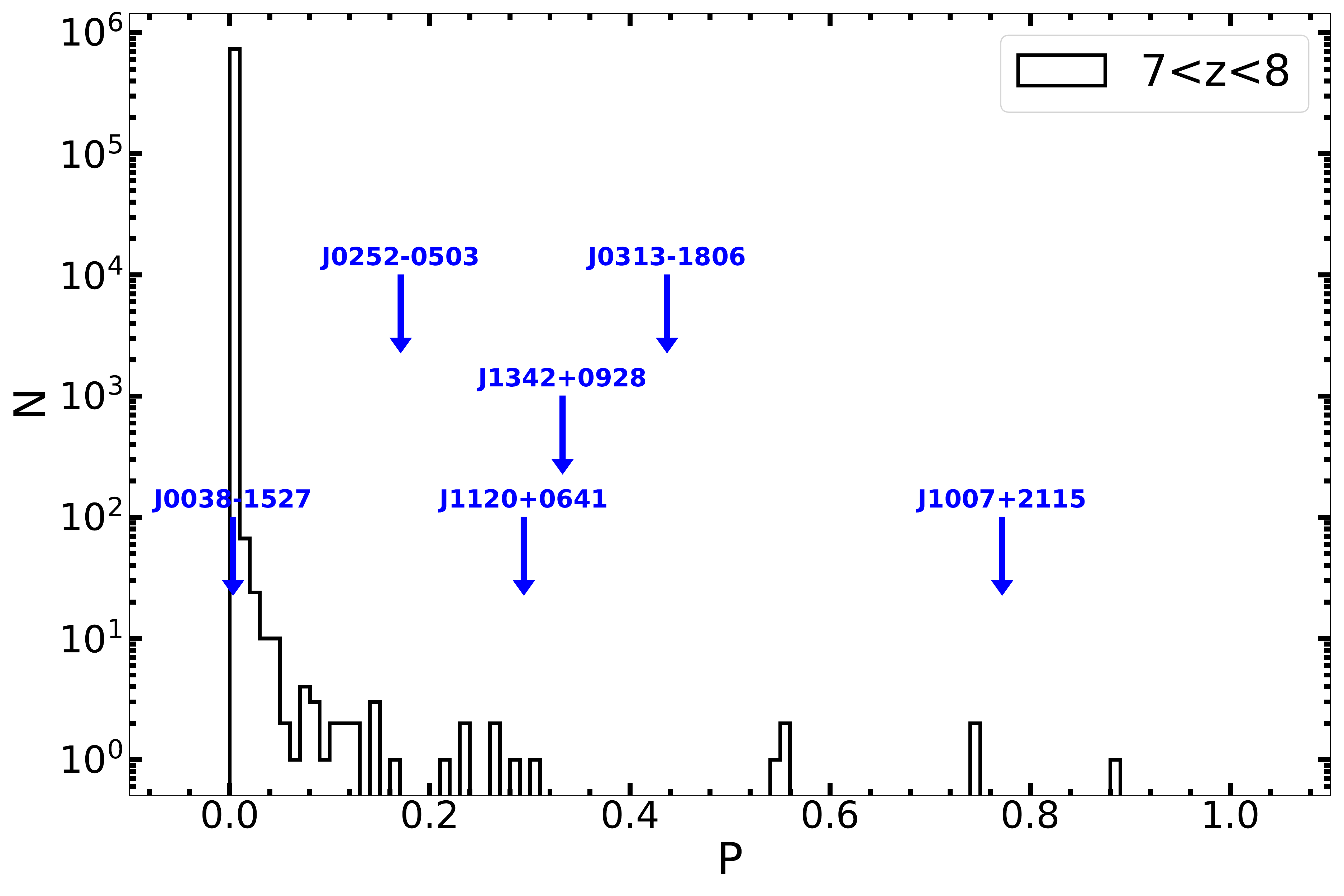}
 \caption{Probability distributions of sources from our VIKING candidate catalog in three different redshift ranges: $6\le z \le 6.5$ (top), $6.5 \le z \le 7$ (central), and $7 \le z \le 8$ (bottom). This catalog has also been used to train the contaminant models, as most of these sources are expected to be contaminants. The downward red arrows highlight the probability of known high-$z$ QSOs in the VIKING survey area. Candidates with $P\sim 0$ are pinpointed with arrows plotted on top of each other. In the bottom panel, downward blue arrows highlight the probability of known $z>7$ QSOs in the entire sky.}
 \label{fig:probabilities}
\end{figure*}

In Fig. \ref{fig:probabilities}, we show the distribution of XDHZQSO quasar probabilities for the sources we classified in the VIKING survey area. 
Since the catalog is expected to contain mostly contaminant sources, the probability distribution is peaked at zero in each redshift bin, with a few exceptions at higher probabilities that represent our best candidate quasars for future spectroscopic confirmation. It is also apparent that the number of the best candidates
%% JFH how are these "best candidates" defined, provide a rough probability threshold to show the reader what you are 
%% referring to. 
%% RN done
for spectroscopic follow-up 
%% JFH followups is not a word -- follow-up
%% RN done
(i.e. those with $P_{\rm QSO}>0.1$) decreases as the redshift increases. This
results from the combination of two factors: 1) the number density of QSOs decreases as redshift increases, 2) the overlap in the relative-flux-relative-flux
space between the higher-$z$ QSOs and the contaminants is larger, in particular in the $6.5\le z \le 7$ range (see Fig. \ref{fig:model_comparison}).
%% JFH Can you label the arrows in these figures with the source name? I think that is easy and would help in interpreting
%% Fig. 12. 
%% RN done

\subsection{Completeness and efficiency computation}\label{completeness}

To select high-$z$ 
QSO candidates for spectroscopic follow-up confirmation, we defined a probability threshold ($P_{\rm th}$) 
that effects a balance between contamination and
completeness: this threshold should be small enough to avoid missing many high-$z$ QSOs, so that the sample completeness is high, and it should be large enough to keep the number of contaminants low to increase the efficiency of the selection method. From a practical perspective, the completeness can be seen as a proxy for
the expected fraction of recovered high-$z$ QSOs as a function of the probability threshold, in a certain sky area, while the efficiency is a proxy for 
the expected spectroscopic confirmation efficiency of the candidates at the telescope.

%%JFH2 Added a new paragraph here. 
%% RN done
The completeness (C) is defined as:
%% JFH use {\rm }for subscripts in math mode that are not mathematical symbols. 
%% RN done
\begin{equation}\label{eq:comp}
%% JFH = ---> \equiv for a definition 
%% RN done
  C\equiv \frac{N_{\rm Q}(P\ge P_{\rm th})}{N_{\rm Qtot}},
  %% JFH2 "tot" should not be in math mode.
  %% RN done
\end{equation}
where $N_{\rm Q}(P\ge P_{\rm th})$ is the number of high-$z$ QSOs per square
degree with a probability $P\ge P_{\rm th}$, and $N_{\rm Qtot}$ is the total number of QSOs per square degree, while the efficiency (E) is defined as:
\begin{equation}\label{eq:eff}
%% JFH = ---> \equiv
%% RN done
    E \equiv \frac{N_{\rm Q}(P\ge P_{\rm th})}{N_{\rm Q}(P\ge P_{\rm th})+N_{\rm C}(P\ge P_{\rm th})},
\end{equation}
where $N_{\rm C}(P\ge P_{\rm th})$ is the number of contaminants with a probability $P\ge P_{\rm th}$ per square degree. 
%% JFH per square degree (not degrees)
%% RN done
In the limit where the classification of all the sources in our survey is known, we could compute both C and E directly from the VIKING survey area. However, as we do not know the real classification of most of the sources in our sample, we used simulations to compute the completeness and the efficiency of our selection method, as we now describe.
%% JFH our selection method, as we now describe. 
%% RN done

In order to reduce the statistical fluctuations we simulated a large number 
of both high-$z$ QSOs and contaminants. 
High-$z$ QSOs were simulated by sampling the $z\ge 6$ LF from Eq. \ref{eq:LF} (\citealt{Wang19}), using a Markov Chain Monte Carlo (MCMC) approach. Namely, this equation can be interpreted 
as the 2-D probability distribution
of the quasars as a function of redshift and magnitude, and MCMC is a convenient method to generate samples.
%% JFH state the obvious somewhere, namely that the \Phi(M,z)/N_tot can be interpolated as the 2-d probability distribution
%% of the quasars, and MCMC is a convenient method to sample it. 
%% RN done
Again, we expressed the LF as a function of redshift and apparent $J$-band magnitude, by converting the $M_{1450}$ to $J$-band magnitude using the $k$-correction from Eq. \ref{eq:kcorr},
%% JFH You are glossing over some details here, which you have a habit of doing. Recall there is a k-correction needed
%% to relate M_{1450} to J-band apparent magnitude. State what you used for this k-correction. 
%% RN done
and multiplied it by the incompleteness found in \S \ref{sec:model_construction} for the VIKING $J$-band magnitude distribution. We then used the MCMC method to sample the redshift and $J$-band magnitude distributions of 300,000 QSOs with $6\le z\le8$, and $17\le m_J \le 22$. 
%% JFH Somewhere in the text you should quote the number of sources we expect given this luminosity function in each 
%% redshift range, with/without incompleteness. THis need not be here, but it should be done somewhere. 
%% RN done
Given the redshift and $J$-band magnitude of each source, 
%% JFH I think you mean "given the redshift and J-band magnitude of each source, we used our ....
%% RN done
we used our deconvolved quasar models to sample 
%% JFH generate -- sample. 
%% RN done
the noiseless fluxes for the 300,000 simulated QSOs, and added representative
photometric errors according to our noise model
%% JFH this is the first time the noise model in the appendix is being cited, although it has been mentioned in 
%% several other places in the paper. Please correct. 
%% RN done
in Appendix \ref{appendixa}. Then, the simulated QSOs were divided into the three redshift bins adopted previously, and we computed their probability of being quasars using Eq. \ref{eq:bayes}, to derive the $N_{\rm Q}(P\ge P_{\rm th})$ needed for Eq. \ref{eq:comp} and Eq. \ref{eq:eff}.

%%JFH2 added a paragraph break here.
%% RN done
To simulate the contaminants, we drew 100 million $17\le m_J \le 22$ sources from the $J$-band magnitude distribution of the contaminant training catalog (upper-left panel Fig. \ref{fig:prior_dist}). We again sampled 
the deconvolved contaminant models to generate the noiseless fluxes for our simulated sources, and added the errors as explained in Appendix \ref{appendixa}.
Then, we evaluated the probability that these synthetic sampled ``sky" objects
are quasars using Eq. \ref{eq:bayes}, which is needed to determine the $N_{\rm C}(P\ge P_{\rm th})$ term from Eq. \ref{eq:eff}.
Finally, we rescaled the numbers of simulated contaminants and high-$z$ QSOs
to reflect the prior number count distributions shown in Fig. \ref{fig:prior_dist},
and we used Eq. \ref{eq:comp} and Eq. \ref{eq:eff} to compute the completeness and efficiency, down to a $J$-band magnitude of 21.5. This magnitude limit was introduced since it is representative of what can be realistically confirmed
with a near-IR instrument on an 8m class telescope in a reasonable exposure time, and is also close to the 5$\sigma$ limit of the VIKING data we use.
Fainter objects would require longer exposure times and excellent observing
conditions making them much more challenging to spectroscopically confirm. 

\begin{figure}
 \centering
 \includegraphics[height=7cm, width=9cm]{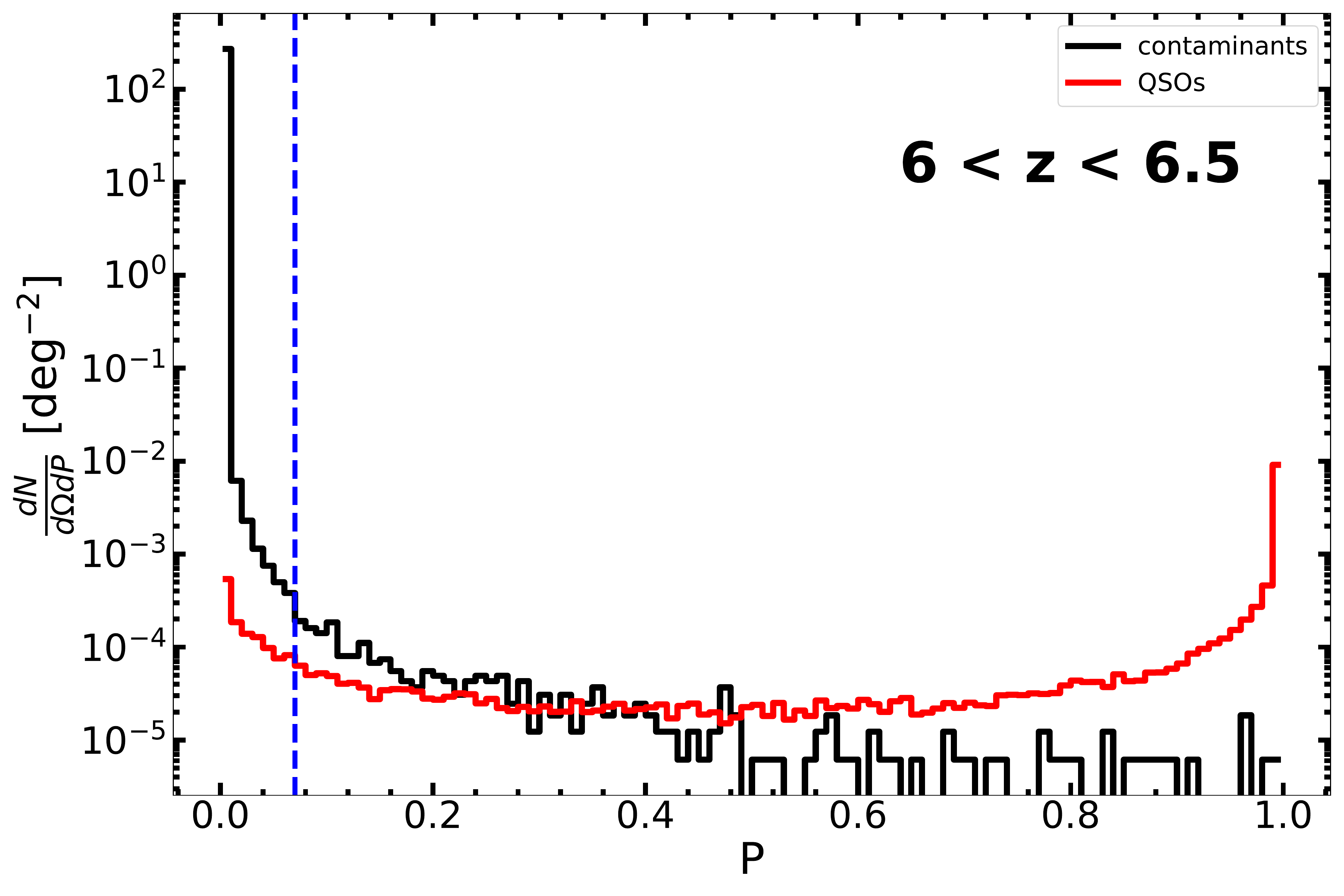}
 \includegraphics[height=7cm, width=9cm]{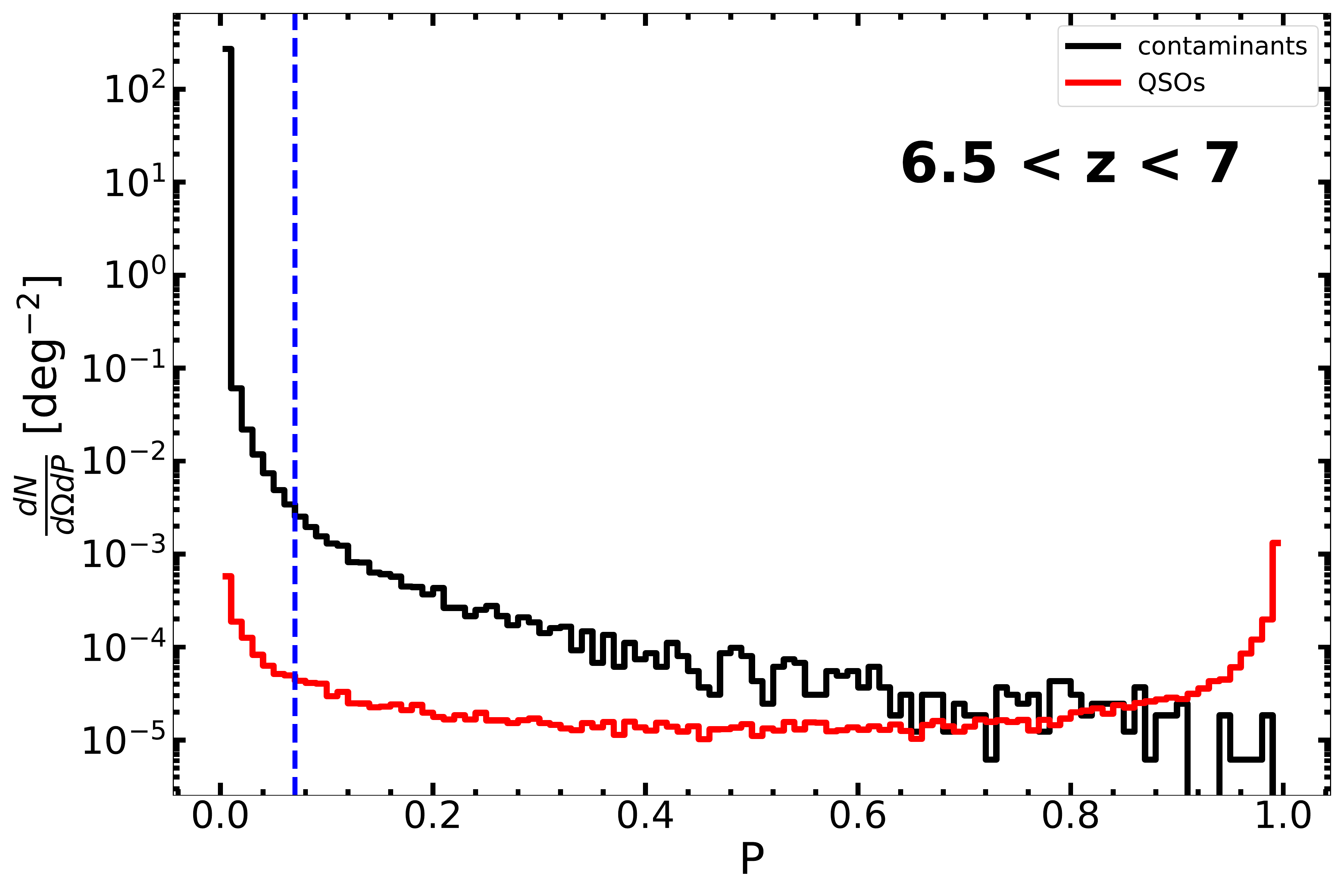}
 \includegraphics[height=7cm, width=9cm]{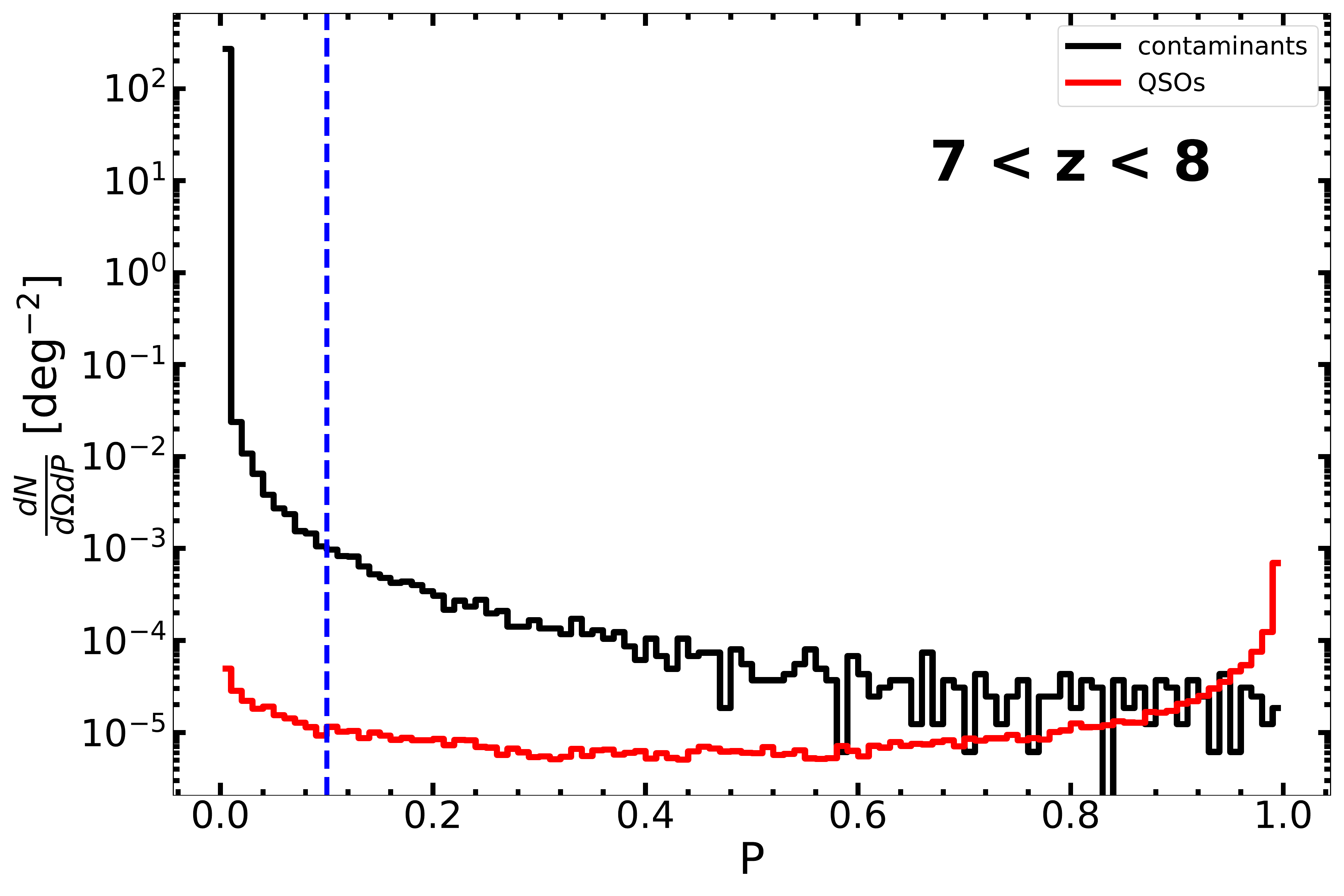}
 \caption{Probability distributions of simulated contaminants (black) and high-$z$ QSOs (red) per square degrees, in three different redshift ranges: $6\le z \le 6.5$ (top), $6.5 \le z \le 7$ (central), and $7 \le z \le 8$ (bottom). The blue dashed vertical line marks our adopted probability threshold.}
 \label{fig:prob_simulations}
\end{figure}

\begin{figure*}
 \centering
 \includegraphics[height=7cm, width=9cm]{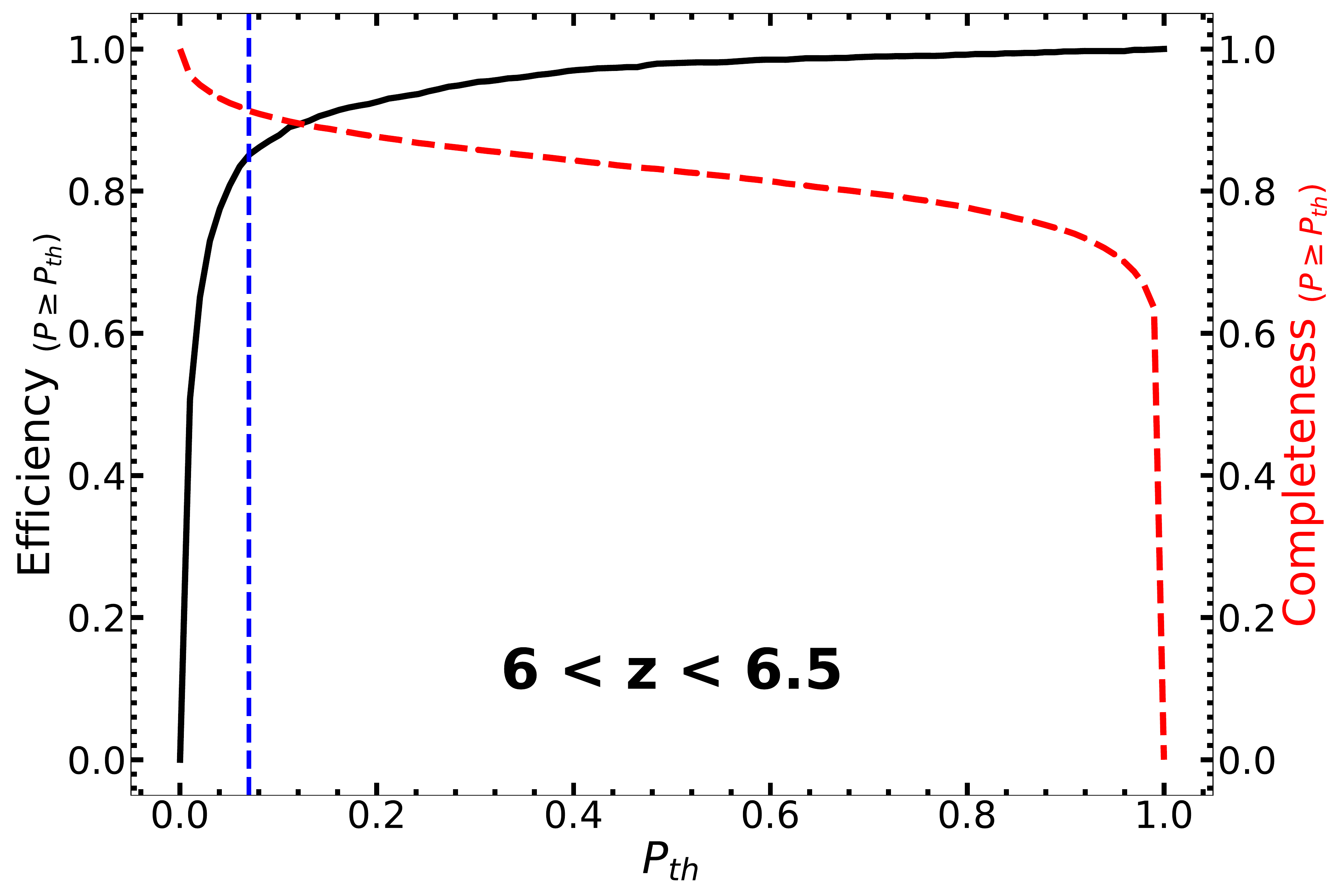}
 \includegraphics[height=7cm, width=9cm]{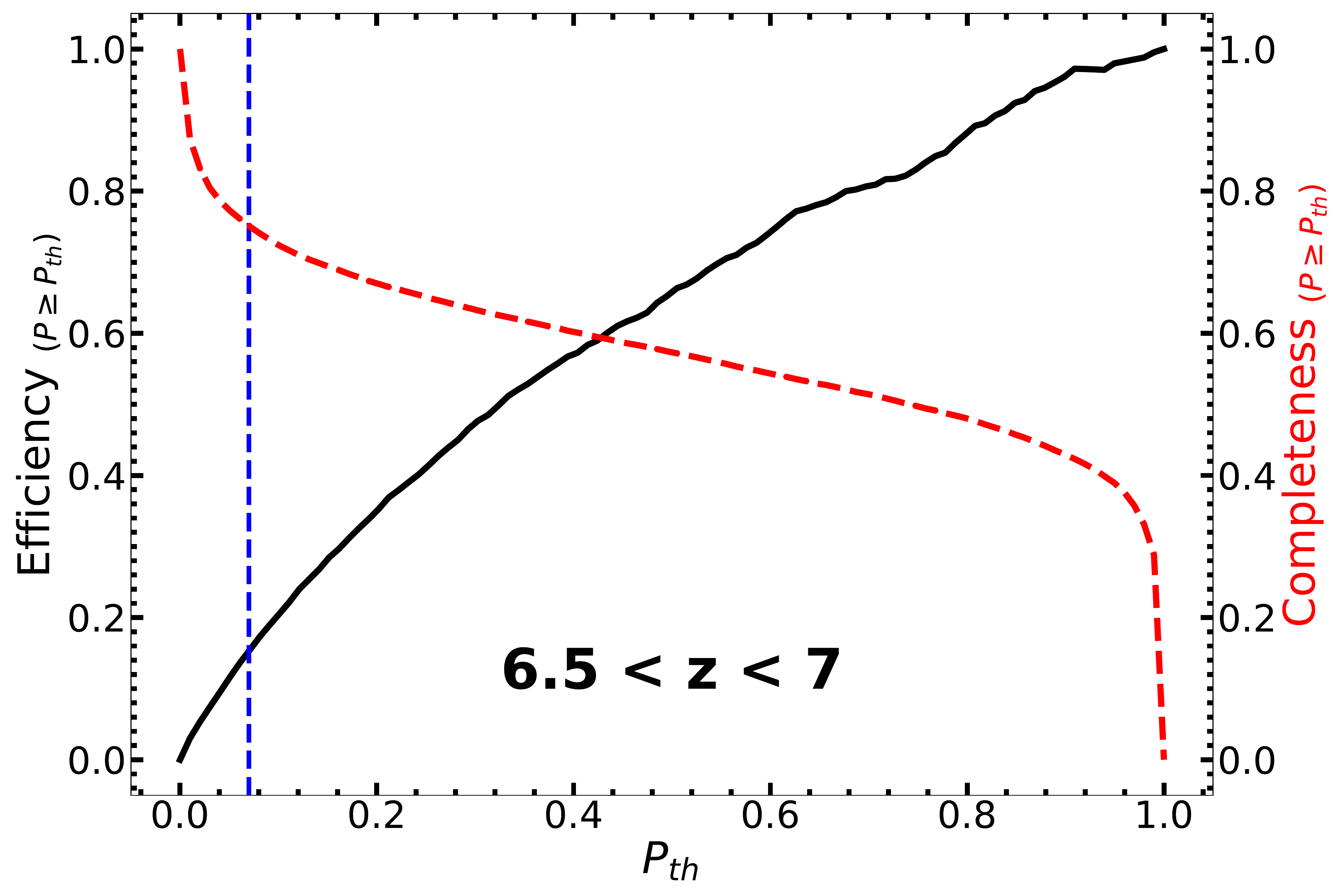}
 \includegraphics[height=7cm, width=9cm]{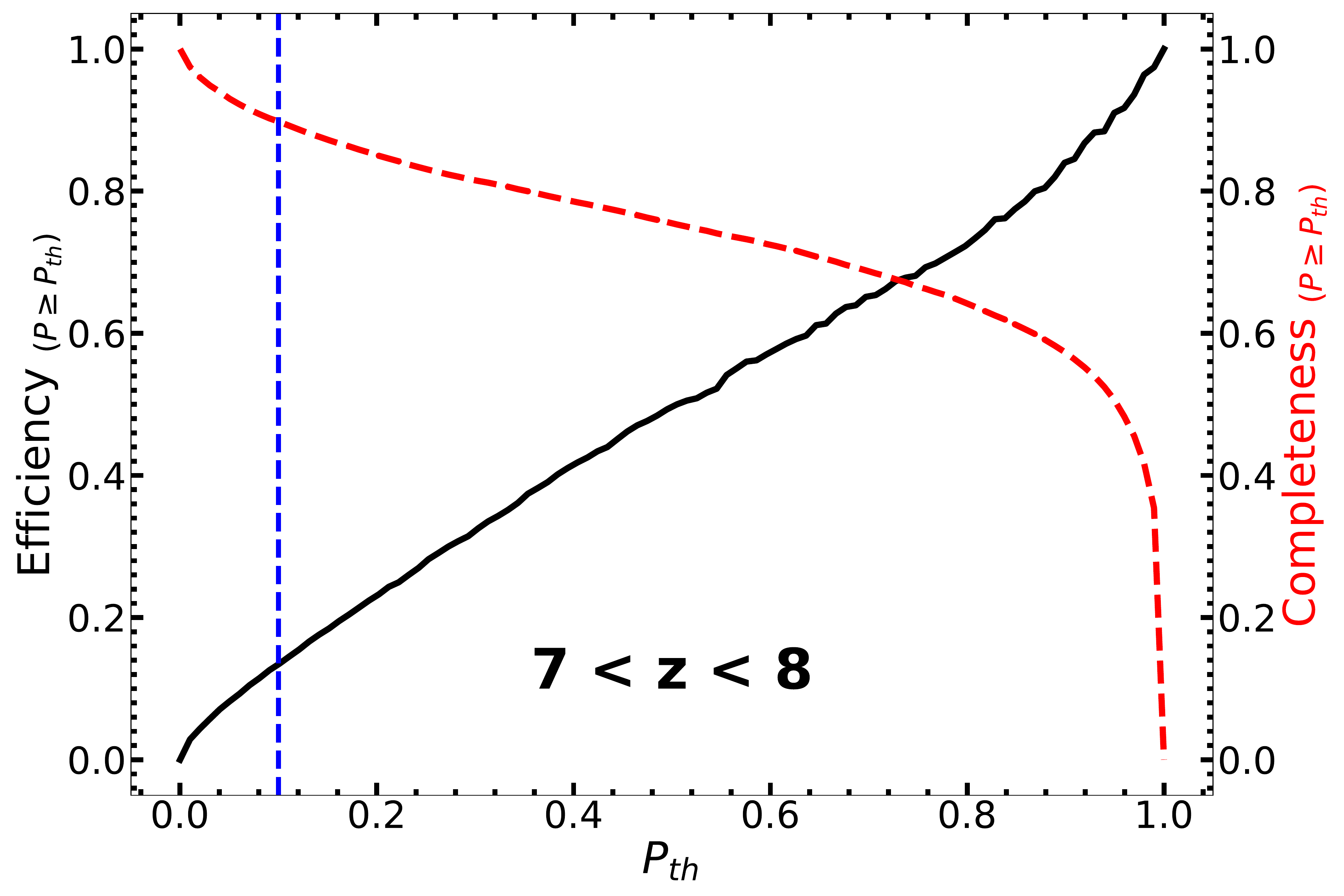}
 \caption{Efficiency (black solid line) and completeness (red dashed line) of the XDHZQSO selection method as a function of probability threshold ($P_{\rm th}$). The three panels show the results for the three redshift bins: $6\le z \le 6.5$ (top),  $6.5\le z \le 7$ (central), and  $7\le z \le 8$ (bottom). The blue dashed vertical line marks our adopted probability threshold. It is apparent that lowering the threshold will always increase the incompleteness
 but this comes at the cost of lower efficiency, thus increasing the number of contaminants selected for spectroscopic follow-up. It is also evident that both the efficiency and completeness are lower at $6.5\le z \le 7$, and  $7\le z \le 8$, where the QSO properties largely overlap with the contaminant distribution (see the overlap between the red and green over the the black contours in Fig. \ref{fig:model_comparison}).}
 \label{fig:eff_comp_vs_p}
\end{figure*}

In Fig. \ref{fig:prob_simulations} we display the number count distribution of quasar probabilities, $dN\slash d\Omega\slash dP$, 
for simulated QSOs and contaminants. This 
quantity is defined such that the integral over probability $P$ yields
the number of objects per square degree. 
Fig. \ref{fig:eff_comp_vs_p} shows the efficiency (black) and the completeness (red) of our selection method as a function of the probability threshold ($P_{\rm th}$), in the three redshift bins: $6\le z \le 6.5$ (top),  $6.5\le z \le 7$ (central), and  $7\le z \le 8$ (bottom). It is apparent that lowering the threshold will always increase the completeness, but this comes at the cost of a lower efficiency, thus increasing the number of contaminants that are
spectroscopically followed up. It is also evident that the completeness and efficiency are generally higher in the $6\le z \le 6.5$ range, where the overlap between the QSO and contaminant relative-flux distributions is smaller compared to the $6.5\le z \le 7$, and $7\le z \le 8$ cases (i.e. the red and green contours
overlap the black contours in Fig. \ref{fig:model_comparison} more than the blue contours). 
%% JFH Nobody has ever seen plots like this before and you are hardly providing any commentary on their apperance
%% and general behavior. 
%% You also need to make some general statements like lowering the threshold will always increase the incompleteness
%% but this comes at the cost of lower efficiency, etc. 
%% RN done

\begin{figure}
 \centering
 \includegraphics[height=7cm, width=9cm]{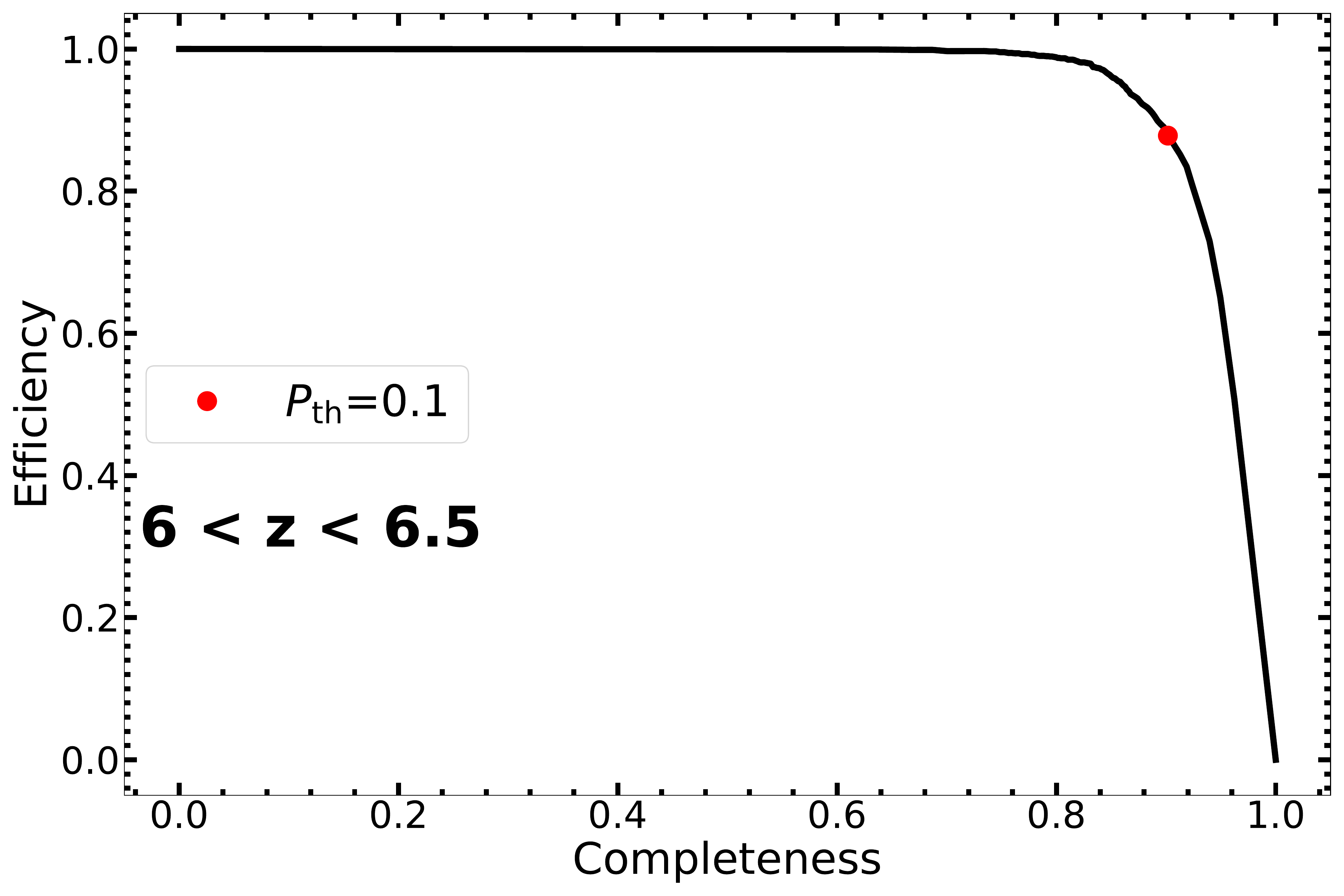}
 \includegraphics[height=7cm, width=9cm]{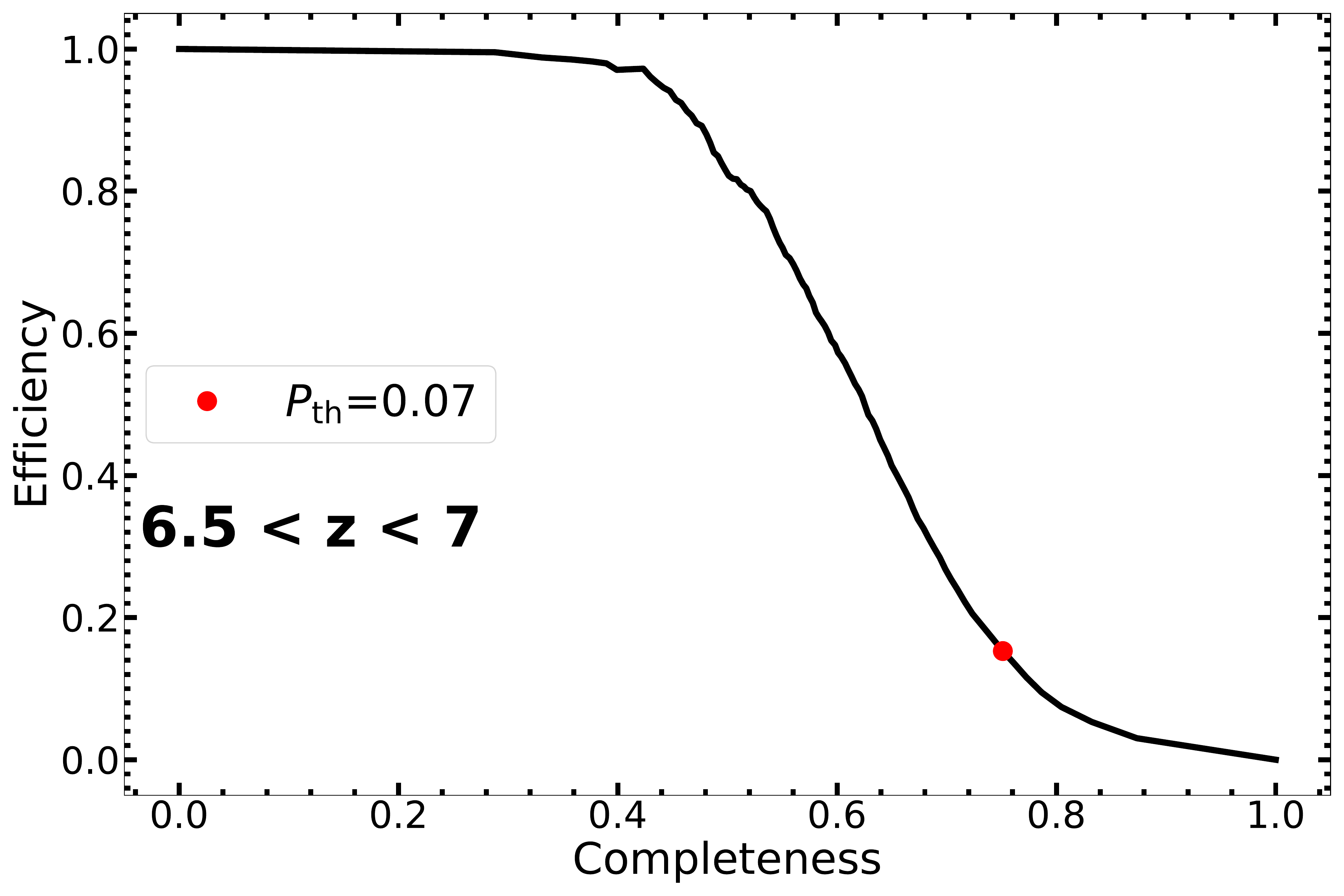}
 \includegraphics[height=7cm, width=9cm]{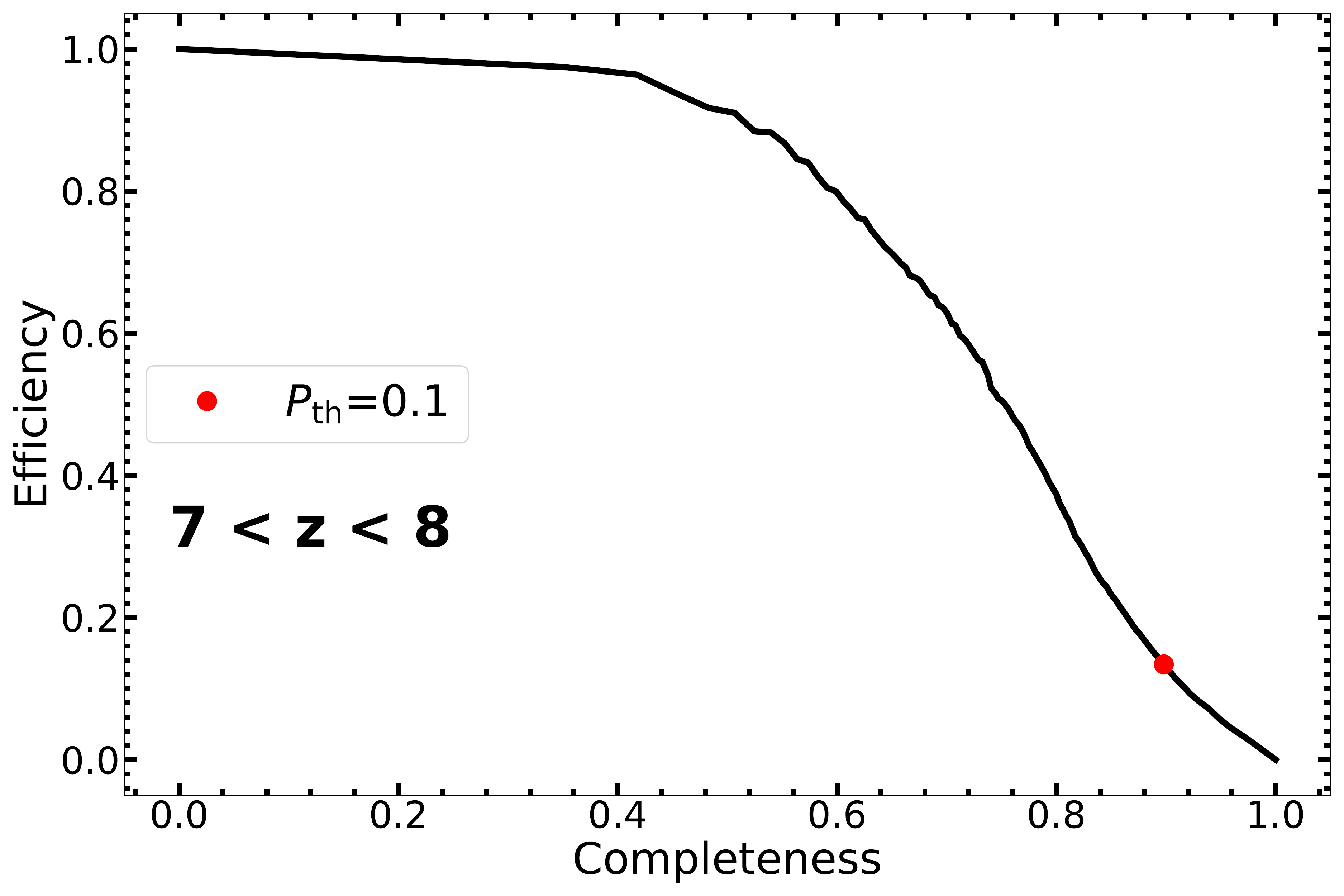}
 \caption{Efficiency vs completeness in the three redshift bins: $6\le z \le 6.5$ (top),  $6.5\le z \le 7$ (central), and  $7\le z \le 8$ (bottom). The red point marks the efficiency and completeness at the value of the chosen probability threshold ($P_{\rm th}$). The low overlap between the $6\le z \le 6.5$ QSO and contaminant contours allows us to work with high values of efficiency (88\%) and completeness (90\%). However, at $6.5\le z \le 7$ and  $7\le z \le 8$ the overlap with the contaminant properties is so large that we are forced to work at a lower efficiency (15\%) to have a high completeness ($\ge75\%$).}
 \label{fig:eff_comp}
\end{figure}

Since the expected number density of high-$z$ QSOs is very low, the choice of the $P_{\rm th}$ is mostly determined by the need to have a high completeness to avoid missing the coveted
highest-redshift sources. In fact, by integrating the LF in Eq. \ref{eq:LF} down to $J=21.5$, we expect to find $\approx 15$, $\approx 5$, and $\approx 2$ QSOs in the ranges
$6\le z\le 6.5$, $6.5\le z\le 7$, and $7\le z\le 8$, respectively, in the $1076$ deg$^2$ VIKING survey area. 
%% JFH our catalog --> in the XXX deg^2 VIKING survey area.
%% RN done
Recovering this small number of expected sources would require a relatively
high completeness (possibly $C\approx 90\%$). 
%% JFH Provide these numbers for all redshifts bins for VIKING somewhere. Consider maybe a table. 
%% RN it is already present a table below
As such, we chose use the completeness as the main criterion for setting the
probability threshold $P_{\rm th}$, whereas the efficiency plays a pivotal role in setting $P_{\rm th}$ when a high completeness corresponds to $E<10\%$. To visualize the tradeoff between completeness and efficiency (both of which are parameterized by $P_{\rm th}$), we plot
%% JFH For display purposes --> to visualize the tradeoff between completeness and efficiency (both of which are 
%% parameterized by P_th), we plot.....
%% RN done
in Fig. \ref{fig:eff_comp} the efficiency as a function of the completeness for the three redshift bins.

In the $6\le z \le 6.5$ range, the $90\%$ completeness requirement corresponds to $P_{\rm th}=0.1$ and $E=88\%$ (Fig. \ref{fig:eff_comp}, top panel). In the $6.5\le z \le 7$ range the high completeness requirement ($C= 90\%$) cannot be achieved without lowering the efficiency to an unacceptable value ($E\approx10^{-3}\%$; see Fig. \ref{fig:eff_comp}, central panel),
%% JFH2 Can you please quote the number even if it is 1e-4%. The point is
%% that writing zero is not very informative. 
%% RN done
while a $75\%$ completeness (achievable with $P_{\rm th}=0.07$)
%% JFH2 You are quoting P and P_th as a fraction throughout the text
%% and on the figures so please don't change the convention now to
%% percent!
%% RN done
corresponds to
$E\approx15\%$, which is a more reasonable efficiency value to work with.
For the $7\le z\le 8$ range, 
we want a completeness of $90\%$ to avoid missing the $\approx2$ expected $z>7$ QSOs. This requirement corresponds to $\approx 15\%$ efficiency, 
and can be achieved with $P_{\rm th}= 0.1$. 
%% JFH2 I think this next sentence makes more sense after you discus the
%% numbers of the 7 < z < 8 bin. 
%% RN done
The very low value of efficiency in the two highest redshift ranges is caused by the large overlap between the $6.5\le z \le 8$ QSOs and the contaminant models, as is apparent in Fig. \ref{fig:model_comparison} (see the larger overlap
of the green and red with the black contours). Consequently, also the number of QSO candidates with probability above the threshold in these redshift ranges is lower compared to the $6\le z \le 6.5$ range.

%% JFH2 Probably we should add a discussion of the VIKING survey to this
%% paragraph, and in particular quote how many z > 7 sources or z > 6.5
%% sources they targeted and found. Also, a lot of this stuff below could
%% rather go into a comparison to previous work section. 
%% RN to be implemented
To summarize, we report in Table \ref{tab:thresholds} the three probability thresholds derived from our completeness and efficiency analysis, and the corresponding completeness, efficiency, and number of candidates $N(P_{\rm QSO}\ge P_{\rm th})$  with $P_{\rm QSO}\ge P_{\rm th}$ that are selected for future spectroscopic follow-up.
%% JFH2 Include the symbolic N(P_QSO > P_th) here to be consistent with your
%% table. 
%% RN done
For the $7\le z \le 8$ range, we obtain an efficiency that is $15\%$, whereas quasar selections based on color-cuts work at percent level efficiency in
this redshift range (\citealt{Banados18}; \citealt{Wang21}).
%% JFH2 Provide references for this statement!!
%% RN done
The higher efficiency that we derive results from the combination of two primary factors: 1) our probabilistic density estimation takes advantage of the full feature space (all flux ratios) at once without strict boundaries, making it more effective and inclusive than simple color-cuts, and 2) our effort to compile as much panchromatic photometry as possible improves the efficiency, relative to
previous efforts (\citealt{Mortlock11}; \citealt{Banados18}; \citealt{Yang20b}; \citealt{Wang21}), to select $z > 7$ quasars using color cuts.
%% JFH2 I think you need to be more specific here and say "to select z > 7
%% quasars using color cuts and provide references. 
%% RN done
On the other hand, an efficiency of $15\%$ for the $6.5\le z \le 7$ range is lower compared to some color-cut selections performed in the past (i.e., \citealt{Banados16}). This likely results from the fact that our study does not include the PS1-$zy$ filters, which greatly improves the selection of $6.5\le z \le 7$ QSOs, since in this redshift range the Ly$\alpha$ line enters the PS1-$y$ filter and drops out of the PS1-$z$ filter, while the broader DECaLS-$z$ filter covers
both the aforementioned PS1 filters. In a future study we plan to include the PS1-$zy$ filters to improve our selection efficiency for this particular redshift range.
%% JFH followups --> follow-up
%% RN done
\begin{table*}
  %% JFH2 This table needs a caption.  You might also quote the total number of expected quasars from the LF in each bin in this table that you would recover as well as the total number expected. 
  %% RN done
  \centering
  \captionsetup{justification=centering, labelsep = newline}
      \caption[]{Number of selected candidates in the three redshift bins}
      \begin{adjustbox}{center, max width=\textwidth}
         \begin{tabular}{c c c c c c c}
            \hline
            \hline \rule[0.7mm]{0mm}{3.5mm}
            $z$ range & $P_{\rm th} \, [\%]$& $C \, [\%]$& $E \, [\%]$ & $N(P_{\rm QSO}\ge P_{\rm th})$ & $N_{\rm exp}$ & $N_{\rm rec}$\\
            \hhline{~~~~~~}
            \hline \rule[0.7mm]{0mm}{3.5mm}
            $6.0-6.5$ & 10 & 90 & 88 & 14 & 15 & 10\\
            \rule[0.7mm]{0mm}{3.5mm}
            $6.5-7.0$ & 7 & 75 & 15 & 27 & 5 & 1\\
            \rule[0.7mm]{0mm}{3.5mm}
            $7.0-8.0$ & 10 & 90 & 15 & 23 & 2 & 2\\[2pt]
            \hline 
         \end{tabular}
        \end{adjustbox}
        \begin{tablenotes}\small{
       \item Summary of the probability threshold ($P_{\rm th}$) adopted in each redshift bin to select high-$z$ QSO candidates for spectroscopic follow-up, and the corresponding completeness (C), efficiency (E), and number of candidates selected ($N(P_{\rm QSO}\ge P_{\rm th})$). The last two columns represent the number of QSOs expected ($N_{\rm exp}$) according to our adopted LF (Eq. \ref{eq:LF}) down to $J=21.5$, and how many of them we expect to recover among our candidates ($N_{\rm rec}$).}
    \end{tablenotes}
         \label{tab:thresholds}
\end{table*} 

%% JFH Somewhere in the text, either here or in the Discussion you need to provide some commentary on why you are
%% deducing an efficiency of 17% at z = 7-8, whereas other workers are working at percent level efficiency. The 
%% answer is complicated, but I think it is 1) color cuts are not that effective, and 2) they are using limited filters
%% so the efficiency is anyway going to be lower compared to what you find. 
%% RN done

%% JFH Also add a comment that since you don't have PS1y, the 6.5-7 efficiency is lower than what other workers find
%% and that this could be added. 
%% RN done

%% JFH Change this title to "Classification of known high-$z$ QSOs in the VIKING Survey Area", or something like that
%% to distinguish from the z > 7 section below. 
%% RN done
\section{Classification of Known high-$z$ Quasars}\label{sec:known_qsos}
By integrating the $z=6.7$ LF from \citet{Wang17} in the $17\le J\le21.5$ range, we expect to find $\approx 21$ ($\approx 28$) QSOs at $6\le z\le 6.5$, $\approx 7$ ($\approx 9$) QSOs at $6.5\le z\le 7$, and $\approx 3$ ($\approx 4$) QSOs at $7\le z\le 8$, depending on whether (or not) we consider the effect of the
$J$-band photometric incompleteness in the VIKING survey. Thus,
after performing the spectroscopic follow-up of the targets with $P\ge P_{\rm th}$,  we expect to discover high-$z$ QSOs among our candidates with numbers consistent with these
estimates.

Past works already studied the VIKING area and searched for $z\ge 6$ QSOs (e.g., \citealt{Venemans13}, \citeyear{Venemans15}; \citealt{Barnett21}). For example, both \citet{Venemans13} and \citet{Barnett21} used the $ZYJHK_s$ filters from the VIKING survey to find $z>6.5$ QSOs: \citet{Venemans13} applied color-cuts and found three new QSOs, while \citet{Barnett21} selected four known QSOs and 17 QSO candidates using the BMC method, but no new QSOs were found. Other QSOs where found in the VIKING footprint from past works, as they searched for high-$z$ QSOs in other surveys that partially overlap with the VIKING area: i.e., the CFHTLS (\citealt{Willott09}), the Pan-STARRS1 (\citealt{Banados16}), the VST-ATLAS (\citealt{Carnall15}), the DELS (\citealt{Wang17}), and the HSC (\citealt{Matsuoka16}; \citeyear{Matsuoka18a}; \citeyear{Matsuoka18b}; \citeyear{Matsuoka18c}; \citeyear{Matsuoka19a}; \citeyear{Matsuoka19b}).
So, we expect to have some known high-$z$ QSOs in our VIKING dataset, and to recover them among our candidates. In \S \ref{missed_qsos}, we provide a summary of the known QSOs that are covered within our search area but that are not in our VIKING dataset due to our selection criteria.
Then, we describe the performance of XDHZQSO in recovering and classifying both the known high-$z$ QSOs in the VIKING survey area (\S \ref{known_qsos}), as
well as the known $z>7$ QSOs (\S \ref{z7_qsos}) over the entire sky.
%% JFH2 Provide some qualifier for the known z >7 QSOs, i.e. over the entire
%% sky or brighter than J = 21.5 over the entire sky or something. 
%% RN done

%% JFH I am confused by the appearance of this section here and do not really follow the logic of putting it here at the 
%% end. State these overall statistics at the beginning of this section. Then explain why you restrict to the sources that
%% you do. Provide numbers and summaries all along the way. Then it would be more clear than tacking this on at the end. 
%%
%% I think you can just state at the outset that 31 QSOs at z > 6 are in the VIKING area, but imposing our selection criteria
%% reduces this number to XX, where XX are lost because they are not above the S/N > 5  in J limit and another XX are lost
%% because they are not covered in all bands. That leaves XX quasars, and you can proceed to described the stats on those. 
%% RN done
\subsection{Missed high-$z$ QSOs}\label{missed_qsos}
From past works (\citealt{Willott09}; \citealt{Venemans13}; \citealt{Banados16}; \citealt{Matsuoka16}; \citeyear{Matsuoka18a}; \citeyear{Matsuoka18b}; \citeyear{Matsuoka18c}; \citeyear{Matsuoka19a}; \citeyear{Matsuoka19b}), we identified 32 known $z\ge6$ QSOs in the DECaLS+VIKING area.
%% JFH2 seems weird to not have references for past works??. 
%% RN done
However, the imposition of our selection criteria reduced this number in our final VIKING area dataset, as 20 QSOs are lost because they do not satisfy ${\rm SNR}(J)>5$, and another four QSOs are not selected as they do not have data in all the bands considered in our study. That leaves eight known $ z > 6$ quasars in the VIKING area dataset whose probabilistic classification is described in the following section.

%(CFHQS J021627-045534 $z=6.01$; \citealt{Willott09}), HSC J1152+0055 ($z=6.37$; \citealt{Matsuoka16}), PSOJ159.2257-02.5438 ($z=6.38$; \citealt{Banados16}), and VIK J010953.13-304726.3 ($z=6.75$; \citealt{Venemans13}). The other 21 were either discovered in the HSC survey (20 QSOs; \citealt{Matsuoka16}; \citeyear{Matsuoka18a}; \citeyear{matsuoka18b}; \citeyear{matsuoka18c}; \citeyear{Matsuoka19a}; \citeyear{Matsuoka19b}), or in the CFHQ survey (one QSO; \citealt{Willott09}), and are too faint (${\rm SNR}(J)<5$) for our selection criteria.

\subsection{Classification of known high-$z$ QSOs in the VIKING Survey Area}\label{known_qsos}
%% JFH My advice is to make section 5.2, 5.3 and 6.1 all their own section which is "Classification of Known high-$z$ Quasars"
%% RN done

%% JFH Printing numbers in text like this is not an effective way to display data. I suggest you just make a table
%% to get this point across. Then you can list some relevant information, z, J_mag, P_QSO, reference, etc. 
%% RN done
Among the classified sources there are eight known high-$z$ QSOs that were found in the VIKING survey area from past works: DELS J1217+0131 ($z=6.17$; \citealt{Banados16}; \citealt{Wang17}), ATLAS J025.6821-33.4627 ($z=6.31$, hereafter J0142-3327; \citealt{Carnall15}), HSC J1137+0045 ($z=6.4$; \citealt{Matsuoka19a}), J0148-2826 ($z=6.54$; \citealt{Yang20a}), HSC J0921+0007 ($z=6.56$; \citealt{Matsuoka18b}), VIK J0305-3400 ($z=6.604$; \citealt{Venemans13}), DELS J1048-0109 ($z=6.63$; \citealt{Wang17}), HSC J1205-0000 ($z=6.74$; \citealt{Matsuoka16}). These sources and their main properties are listed in Table \ref{tab:known_qsos}, while their probabilities of being high-$z$ QSOs are pinpointed with red arrows in Fig. \ref{fig:probabilities}.
\begin{table*}
  \captionsetup{justification=centering, labelsep = newline}
  \caption[]{Known QSOs in the VIKING survey area}
      \begin{adjustbox}{center, max width=\textwidth}
         \begin{tabular}{l c c c l}
            \hline
            \hline \rule[0.7mm]{0mm}{3.5mm}
            Name & $z$ & $J$ & $P_{\rm QSO}$ & Ref. \\
            \hhline{~~~~}
            \hline \rule[0.7mm]{0mm}{3.5mm}
            DELS J1217+0131 & 6.17 & $21.28\pm0.14$ & 0.06\% & \citet{Banados16}; \citet{Wang17} \\
            \rule[0.7mm]{0mm}{3.5mm}
            ATLAS J025.6821-33.4627 & 6.31 & $19.02\pm0.02$ & 99.9\% & \citet{Carnall15}\\
            \rule[0.7mm]{0mm}{3.5mm}
            HSC J1137+0045 & 6.4 & $21.51\pm0.20$ & $3\times 10^{-7}\%$ & \citet{Matsuoka19a}\\
            \rule[0.7mm]{0mm}{3.5mm}
            J0148-2826 & 6.54 & $21.09\pm0.13$ & 0.01\% & \citet{Yang20a}\\
            \rule[0.7mm]{0mm}{3.5mm}
            HSC J0921+0007 & 6.56 & $20.9\pm0.26$ & $10^{-4}\%$ & \citet{Matsuoka18b}\\
            \rule[0.7mm]{0mm}{3.5mm}
            VIK J0305-3400 & 6.61 & $20.07\pm0.09$ & 24.5\% & \citet{Venemans13}\\
            \rule[0.7mm]{0mm}{3.5mm}
            DELS J1048-0109 & 6.63 & $20.99\pm0.12$ & 0.04\% & \citet{Wang17}\\
            \rule[0.7mm]{0mm}{3.5mm}
            HSC J1205-0000 & 6.75 & $21.95\pm0.21$ & $10^{-11}$\% & \citet{Matsuoka16}\\[2pt]
            \hline 
         \end{tabular}
        \end{adjustbox}
         \label{tab:known_qsos}
\end{table*}

\begin{figure*}
 \centering
 \includegraphics[height=18cm, width=18cm]{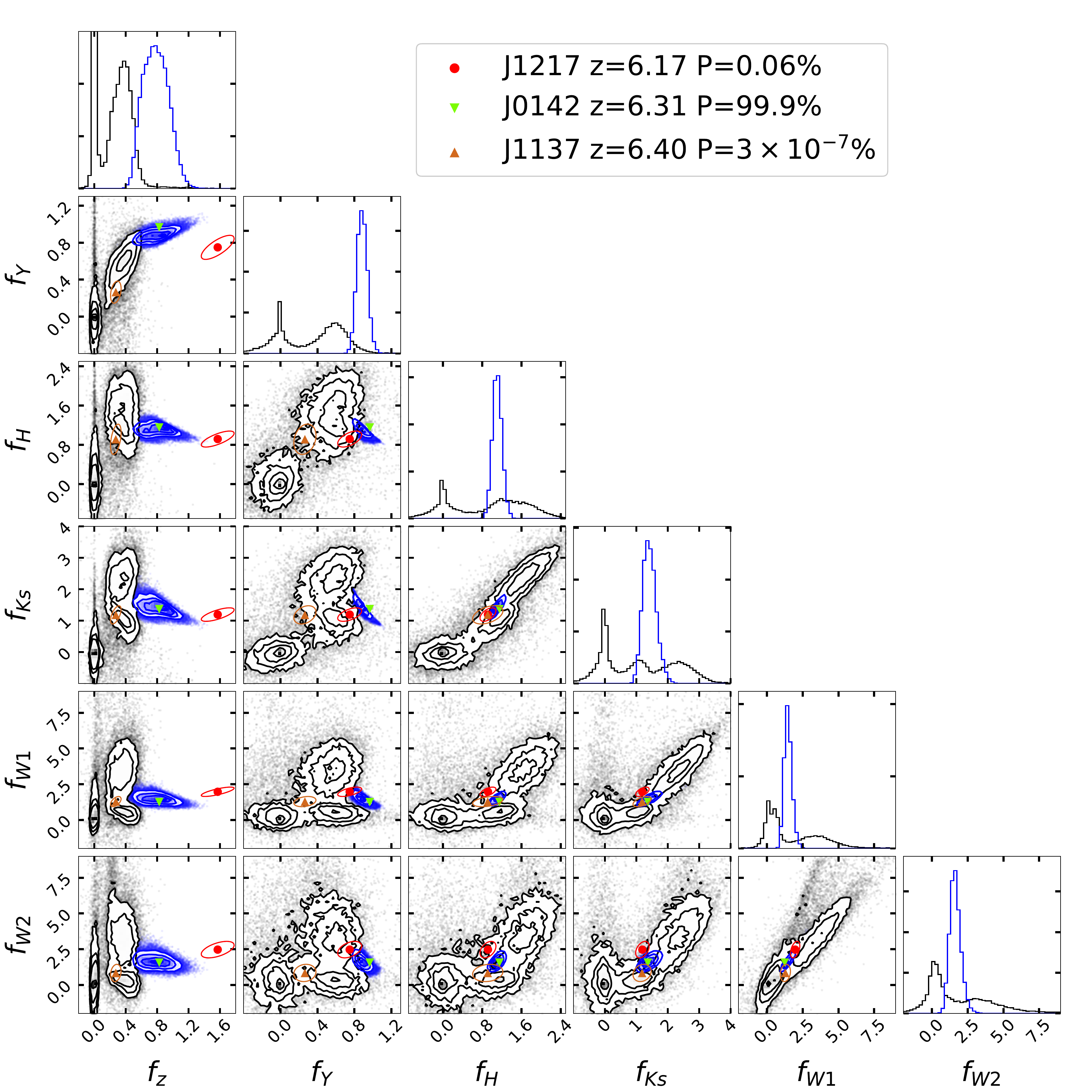}
 \caption{Deconvolved relative-flux relative-flux contours for the simulated contaminants (black) and $6\le z \le6.5$ QSOs (blue), compared to the properties of the known $6\le z \le6.5$ QSOs from the VIKING survey area. The probability threshold to select these sources with our method is $P_{\rm th}=0.1$. It is apparent that both J1217 and J1137 are ``off'' from the QSO contours in the $f_z$ sub-plots, while J1137 is also ``off'' in the $f_Y$ sub-plots, thus lowering their probabilities of being classified as high-$z$ QSOs.}
 \label{fig:known_6_65}
\end{figure*}

In the range $6\le z \le6.5$, our models are able to correctly classify one known QSO out of three,
%% JFH2 State out of how many!
%% RN done
J0142-3327 ($P_{\rm QSO}\approx99.9\%$), but our selection threshold in this redshift range ($P_{\rm th}=10\%$) does not allow us to recover DELS J1217+0131 ($P_{\rm QSO}\approx0.06\%$), and HSC J1137+0045 ($P_{\rm QSO}\sim10^{-7}\%$). The probability of these three quasars are also reported in Table \ref{tab:known_qsos} and shown in Fig. \ref{fig:probabilities} (upper panel).
%% JFH Reference Figure 8 here. Also add the Table I request and reference that. 
%% RN done
The low probability of the latter one is not surprising, considering that HSC J1137+0045 is a very faint QSO ($J=21.51$ and ${\rm SNR}(J)=5.4$),
%% JFH2 Quote also the S/N ratio in J-band in VIKING
%% RN done
selected from the Hyper Suprime-Cam (HSC) Subaru Strategic Program (SSP) survey (\citealt{Aihara18}), and that apparently lacks strong Ly$\alpha$ in emission (\citealt{Matsuoka19a}). However, to better understand the low probability values obtained for these two quasars, we compared their photometric properties with those sampled from our XD deconvolved models. 
%% JFH You need to provide some sentences on how Figure 12-14 are made and how they should be interpreted. You always
%% err on the side of providing very little information!  But then the reader is confused why you simulated 10,000 
%% copies of each source. 
%% 
%% 1) The reason we are creating 10000 copies of each contaminant and 10000 QSOs is that the contaminant and quasar
%% models are magnitude dependent. Thus formally, we would need to show a plot for each object. However, given 
%% that these magnitude dependencies are subtle, we chose to simply simulate 10,000 copies of sources at each 
%% magnitude and aggregate them onto a single plot. 
%%2) To visualize the probability of selecting a known quasar we contours/samples drawn from the "deconvolved" i.e. noise 
%% free XD models and overplot the relative flux measurements of the real quasars, with ellipses indicating their 
%% (covariant) 1\sigma errors. 
%% RN done
For each of the three known $6\le z \le6.5$ QSOs, we simulated 10,000 contaminants and 10,000 $6\le z \le6.5$ QSOs, using the XDHZQSO models in the magnitude bins that include the $J$-band magnitudes of the three QSOs. To visualize the probability of selecting a known quasar, we draw samples from the ``deconvolved" (i.e., noise free) XDHZQSO contaminant and quasar models, and overplot the relative flux measurements of the real quasars, with ellipses indicating their (covariant) 1$\sigma$ errors.
This is shown in Fig. \ref{fig:known_6_65}, where we plot the deconvolved relative-flux relative-flux contours for the simulated contaminants (black) and $6\le z \le6.5$ QSOs (blue), compared to the properties of the known $6\le z \le6.5$ QSOs from the VIKING survey area. The reason we are creating 10,000 copies of contaminant and 10,000 of QSOs for each known high-$z$ QSO is that the contaminant and quasar models are magnitude dependent. Thus formally, we would need to show a plot for each object, where we compare its properties with those from the sampled contaminants and QSOs. However, given that these magnitude dependencies are subtle, we chose to simply simulate 10,000 copies of sources at each magnitude and aggregate them onto a single plot. It is apparent that in some sub-plots of Fig. \ref{fig:known_6_65} (especially those with $f_z$ and $f_Y$), the relative fluxes of both HSC J1137+0045 and DELS J1217+0131 are not consistent with the simulated $6\le z \le6.5$ QSOs relative flux distributions (blue contours), consequently lowering the classification probability of these two objects. Considering that HSC J1137+0045 is a QSO that apparently lacks strong Ly$\alpha$ in emission (\citealt{Matsuoka19b}), while DELS J1217+0131 exhibits a strong Ly$\alpha$ emission line (\citealt{Wang17}), we conclude that the properties of the ``simqso'' simulated high-$z$ QSOs, that have been used for the training of our XDHZQSO QSO models, are too rigid to include these two sources.
%% JFH2 Do we have any undertsanding of why this is the case??
%% RN done
%% JFH I would add a few more sentences on what filters in particular are "off" for these HSC sources. According to what
%% you are showing here, your failure to detect them seems to have very little to due with noise and how faint they are. 
%% RN done

\begin{figure*}
 \centering
 \includegraphics[height=18cm, width=18cm]{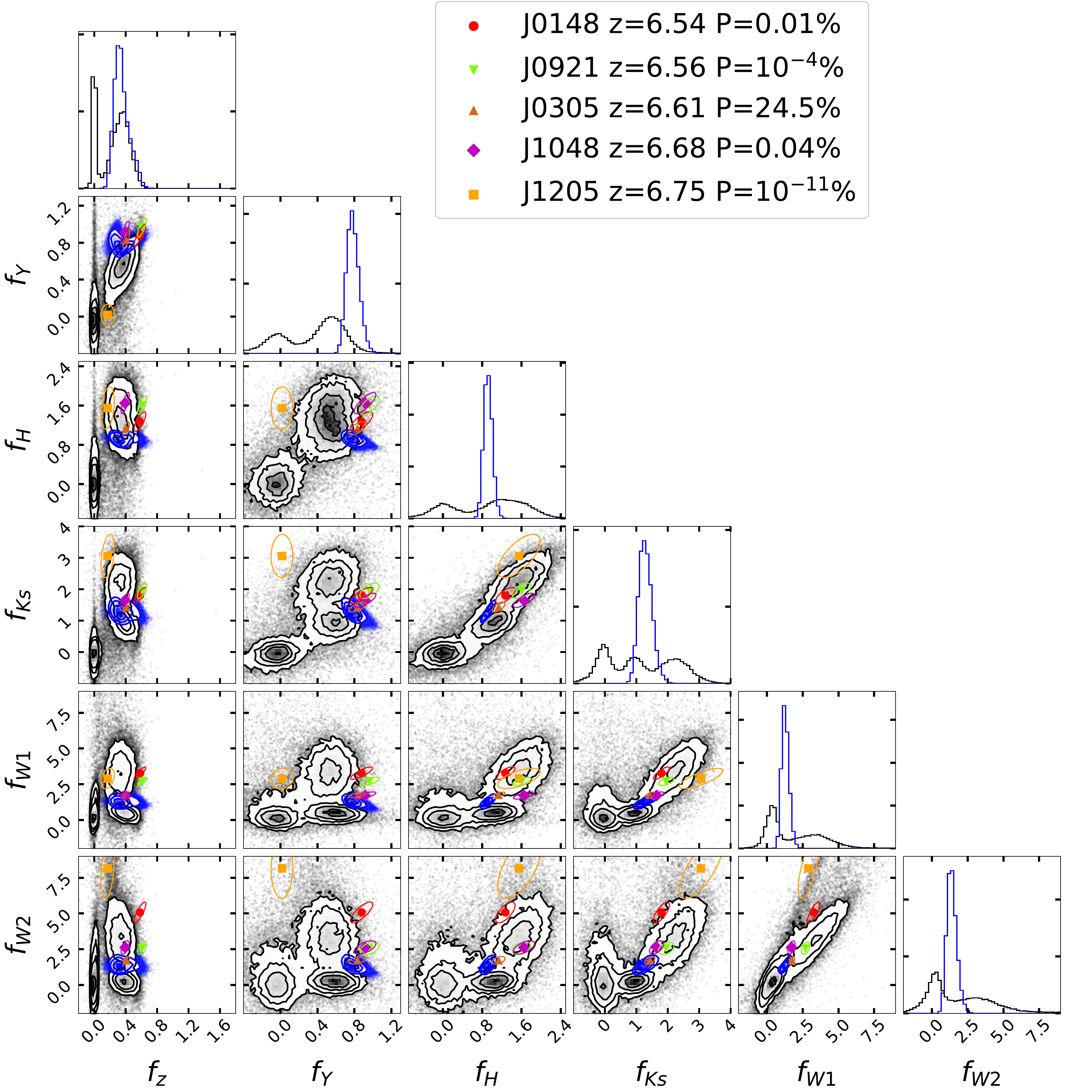}
 \caption{Same as Fig. \ref{fig:known_6_65} but in the $6.5\le z \le7$ bin. The probability threshold to select these sources with our method is $P_{\rm th}=0.07$. The four QSOs with $P_{\rm QSO}\le10^{-2}\%$ are inconsistent with the deconvolved QSO model properties (blue contours) in the following sub-plots: 1) J0148 is inconsistent with panels showing $f_{H}$, $f_{W1}$, and $f_{W2}$, 2) J0921 is inconsistent with panels showing $f_{W1}$, and $f_{W2}$, 3) J1048 is not consistent with panels showing $f_{H}$, and $f_{W2}$, and 4) J1205 is not consistent with the QSO distributions in any panel.}
 \label{fig:known_65_7}
\end{figure*}

In the range $6.5\le z \le7$, as reported in Table \ref{tab:known_qsos} and displayed in Fig. \ref{fig:probabilities} (middle panel), our method is able to recover one QSO (based on our $P_{\rm th}=7\%$), VIK J0305-3400 ($P_{\rm th}\approx24.5\%$), while the other four are consistent with being contaminants ($P_{\rm QSO}\le10^{-2}\%$). 
%% JFH reference the figures and tables that I should be looking at here, i.e. Figure 8 middle panel, Figure 13, and the 
%% table you will create. 
%% RN done
Among them, J0921+0007 ($P_{\rm QSO}\sim10^{-4}\%$) is also a HSC selected QSO ($J=20.9$) that has similar optical colors to Galactic brown dwarfs (\citealt{Matsuoka18b}). Adopting the same procedure as described above to generate 10,000
%% JFH Adoping the same procedure as described above....
%% RN done
contaminants and $6.5\le z \le7$ QSOs for each known QSO, we show in Fig. \ref{fig:known_65_7} the deconvolved relative-flux relative-flux
%%JFH2 change to relative-flux relative flux throughout? Or maybe just
%% say somewhere once that we refer to these plots as flux-flux even
%% though they show relative fluxes
%% RN done
contours for the simulated contaminants (black) and high-$z$ QSOs (blue), compared to the properties of the known $6.5\le z \le7$ QSOs from the VIKING survey area. Also in this case, it is apparent that the relative fluxes of the four QSOs with $P_{\rm QSO}\le10^{-2}\%$ are inconsistent with the deconvolved QSO model properties (blue contours in Fig. \ref{fig:known_65_7}) in some sub-plots:
%% JFH be specific which sub-plots
%% RN done
1) J0148-2826 is inconsistent with panels showing $f_{H}$, $f_{W1}$, and $f_{W2}$, 2) HSC J0921+0007 is inconsistent with panels showing $f_{W1}$, and $f_{W2}$, 3) DELS J1048-0109 is not consistent with panels showing $f_{H}$, and $f_{W2}$, and 4) HSC J1205-0000 is not consistent with the QSO distribution in any panel. We provide a more detailed discussion of these discrepancies between real and simulated QSO properties in \S \ref{deviations}.

\subsection{Classification of the $\mathbf{z\ge 7}$ QSOs}\label{z7_qsos}
While we tested in \S \ref{known_qsos} the ability of our models to recover the known $6 \le z \le 7$ QSOs in the VIKING survey area,
%% JFH in our catalog --> in the VIKING survey area or something like that. Adopt a consistent term for this. 
%% RN done
testing our classification models for the highest redshift range was not possible as there are no known $z> 7$ QSOs in the VIKING footprint. Therefore, we applied our method to the $z> 7$ QSOs that have been discovered so far over the entire sky, using published photometric measurements.
%% JFH using published photometric measurements. 
%% RN done

%% JFH Please put these in a summary table. 
%% RN done
There are, at the time of writing, a total of eight known $z>7$ QSOs: J2356+0017 ($z=7.01$; \citealt{Matsuoka19a}), J0252-0503 ($z=7.02$; \citealt{Yang19}), J0038-1527 ($z=7.021$; \citealt{Wang18}), J1243+0100 ($z=7.07$; \citealt{Matsuoka19b}), J1120+0641 ($z=7.085$; \citealt{Mortlock11}), J1007+2115 ($z=7.515$; \citealt{Yang20b}), J1342+0928 ($z=7.541$; \citealt{Banados18}), and J0313-1806 ($z=7.642$; \citealt{Wang21}).
To classify them, we first collected the photometric data in the seven bands of interest (DECaLS-$z$, VIKING-$YJHK_S$, and WISE-$W1W2$) from the literature, when available. Since some of these sources have public NIR data coming from the Wide Field Infrared Camera (WFCAM) for the UK Infrared Telescope (UKIRT), we used the transformation equations between VISTA and WFCAM derived by \citet{Gonzales18}, to convert the UKIRT magnitudes into the VIKING ones.
For the missing flux measurements, we performed forced photometry. 
Since J0313-1806 has no photometric measurements in the $Y$ and $H$ bands, we used synthetic photometry computed by integrating the observed spectrum of this source from \citet{Wang21} against the respective filter curves. 
However, we excluded from our classification list both J2356+0017 and J1243+0100, as they are too faint (${\rm SNR}(J)<5$)
%% JFH2 It seems more fair to exclude them if they have S/N < 5.0. I'm not sure
%% I agree these should be excluded if they made it into our catalog limit.
%% This would seem to suggest we are cherry picking. 
%% RN done
to make it into our catalog.
%% JFH Too faint to be selected by our method makes it sound like our method does not work which is inaccurate. How faint
%% are these sources anyway, I don't think I agree with your assesment that they are too faint. You too often use that excuse
%% whereas in reality the contour plots tell a different story. Do not say things like "too faint" without quoting numbers. 
%% Too faint to me would mean S/N J in VIKING < 5 but you should be very clear when you say things like this. 
%% RN done
Finally, we used our XDHZQSO models to classify the remaining six sources following the same procedure described in \S \ref{sec:qso_selection}. In Table \ref{tab:known_qsos7} we summarize the properties and results from our classification of these six $z\ge7$ QSOs.
\begin{table*}
  \captionsetup{justification=centering, labelsep = newline}
  \caption[]{Known $z\ge7$ QSOs classified by our XDHZQSO method}
      \begin{adjustbox}{center, max width=\textwidth}
         \begin{tabular}{l c c c l}
            \hline
            \hline \rule[0.7mm]{0mm}{3.5mm}
            Name & $z$ & $J$ & $P_{\rm QSO}$ & Ref. \\
            \hhline{~~~~}
            \hline \rule[0.7mm]{0mm}{3.5mm}
            J0252-0503 & 7.02 & $21.13\pm0.07$ & 17.1\% & \citet{Yang19} \\
            \rule[0.7mm]{0mm}{3.5mm}
            J0038-1527 & 7.021 & $20.63\pm0.08$ & 0.4\% & \citet{Wang18}\\
            \rule[0.7mm]{0mm}{3.5mm}
            J1120+0641 & 7.085 & $21.22\pm0.17$ & 29.4\% & \citet{Mortlock11}\\
            \rule[0.7mm]{0mm}{3.5mm}
            J1007+2115 & 7.515 & $21.14\pm0.18$ & 77.2\% & \citet{Yang20a}\\
            \rule[0.7mm]{0mm}{3.5mm}
            J1342+0928 & 7.541 & $21.24\pm0.02$ & 33.3\% & \citet{Banados18}\\
            \rule[0.7mm]{0mm}{3.5mm}
            J0313-1806 & 7.642 & $20.92\pm0.13$ & 43.7\% & \citet{Wang21}\\[2pt]
            \hline 
         \end{tabular}
        \end{adjustbox}
         \label{tab:known_qsos7}
\end{table*}
%% JFH following the same procedure described above. 
%% RN done

\begin{figure*}
 \centering
 \includegraphics[height=18cm, width=18cm]{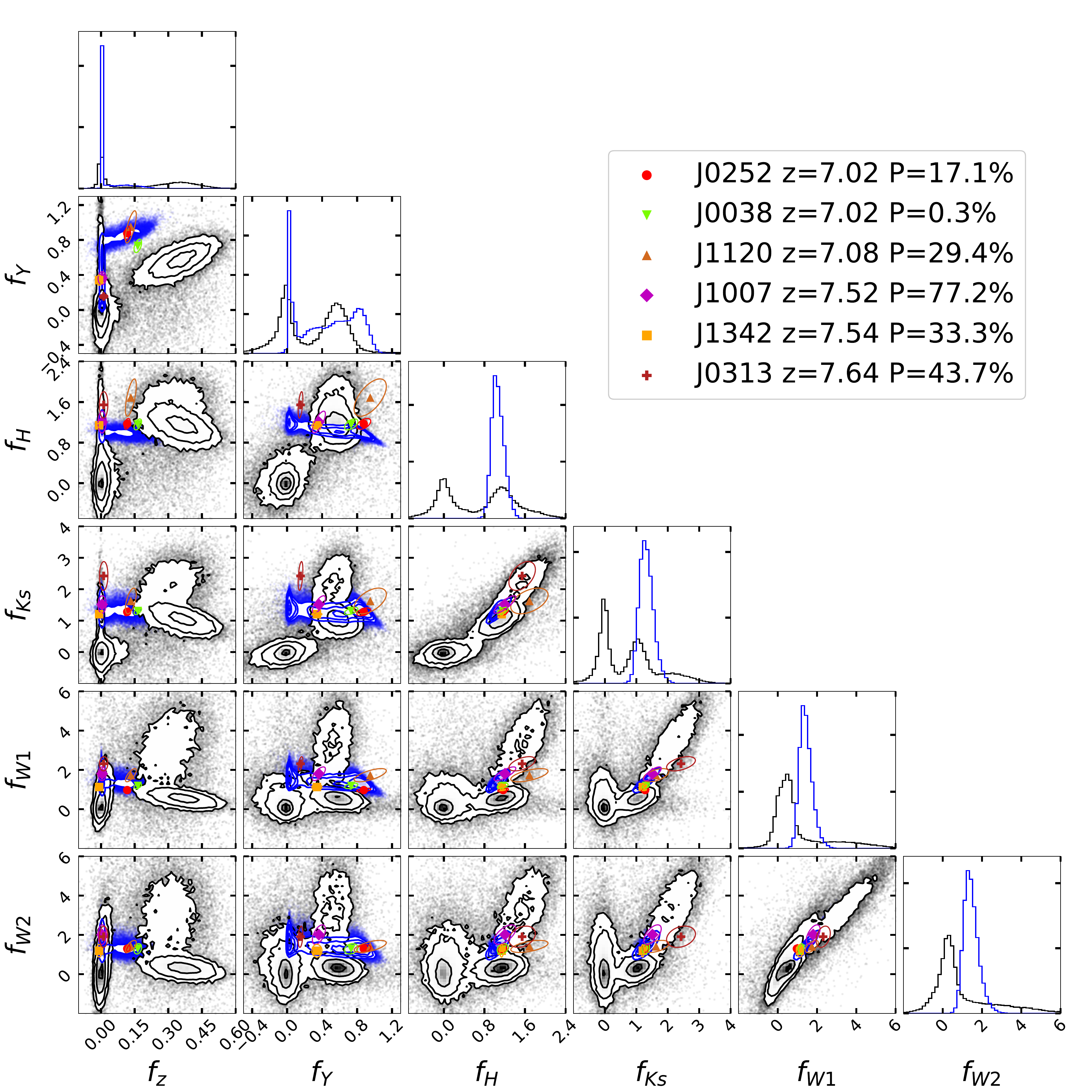}
 \caption{Deconvolved relative-flux relative-flux contours for the simulated contaminants (black) and $7\le z \le9$ QSOs (blue), compared to the properties of the known $z \ge7$ QSOs to date. The probability threshold to select these sources with our method is $P_{\rm th}=0.1$. It is apparent that J0038-1527 deviates from the quasar locus indicated by the blue contours in the sub-plot displaying $f_z$ vs. $f_Y$, with the effect of lowering its QSO classification probability.}
 \label{fig:known_7}
\end{figure*}

Based on our defined probability threshold for the $z\ge 7$ range ($P_{\rm th}=10\%$), we are able to recover five QSOs: J0252-0503 ($P_{\rm QSO}=17.1\%$), J1120+0641 ($P_{\rm QSO}=29.4\%$), J1007+2115 ($P_{\rm QSO}=77.2\%$), J1342+0928 ($P_{\rm QSO}=33.3\%$), J0103-1806 ($P_{\rm QSO}=43.7\%$). However, we fail to select J0038-1527 ($P_{\rm QSO}=0.3\%$).
%% JFH misclassified at lower-z makes it sound like you assigned them a higher probability in a different redshift bin. 
%% Be more clear or just omit lower-z or just state "two of the z \sim 7 sources"
%% RN done 
J0038-1527 exhibits strong broad absorption line (BAL) features (\citealt{Wang18}), that can alter its colors,
%% JFH emitted properties --> alter its colors
%% RN done
making it different compared to our $7\le z \le 8$ QSO models, which do not
attempt to model BAL absorption. As in \S \ref{known_qsos}, we simulated a
large number of contaminants and $7\le z \le 8$ QSOs, and compare their relative fluxes with those from the real $z>7$ QSOs in Fig. \ref{fig:known_7}. It is evident that J0038-1527 deviates
%% JFH state the colors of the symbols parenthetically so that the reader knows where to look
%% RN done
from the blue contours (deconvolved $7\le z \le 8$ QSO models) in the sub-plot displaying $f_z$ vs. $f_Y$, 
%% JFH "some sub-plots" is ambigious. Get into the habit of always telling your reader exactly where to look or what
%% you refer to. Your plots have 15 panels in them, so it is never obvious what you are referring to. 
%% RN done
as the absorption from the BALs impacts the $Y$-band flux.
%% JFH2 This would be slightly more compelling if you stated that the
%% the BAL absorption impacts the Y or Z band flux hence explaining
%% why it deviates there. 
%% RN done
Again, we discuss the deviations of the real QSO properties from the expected simulated ones in \S \ref{deviations}. 

\section{Discussion}\label{sec:discussion}

In \S \ref{sec:known_qsos} we showed that XDHZQSO
%% JFH2 Our code is not so descriptive. How about XDHIZQSO
%% RN done
is able to recover
%% JFH2 In this sentence below state the number of quasars recovered and the
%% total number that make it into our catalog, i.e. in the survey area,
%% S/N > 5 (or whatever cut we are using for this?), and no missing data. 
%% RN done
two $6 \le z \le 7$ QSOs out of eight that passed our selection criteria (i.e., ${\rm SNR}(J)\ge5$, ${\rm SNR}(g,r)<3$, and no missing data) and made into our catalog: we select
one QSO at $6\le z\le 6.5$ (ATLAS J025.6821-33.462 at $z=6.31$), and one QSO at $6.5\le z\le 7$ (VIK 0305-3400 at $z=6.61$). The application of XDHZQSO on the $7\le z\le 8$ QSOs found in the entire sky, that meet our selection criteria, allow us to recover five out of six $7\le z\le 8$ QSOs
%% JFH2 for this state the number considered meeting our cuts
%% RN done
(J0252-0503 at $z=7.02$, 1120+0641 at $z=7.085$, J1007+2115 at $z=7.515$, J1342+0928 at $z=7.541$, and J0313-1806 at $z=7.642$). In \S \ref{deviations} we discuss the limitations of our selection technique that could explain our failure to select
%% JFH2 explain the non-selection of --> explain our failure to select
%% RN done
of some of the known high-z QSOs in the VIKING area, while in \S \ref{comparison} we provide a comparison between our code and other probabilistic classification methods.

%% JFH2 Typeo here, I think it is XDHIZQSO
%% RN done
\subsection{Limitations of the XDHZQSO selection method}\label{deviations}

%% JFH Provide an executive summary of how many quasars that satisfied your selection criteria that you actually 
%% recovered. Break that down by redshift bin as well. 
%% RN done
In \S \ref{known_qsos} and \S \ref{z7_qsos} we showed that our method is only able to recover some of the known high-$z$ QSOs. In fact, there are several reasons that can lead to the failure to select
%% JFH2 non-selection --> failure to select
%% RN done
a source, and all of them involve the source properties and corresponding errors being more consistent with the XDHZQSO contaminant models rather than the high-$z$ QSO ones. Here we discuss the possible causes that lead to the non-selection of some of the known $z>6$ sources:

\begin{itemize}
    %% JFH Bad data measurement is not proper english. Either it is "bad data" or a "bad measurement" not both. 
    %% RN done
\item {\it Noisy data or photometric variability.} In the case of a source with large photometric errors,
  %% JFH2  large data errors --> large photometric errors or noisy flux measuremens
 %% RN done
  our method naturally degrades its probability of belonging to high-$z$ QSOs class if the data uncertainties imply that the object overlaps with the contaminant class. On the other hand, this limitation is not afflicting other selection methods. In fact, a color-selection technique that does not use photometric errors
%% JFH2 data errors --> photometric errors
%% RN done
  could select a noisy object, whereas XD would spread that probability out,
  %% JFH2 This explanation would be more compelling if you could indicate
  %% which sources you thought fit this explanation. Also, you might
  %% add that taking errors into account is a feature not a flaw of our
  %% method. In other words, not taking into account errors will generally
  %% result in an overall lower efficiency then taking the errors into
  %% account which is the more optimal approach. 
  %% RN done
  meaning it might be more likely to be classified as a contaminant if, given the errors, it significantly overlaps the contaminant locus. However, we stress that taking errors into account is a feature not a flaw of our method (i.e., not taking into account errors will generally result in an overall lower efficiency then taking them into account, which is the more optimal approach). 
  %%JFH2 meaning it might be more likely to be classified as a contaminant
  %% if, given the errors, it significantly overlaps the contaminant locus
  %% RN done
  Furthermore, since many surveys were performed at different epochs,
  intrinsic variability of sources could also play a role in lowering the computed probabilities (see \citealt{RossCross20} for a study of the variability of $5<z<7$ quasars). However, since the variability of these objects is supposed to be small (at most 10\% given low-$z$ structure functions; e.g., \citealt{VandenBerk04}; \citealt{Kelly09}; \citealt{Schmidt10}), we argue that this is probably not the main issue we are facing.
\item {\it Inaccurate models.} Since our method is a classification technique, its validity strongly depends on the correct modelling of the considered classes. If the XDHZQSO models are not a good representation of the underlying deconvolved flux distributions of one or more classes, then the computed probabilities are not reliable. Although, that seems not the case for our contaminant class, as the models are trained with the real data coming from our survey, it can be an issue for our high-$z$ QSO classes. In fact, our quasar models are trained on synthetic photometry determined from simulated QSO spectra whose properties are consistent with the mean spectrum of low-$z$ luminous QSOs (\citealt{McGreer13}).
%% JFH2 cite McGreer here again just for good measure. 
%% RN done
  However, these simulated quasar spectra could not well represent the intrinsic relative flux scatter of all the luminous QSOs, or the properties of peculiar sources such as Broad Absorption Line QSOs (BALQSOs). For example, J0038-1527 is a BALQSO (\citealt{Wang18}), and its $Y$-band relative flux is lower than expected compared to objects with similar redshift and luminosity
%% JFH2 similar redshift and luminosity
%% RN done
  (see Fig. \ref{fig:known_7}). Furthermore, in the sub-panels showing $H$, $K$, $W1$, and $W2$ bands in Figs. \ref{fig:known_6_65}, \ref{fig:known_65_7}, \ref{fig:known_7} it is apparent that our XDHZQSO QSO models are too rigid,
  as the simulated QSO deconvolved density distributions (blue contours) appear too little scatter as compared to the real QSOs
  %% JFH2 too narrow --> too little scatter as compared to the real QSOs
  %% RN done
  to be a good representation of the intrinsic QSO scatter. For the $W1W2$-bands, there could be also
  source confusion/deblending errors in the photometry since we just performed aperture photometry, without taking into account the large unWISE ($\approx6"$) point spread function.
  %% JFH2 For the above sentence you need to clearly state which set
  %% of panels you are referring to. In my opinion, the model looks a
  %% bit to rigid in the redder filters like H, K, W1 and W2, but
  %% please try to be specific. Making such general statements without
  %% telling people where to look is not that helpful. I think you
  %% should also add a sentence here indicating that there could be
  %% source confusion/deblending errors in the WISE photometry since
  %% we jus did aperture photometry, without using something more
  %% sophisticated like tractor.
  %% RN done
  %% JFH2 I don't thikn this is a "Probably" given your plots. I would just
  %% conclude that is the case. 
  %% RN done
  A model that better reproduces the full distribution
  %% JFH2 a more careful modelling --> I don't think we were careless. Rather
  %% just say that a model that better reproduces the full distribution of
  %% relative fluxes
  %% RN done
  of the relative fluxes of the luminous QSOs at low-$z$ would provide a better classification of our sources. Therefore, our conclusion is that the ``simqso'' simulator was designed for color-cuts, but it is not up to the demands of a density estimation method.
\end{itemize}
%% JFH Emphasize that these represent shortcomings of the QSO model and
%% not the XD-QSO selection technique. 
%% RN done

Our current simulated quasar sample fails to capture the full spectral diversity of the observed quasar population, which is important for the density estimation method. 
Hence, to improve on our quasar selection, we have to move beyond modeling average quasar properties, for which ``simqso'' was originally designed, but rather capture the full relative flux distribution 
%% JFH2 variety --> full relative flux distribution 
%% RN done
of the full population.
In the future, we plan to mitigate these limitations by carefully modelling of the relative fluxes of QSOs 
%% JFH Stop saying "peculiar QSOs". Just leave it at the model is too rigid. Given that you are missing like at least
%% half the sources it seems silly to call them peculiar -- can half the population be peculiar. Just state that the 
%% McGreer simulator was designed for color-cuts, but it is not up to the demands of a density estimationg method. 
%% RN done
using empirical data coming from the SDSS and BOSS surveys, which would
capture the full distribution of quasar SEDs and hence relative fluxes.  
%% JFH our models will empirically capture the full distribution of quasar SEDs and hence relative fluxes 
%% RN done
%%easing the QSO modelling limitation.

%% JFH2 Should this be probabilistic classification methods?
%% RN done
\subsection{Comparison with other probabilistic classification methods}\label{comparison}

Compared to other probabilistic classification methods, our approach has two main advantages:
\begin{enumerate}
\item Our method accounts for the photometric errors
%%JFH2 data errors --> photometric errors
%% RN done
  by convolving the underlying density distribution
  %% JFH2 underlying mdoel --> underlying density distribution 
  %% RN done
  with the object’s uncertainties, assuming that the relative-flux uncertainties are Gaussian. While this approach is required to correctly estimate the probability that a noisy object is a member of given class, standard random forest methods ignore the photometric errors (e.g., \citealt{Schindler17}; \citealt{Wenzl21}), thus not utilizing all the information contained in the data. For bright sources this should not be so problematic given the small associated uncertainties.  However, at high-$z$ we have to take into account that: 1) QSOs dropout of optical bands (e.g., $grz$) and so we need to accurately treat low signal to noise dropout fluxes, and 2) QSOs are rare at high-$z$ and the LFs rise with decreasing flux. So, to build up statistics, the majority of targets will always be near the flux limits of our data,
  %% JFH2 Two things 1) the sources dropout of bands e.g. z, y
  %% etc. and so we need to accurately treat low signal to noise
  %% dropout fluxes, and 2) quasars are rare at high-z and LFs rise
  %% with decreasing flux. To build up statistics the majority of
  %% targets will always be near the flux limits of our data.
  %% RN done
  while the inclusion of the photometric errors in the analysis of fainter sources would prevent the overly optimistic identification of contaminants as high-$z$ QSO candidates.
\item The BMC method (\citealt{Mortlock12}) is also Bayesian and is directly analogous to what we are doing, with the caveat that they assume perfect knowledge of the contaminant models based on templates and priors (number counts), which are unlikely to be correct in detail. 
  Instead, our model for the contaminant class is purely empirical and does not need to construct SED models for the mean properties of each possible contaminant.
  %% JFH2 I might start this bullet point by pointing out that BMC also
  %% being Bayesian is directly analogous to what we are doing with the
  %% caveat that they assume perfect knowledge of the contaminant
  %% models based on templates and priors (number counts) which are 
  %% unlikely to be correct in detail. 
  %% RN done
  This approach is very powerful as it captures the underlying deconvolved distribution of the contaminant using real data, and includes all the kind of possible contaminants without the need of modelling them. On the contrary, the BMC method
  %% JFH2 I am pretty sure Matsuoka uses the Mortlock method, so I
  %% would cite him here. I would make sure to distinguish mortlock
  %% for coming up with the method but also cite Matsuoka and Barnett for using
  %% it. 
  %% RN done
  requires a perfect knowledge of both the properties and the type of contaminants, whose feasibility has not been yet demonstrated. For example, even if brown dwarfs
%% JFH2 I think it is brown dwarfs
%% RN done
  and early type galaxies are the majority among the contaminants, also Type-2 QSOs, reddened low-$z$ QSOs, and FeLoBAL QSOs could also contaminate the high-$z$ selection, whereas constructing models for the number density and colors of all these sources would be a daunting task.
\end{enumerate}

\section{Conclusion}\label{sec:conclusions}

%% JFH Somewhere I think you need a comparison to other probabilstic methods that illustrates the main advantages of our
%% approach 1) principled accounting of errors (random forests cannot do this), 2) empirical contaminant model (Bayesian 
%% model selection fails here, and hence could be way off), 3) no need for a spectroscopic training set since quasars are 
%% such rare objects (random forests fails here, and this also mitigates the need of using templates and object classes in 
%% Bayesian model selection since basically we don't care what the contaminants are for our method. For example, it is known 
% anecedotally that Type-2 QSOs contaminate selection as do reddened low-z QSOS and FeLoBals, as well as compact red galaxies whereas constructing models for the number density and colors of all these sources is a daunting task). 
%% RN done

In this paper we described the application of the XDHZQSO method to select high-$z$ ($6\le z\le 8$) QSOs. Our approach is based on density estimation in the high-dimensional space inhabited by the optical-IR photometry. The main idea is that quasars and the far more abundant contaminants (cool dwarf stars, red galaxies, lower-z reddened or absorbed QSOs)
%% JFH2 lower-z reddened or absorbed QSOs
%% RN done
inhabit different regions of this space. Thus, probability density ratios yield the probability that an object is a quasar, which is used to select and prioritize candidates for spectroscopic follow-up, resulting in a fully optimal method. Density distributions are modeled as Gaussian mixtures with principled accounting of errors using the XD algorithm. Compared to other probabilistic selection methods, the great advantage of our approach is that the poorly understood contaminants are modeled fully empirically. 
%% JFH be careful here you might offend by calling them ad hoc. 
%% RN done

High-$z$ quasars were trained on synthetic photometry in three redshift bins ($6\le z\le6.5$, $6.5\le z\le7$, $7\le z\le8$), whereas contaminants were trained on the VIKING ($YJHK_s$) imaging survey combined with deep DECaLS $z$-band and unWISE ($W1W2$), where all sources were required to be $g$ and $r$ dropouts. The combination of depth ($J_{AB}<22$) and wide field (1076 deg$^2$) make this the best panchromatic imaging for training quasar selection until \textit{Euclid} arrives. 

%% JFH2 Somewhere in this paragraph you should make a stronger
%% statement like "Our high efficiencies indicate that the ~1% of
%% recent color-cut based surveys are not necessary" or something along those
%% lines. 
%% RN done
From extensive simulations we determined the threshold ($P > P_{\rm th}$) required to obtain a completeness of $\gtrsim75\%$ in each redshift bin, which results in selection efficiencies $\gtrsim 15\%$. These high efficiencies indicate that the $\approx1\%$ efficiencies of recent color-cut based surveys are not necessary. The required thresholds $P_{\rm th}$ and resulting efficiencies depend on the $z$-bin in question owing to the changing overlap between quasars and contaminants, where the higher redshift bins have lower efficiencies. With the adopted $P_{\rm th}=0.1,0.07,0.1$, we selected 14, 27, and 23 quasar candidates in the range $6\le z\le6.5$, $6.5\le z\le7$, $7\le z\le8$ in the VIKING footprint,
%% JFH2 make clear that this is in the VIKING footprint.
%% RN done
respectively. These targets have been scheduled for optical and NIR spectroscopic follow-up, and the results will be published in a future work (Nanni et al. in prep.). 

%% JFH Provide a more quantitative summary here, i.e. state total number of eligible quasars in the VIKinG area and how
%% many you recover. Provide the same quanttiative summary for the z > 7 sources. 
%% RN done
In the VIKING footprint the there are eight known $6\le z \le 7$ QSOs that meet our catalog criteria, of which
two are selected. Since there are no $z>7$ known QSOs in the VIKING footprint, we applied our method to six out of eight known $z>7$ QSOs in the entire sky (we excluded two $z>7$ QSOs as they do not meet our catalog criteria), and recover five of them.
%% JFH2 This is ambiguously constructed. I think it is always clearer to
%% indicate the total number of sources that meet some criteria and then
%% number of those that you select. I.e. "In the VIKING footprint the
%% there are XX known quasars that meet our catalog criteria of which
%% YY are selected. 
%% RN done
%% JFH2 Well there are some known z > 7 quasrs that I think you didn't
%% consider given that their photometry was not available so make this more
%% clear. 
%% RN done
We argued that the XDHZQSO misses some of these quasars for two reasons: 1) the existing quasar fluxes are noisy so that our model correctly assigns them a low probability, and 2) the inaccuracies in our modeling  of quasars, namely that the synthetic quasar spectra we used do not capture the the scatter in the distribution of relative fluxes. We argued that the first limitation is a feature rather than a flaw in our approach, since we deliver reliable probabilities treating noise, and that this overall will result in higher selection efficiency. As for the second, an empirical model of luminous quasar spectra will definitely improve our classification, which we will pursue in future work. 
%% JFH2 correct classification of our method is poorly worded. How about
%% "We argued that the XDHZQSO misses these quasars for two reasons:
%% the existing quasar fluxes are noisy so that our model correctly assigns
%% them a low probability, and 2) inaccuracies in our modeling  of quasars,
%% namely that the synthetic quasar spectra we used not capture the
%% the scatter in the distribution of relative fluxes 
%% RN done
%% JFH2 We argued tht the first limitation is a feature rather than
%% a flaw in our approach, since we deliver reliable probabilities treating
%% noise, and that this overall will result in higher selection efficiency.
%% As for the second, an empirical model of luminous quasar spectra
%% will definitely improve our classification, which we will pursue
%% in future work (or something like that). 
%% RN done
%% JFH2 Maybe leave out Type-1, I'm not sure it is necessary?
%% RN done

From the integration of the $z=6.7$ LF down to $J=21.5$, we expect to find $\approx15$, $\approx5$, and $\approx2$ QSOs at $6\le z\le6.5$, $6.5\le z\le7$, $7\le z\le8$, respectively, in the VIKING survey area. Considering the completeness we derived in the three redshift ranges and the fact that
%% JFH2 write our three and four instead of using the numerals for small
%% numbers
%% RN done
three, and four $J\le21.5$ QSOs have been already discovered in the VIKING footprint
%% JFH2 our area of study --> the VIKING footprint
%% RN done
at $6\le z\le6.5$, and $6.5\le z\le7$, respectively, we expect to discover $\approx10$, $\approx1$, and $\approx2$ new QSOs at $6\le z\le6.5$, $6.5\le z\le7$, $7\le z\le8$, respectively, with future spectroscopic follow-up of our candidates.
%% JFH2 thanks to our --> with future spectroscopic follow-up of our
%% candidates.
%% RN done

%% JFH2 This is sort of weak. I would say "Future applications of this
%% methodology will focus on two datasets. First, Euclid. Second the UKIDDS
%% dataset is vague and unclear --
%% make your plan with existing data a bit more clear. Maybe quote the
%% rough area witgh similar multi-filter coverage as viking.
%% RN done
Future applications of this methodology will focus on three datasets: UKIDSS, UHS, and \textit{Euclid}.
UKIDSS covers an area of $\approx 4000$ deg$^2$ with similar multi-filter coverage as VIKING ($ZYJHK$), making it the best ground to apply XDHZQSO after VIKING. Instead, UHS covers a larger area ($\approx 12,700$ deg$^2$) but only with three filters ($JHK$). To apply our method to UHS, whose sources have no data in the $Y$-band, we will simply re-score by setting the errors in the bands with no measurements to a large number.
%% JFH2 I don't know that missing data is a common issue. Instead the issue
%% is that large areas of the sky don't actually have all the filters
%% we used here. That is really something different than missing data
%% although effectively it is the same. But missing data sounds like some
%% small fraction of objects don't have measurements. 
%% RN done
%%JFH2 be more specific about the area, the filters you might hypothetically
%% have etc. 
%% RN done
%% JFH2 This does not require improvements, see my point just below. 
%% RN done
%% JFH2 I don't agree with this plan or this description. We do not
%% have to re-train! We just have to re-score and that can be trivially
%% achieved by setting the errors to a large number. 
%% RN done
%% JFH The discussion of Euclid and missing data should be moved to conclusions. 
%% Put the EUCLID plug here, future work with ground based data, etc. 
%% RN done

Finally, the advent of \textit{Euclid} in 2022 will
%% JFH I think you should omit the "bad data measurement" problem since I don't understand what it means and don't 
%% agree that it is an issue. 
%% RN done
provide plenty of optical/IR data with a better separation between high-$z$ QSOs and contaminants properties, as its six-year wide survey will cover $15, 000$ deg$^2$ of extragalactic sky
in four bands: a broad optical band $O$ ($5500-9000$ \AA), and
three NIR bands, $Y$ ($9650-11 920$ \AA), $J$ ($11 920-15 440$ \AA), and
$H$ ($15 440-20 000$ \AA), a depth of 24 mag at 5-$\sigma$ (\citealt{Laureijs11}). The \textit{Euclid}’s wide field IR imaging should enable the discovery of $\sim100$ QSOs at $z>7$, and $\sim25$ beyond the current record of $z= 7.6$, including $\sim8$ beyond $z= 8.0$ (\citealt{Euclid19}). Since no data have been delivered yet from \textit{Euclid}, we will need re-train XDHZQSO on the Euclid photometry to get the contaminant model. 
%%JFH2 Somewhere here explain that one would re-train on the Euclid photometry
%% to get the "contaminant" model. 
%% RN done
Finally, the high efficiencies in finding $z>7$ QSOs reached by XDHZQSO suggest that we can do much more efficient spectroscopic follow-up, while we have a framework to solve the problem of performing low efficiency selection with JWST.
%% JFH2 Can we close on something slightly stronger here, i.e. say that
%% the high efficiencies suggest that we can do much more efficient follow-up
%% and the problem of performing low efficiency selection with JWST is
%% now solved (or we have a framework to solve it). 
%% RN done
\section*{Acknowledgements}
%We acknowledge the referee for a prompt and constructive report.
This work is part of a project that has received funding from the European Research Council (ERC) Advanced Grant program under the European Union’s Horizon 2020 research and innovation programme (Grant agreement No. 885301).
We thank S. Bosman and the ENIGMA group at UCSB for providing useful comments on an initial draft of this paper. We also thank J. Bovy for assistance he provided related to the XD code.

\newpage
\appendix
\section{Covariance computation and application}\label{appendixb}
To construct the contaminant models during the training step, we deconvolved the noisy relative fluxes of our contaminant sources, assuming that the relative-flux uncertainties are Gaussian, and providing the covariance matrix of the uncertainties of the single objects. While the flux measurements in each filter are independent of one another, i.e. their noise is uncorrelated, the relative flux errors are correlated (i.e., they are the ratio of the flux
in a given  band flux and the $J$-band flux). Thus, the covariance of a source with fluxes $\vec{f}=\{f_1,f_2,...,f_N\}$ and errors $\vec{\sigma_{f}}=\{\sigma_{f_1},\sigma_{f_2},...,\sigma_{f_N}\}$ coming 
\vspace{-1mm}
from $N$ filters that include the $J$-band one, can be computed as:
\begin{equation}\label{eq:covariance}
 \begin{aligned}
    %% JFH2 problem with this equation. I'm not following
    %% how the cov([fx/fJ, fy/fJ] = E[[fx/fJ, fy/fJ]. The latter is
    %% just one of the terms if you multiply out the first line. 
    {\rm cov}\left[\frac{f_x}{f_J},\frac{f_y}{f_J}\right] &=
    \EX\left[\left(\frac{f_x}{f_J}-\EX\left[\frac{f_x}{f_J}\right]\right)\left(\frac{f_y}{f_J}-\EX\left[\frac{f_y}{f_J}\right]\right)\right]\\
    &=\EX\left[d\left(\frac{f_x}{f_J}\right)d\left(\frac{f_y}{f_J}\right) \right]\\
    &=\EX\left[\left(\frac{1}{f_J} df_x - \frac{f_x}{f_J^2} df_J\right)\left(\frac{1}{f_J} df_y - \frac{f_y}{f_J^2} df_J\right)\right]\\
    &= \EX\left[\frac{1}{f_J^2} df_x df_y - \frac{f_y}{f_J^3} df_x df_J-\frac{f_x}{f_J^3} df_y df_J + \frac{f_x f_y}{f_J^4} df_J^2 \right]\\
    &= \frac{1}{f_J^2} \EX[df_x df_y] - \frac{f_y}{f_J^3} \left(\EX[df_x df_J]+  \EX[df_y df_J]\right)+ \frac{f_x f_y}{f_J^4} \EX[df_J^2].\\
\end{aligned}
  %% JFH2 I think this would be more clear if you just multiplied out the
  %% various terms and then restrict to the diagonal and off-diagonal terms.
  %% In that sense you can say <df_x df_y> vanishes if x != y and so you
  %% get one expression, versus the other. 
  %% RN done
\end{equation}
In our case, the covariance matrix is:
\begin{align}\label{eq:covariant}
        {\rm cov}\left[\frac{f_x}{f_J},\frac{f_y}{f_J}\right]&=
        \frac{f_x f_y}{f_J^4} \EX[df_J^2]=
    \frac{f_x f_y}{f_J^4} \sigma_{f_J}^2 \quad \quad {\rm for}\quad x \ne y,\\\label{eq:diagonal}
        %{\rm var}\left(\frac{f_x}{f_J}\right)=
        %\left(\frac{1}{f_J}\right)^2 {\rm var}(f_x) + \left(\frac{f_x^2}{f_J^4}\right) {\rm var}(f_J)=\\
        {\rm cov}\left[\frac{f_x}{f_J},\frac{f_y}{f_J}\right]&=
        \left(\frac{1}{f_J}\right)^2 \sigma_{f_x}^2 + \left(\frac{f_x^2}{f_J^4}\right) \sigma_{f_J}^2 \quad \quad \quad {\rm for}\quad x = y.
\end{align}
\begin{figure*}
 \begin{center}
 \includegraphics[height=17cm, width=17cm]
 {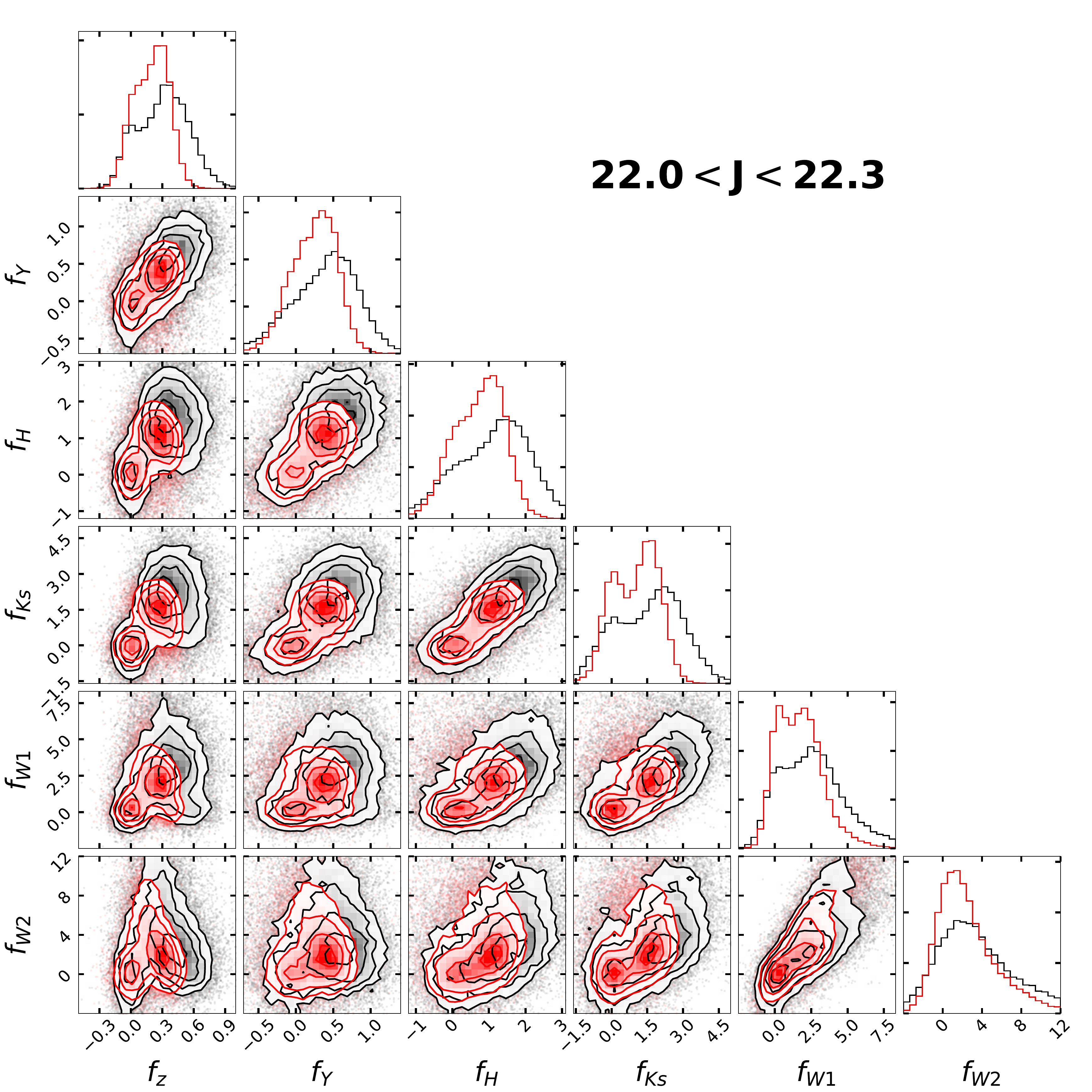}
 \caption{Relative-flux relative-flux contours comparison between the real data (black) and a noise added sample from the deconvolved model (red) generated by the XD code in the $22.0<J<22.3$. Errors have been added as explained in Appendix \ref{appendixa}, while the model was generated providing a covariance matrix in the form of Eq. \ref{eq:covariant} plus \ref{eq:diagonal}. The labelled quantities are relative fluxes (i.e., fluxes in different bands divided by the $J$-band flux). It is apparent that we do not obtain a noisy relative flux distribution that is consistent with the real one.}
 \label{fig:covariant}
 \end{center}
\end{figure*}
\begin{figure*}
 \begin{center}
 \includegraphics[height=17cm, width=17cm]
 {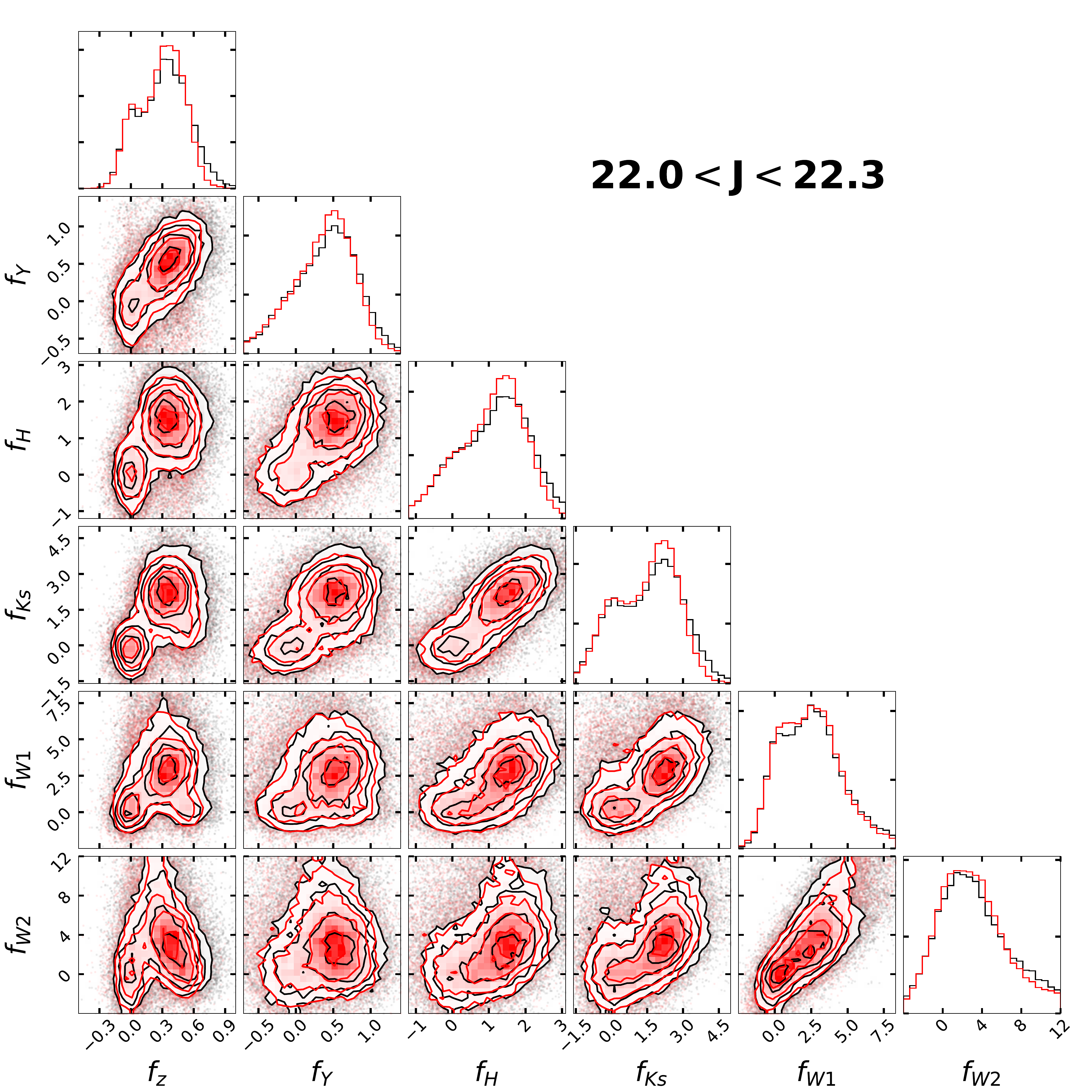}
 \caption{Relative-flux relative-flux contours comparison between the real data (black) and a noise added sample from the deconvolved model (red) generated by the XD code in the $22.0<J<22.3$. Errors have been added as explained in Appendix \ref{appendixa}, while the model was generated providing a diagonal covariance matrix in the form of Eq. \ref{eq:diagonal}. The labelled quantities are relative fluxes (i.e., fluxes in different bands divided by the $J$-band flux). In this case, the two distributions are consistent.}
 \label{fig:diagonal}
 \end{center}
\end{figure*}
At first, to train our contaminant models we provided to the XD code the noisy relative fluxes with covariance matrices computed using Eq. \ref{eq:covariant} and \ref{eq:diagonal}. However, we noticed that for bins whose $J$-band median point is $J_{\rm mp}>21$ (i.e., ${\rm SNR}(J_{\rm mp})<10$) the XD code is not able to correctly deconvolve the contaminants properties. This is apparent in Fig. \ref{fig:covariant}, where we show the comparison between the real data (black contours) and a noise added sample from the deconvolved model (red contours) generated by the XD code in a faint bin ($22.0<J<22.3$, ${\rm SNR}(J_{\rm mp})=5$): it is clear that we do not obtain a noisy relative flux distribution that is consistent with the real one. This deconvolved model was generated after providing a covariance matrix in the form of Eq. \ref{eq:covariant} plus \ref{eq:diagonal}, while we added the errors to the deconvolved sample as described in Appendix \ref{appendixa}. The failure of the XD code to correctly deconvolve the relative fluxes in the limit of faint $J$-band bins (${\rm SNR}(J_{\rm mp})<10$) arises from the violation of our assumption that the relative-flux uncertainties are Gaussian in this regime. In fact, the ratio of noisy quantities is in general not Gaussian distributed, as we assumed in order to use XD. However, this is a good approximation if $f_J$ has small errors relative to $f_X$, whereas as $f_J$ becomes noisier, one will generate progressively stronger tails in $f_X/f_J$. 
%% JFH2 Add one more sentence explaining that the ratio of noisy quantities
%% is in general not Gaussian distrbuted as we have assumed in order to use
%% XD. Howeve, this is a godo approxmation if f_J has small errrors relative
%% to f_X, whereas as f_J becomes noisier, one will generate progressively
%% stronger tails in f_X/f_J. 
%% RN done
To remedy
%% JFH2 I think it is too strong to say we solved it. We did not
%% because it has no easy solution aide from modeling the non-relative
%% fluxes, which would be untractable for other reasons (i.e. power
%% law probability densities). I think you can say to address, remedy,
%% or ameliorate this problem 
%% RN done
this problem, we decided to construct our faint ($J_{\rm mp}>21$) deconvolved contaminant models providing a diagonal covariance: with only elements on the diagonal computed by Eq. \ref{eq:diagonal} and zeros elsewhere. Although, this is not formally the correct approach to deal with non-independent quantities, it simply provides good results during the training step. In Fig. \ref{fig:diagonal} we show the comparison between the real data (black contours) and a noise added sample from the deconvolved model (red contours) generated by the XD code with a diagonal covariance. The $J$-band bin and the real data are the same as those displayed in Fig. \ref{fig:covariant}. In this case it is apparent that after re-adding the errors the noisy simulated distributions are far more consistent with the real ones.

%% JFH2 Errors model --> Noise model. I'm not sure what noise model
%% for the noiseless data means.
%% RN done
\section{Noise model}\label{appendixa}
As described in several parts in this paper,
%%JFH2 well we used this noise model in a bunch of our plots as well, so this
%% is not restricted ot the completeness/efficiency calculation.
%% RN done
we often sampled a huge number of simulated high-$z$ QSOs and contaminants from our XDHZQSO deconvolved models, and finally computed their probabilities of being high-$z$ QSOs based on their simulated properties. However, the sampling of deconvolved models produces noiseless relative fluxes that are not a real representation of the noisy properties usually measured. We explain here our adopted procedure to add the flux uncertainties to the simulated noiseless fluxes.

Lets consider the case of a noiseless sample simulated from a certain $J$-band bin. At first, we convert the noiseless simulated relative fluxes into noiseless fluxes by multiplying the former with the median $J$-band flux of that bin.
%% JFH2 Don't you want to say the bin center or something? Is this problematic
%% for the really wide bins?
%% RN done
Then, for each band used in this work, we divide the noisy fluxes from our VIKING area dataset into 50 bins that roughly contain the same number of sources. For each bin we interpolate the real measurement errors vs the cumulative number of sources distributed in each bin. For each noiseless simulated flux that belongs to a certain flux bin, we draw a random error from the corresponding error-cumulative number of sources distribution.
%% JFH2 something went wrong in the sentence above. You saple or draw from the
%% distribution, not from the interpolation. I know you interpolate to do this
%% in practice but sampling from the interpolation does not make sense. 
%% RN done
This error is finally added to the noiseless flux using a normal distribution centered on the noiseless flux and $\sigma$ equal to the sampled error. In this way, we can add the real errors coming from our VIKING area dataset to our simulated noiseless fluxes: i.e., we capture the distribution of the noise at a given flux level, instead of simply using its mean value.
%% JFH2 Make it clear why you do this, i.e. you captue the distribution of
%% the noise at a given flux level, not just the mean value. 
%% RN done

%% JFH Put the covariance calculation in the appendix. 
%% RN done

\bibliography{high-z_QSO_selection}

\begin{thebibliography}{}
\makeatletter
\relax
\def\mn@urlcharsother{\let\do\@makeother \do\$\do\&\do\#\do\^\do\_\do\%\do\~}
\def\mn@doi{\begingroup\mn@urlcharsother \@ifnextchar [ {\mn@doi@}
  {\mn@doi@[]}}
\def\mn@doi@[#1]#2{\def\@tempa{#1}\ifx\@tempa\@empty \href
  {http://dx.doi.org/#2} {doi:#2}\else \href {http://dx.doi.org/#2} {#1}\fi
  \endgroup}
\def\mn@eprint#1#2{\mn@eprint@#1:#2::\@nil}
\def\mn@eprint@arXiv#1{\href {http://arxiv.org/abs/#1} {{\tt arXiv:#1}}}
\def\mn@eprint@dblp#1{\href {http://dblp.uni-trier.de/rec/bibtex/#1.xml}
  {dblp:#1}}
\def\mn@eprint@#1:#2:#3:#4\@nil{\def\@tempa {#1}\def\@tempb {#2}\def\@tempc
  {#3}\ifx \@tempc \@empty \let \@tempc \@tempb \let \@tempb \@tempa \fi \ifx
  \@tempb \@empty \def\@tempb {arXiv}\fi \@ifundefined
  {mn@eprint@\@tempb}{\@tempb:\@tempc}{\expandafter \expandafter \csname
  mn@eprint@\@tempb\endcsname \expandafter{\@tempc}}}

\bibitem[\protect\citeauthoryear{{Aihara} et~al.,}{{Aihara}
  et~al.}{2018}]{Aihara18}
{Aihara} H.,  et~al., 2018, \mn@doi [\pasj] {10.1093/pasj/psx066}, \href
  {https://ui.adsabs.harvard.edu/abs/2018PASJ...70S...4A} {70, S4}

\bibitem[\protect\citeauthoryear{{Ba{\~n}ados} et~al.,}{{Ba{\~n}ados}
  et~al.}{2016}]{Banados16}
{Ba{\~n}ados} E.,  et~al., 2016, \mn@doi [\apjs] {10.3847/0067-0049/227/1/11},
  \href {https://ui.adsabs.harvard.edu/abs/2016ApJS..227...11B} {227, 11}

\bibitem[\protect\citeauthoryear{{Ba{\~n}ados} et~al.,}{{Ba{\~n}ados}
  et~al.}{2018}]{Banados18}
{Ba{\~n}ados} E.,  et~al., 2018, \mn@doi [\nat] {10.1038/nature25180}, \href
  {https://ui.adsabs.harvard.edu/abs/2018Natur.553..473B} {553, 473}

\bibitem[\protect\citeauthoryear{{Baldwin}}{{Baldwin}}{1977}]{Baldwin77}
{Baldwin} J.~A.,  1977, \mn@doi [\apj] {10.1086/155294}, \href
  {https://ui.adsabs.harvard.edu/abs/1977ApJ...214..679B} {214, 679}

\bibitem[\protect\citeauthoryear{{Barnett}, {Warren}, {Cross}, {Mortlock},
  {Fan}, {Wang}  \& {Hewett}}{{Barnett} et~al.}{2021}]{Barnett21}
{Barnett} R.,  {Warren} S.~J.,  {Cross} N.~J.~G.,  {Mortlock} D.~J.,  {Fan} X.,
   {Wang} F.,   {Hewett} P.~C.,  2021, \mn@doi [\mnras]
  {10.1093/mnras/staa3808}, \href
  {https://ui.adsabs.harvard.edu/abs/2021MNRAS.501.1663B} {501, 1663}

\bibitem[\protect\citeauthoryear{{Betoule} et~al.,}{{Betoule}
  et~al.}{2014}]{Betoule14}
{Betoule} M.,  et~al., 2014, \mn@doi [\aap] {10.1051/0004-6361/201423413},
  \href {https://ui.adsabs.harvard.edu/abs/2014A&A...568A..22B} {568, A22}

\bibitem[\protect\citeauthoryear{{Bovy}, {Hogg}  \& {Roweis}}{{Bovy}
  et~al.}{2011a}]{Bovy11a}
{Bovy} J.,  {Hogg} D.~W.,   {Roweis} S.~T.,  2011a, \mn@doi [Annals of Applied
  Statistics] {10.1214/10-AOAS439}, \href
  {https://ui.adsabs.harvard.edu/abs/2011AnApS...5.1657B} {5, 1657}

\bibitem[\protect\citeauthoryear{{Bovy} et~al.,}{{Bovy} et~al.}{2011b}]{Bovy11}
{Bovy} J.,  et~al., 2011b, \mn@doi [\apj] {10.1088/0004-637X/729/2/141}, \href
  {https://ui.adsabs.harvard.edu/abs/2011ApJ...729..141B} {729, 141}

\bibitem[\protect\citeauthoryear{{Bovy} et~al.,}{{Bovy} et~al.}{2012}]{Bovy12}
{Bovy} J.,  et~al., 2012, \mn@doi [\apj] {10.1088/0004-637X/749/1/41}, \href
  {https://ui.adsabs.harvard.edu/abs/2012ApJ...749...41B} {749, 41}

\bibitem[\protect\citeauthoryear{Buitinck et~al.,}{Buitinck
  et~al.}{2013}]{sklearn_api}
Buitinck L.,  et~al., 2013, in ECML PKDD Workshop: Languages for Data Mining
  and Machine Learning. pp 108--122

\bibitem[\protect\citeauthoryear{{Carnall} et~al.,}{{Carnall}
  et~al.}{2015}]{Carnall15}
{Carnall} A.~C.,  et~al., 2015, \mn@doi [\mnras] {10.1093/mnrasl/slv057}, \href
  {https://ui.adsabs.harvard.edu/abs/2015MNRAS.451L..16C} {451, L16}

\bibitem[\protect\citeauthoryear{{Davies} et~al.,}{{Davies}
  et~al.}{2018}]{Davies18}
{Davies} F.~B.,  et~al., 2018, \mn@doi [\apj] {10.3847/1538-4357/aad6dc}, \href
  {https://ui.adsabs.harvard.edu/abs/2018ApJ...864..142D} {864, 142}

\bibitem[\protect\citeauthoryear{{Davies}, {Hennawi}  \& {Eilers}}{{Davies}
  et~al.}{2019}]{Davies19}
{Davies} F.~B.,  {Hennawi} J.~F.,   {Eilers} A.-C.,  2019, \mn@doi [\apjl]
  {10.3847/2041-8213/ab42e3}, \href
  {https://ui.adsabs.harvard.edu/abs/2019ApJ...884L..19D} {884, L19}

\bibitem[\protect\citeauthoryear{{Dayal}, {Rossi}, {Shiralilou}, {Piana},
  {Choudhury}  \& {Volonteri}}{{Dayal} et~al.}{2019}]{Dayal19}
{Dayal} P.,  {Rossi} E.~M.,  {Shiralilou} B.,  {Piana} O.,  {Choudhury} T.~R.,
   {Volonteri} M.,  2019, \mn@doi [\mnras] {10.1093/mnras/stz897}, \href
  {https://ui.adsabs.harvard.edu/abs/2019MNRAS.486.2336D} {486, 2336}

\bibitem[\protect\citeauthoryear{{Dey} et~al.,}{{Dey} et~al.}{2019}]{Dey19}
{Dey} A.,  et~al., 2019, \mn@doi [\aj] {10.3847/1538-3881/ab089d}, \href
  {https://ui.adsabs.harvard.edu/abs/2019AJ....157..168D} {157, 168}

\bibitem[\protect\citeauthoryear{{Euclid Collaboration} et~al.,}{{Euclid
  Collaboration} et~al.}{2019}]{Euclid19}
{Euclid Collaboration} et~al., 2019, \mn@doi [\aap]
  {10.1051/0004-6361/201936427}, \href
  {https://ui.adsabs.harvard.edu/abs/2019A&A...631A..85E} {631, A85}

\bibitem[\protect\citeauthoryear{{Fan} et~al.,}{{Fan} et~al.}{2001}]{Fan01}
{Fan} X.,  et~al., 2001, \mn@doi [\aj] {10.1086/324111}, \href
  {https://ui.adsabs.harvard.edu/abs/2001AJ....122.2833F} {122, 2833}

\bibitem[\protect\citeauthoryear{{Gaskell}}{{Gaskell}}{1982}]{Gaskell82}
{Gaskell} C.~M.,  1982, \mn@doi [\apj] {10.1086/160481}, \href
  {https://ui.adsabs.harvard.edu/abs/1982ApJ...263...79G} {263, 79}

\bibitem[\protect\citeauthoryear{{Glikman}, {Helfand}  \& {White}}{{Glikman}
  et~al.}{2006}]{Glikman06}
{Glikman} E.,  {Helfand} D.~J.,   {White} R.~L.,  2006, \mn@doi [\apj]
  {10.1086/500098}, \href
  {https://ui.adsabs.harvard.edu/abs/2006ApJ...640..579G} {640, 579}

\bibitem[\protect\citeauthoryear{{Gonz{\'a}lez-Fern{\'a}ndez}
  et~al.,}{{Gonz{\'a}lez-Fern{\'a}ndez} et~al.}{2018}]{Gonzales18}
{Gonz{\'a}lez-Fern{\'a}ndez} C.,  et~al., 2018, \mn@doi [\mnras]
  {10.1093/mnras/stx3073}, \href
  {https://ui.adsabs.harvard.edu/abs/2018MNRAS.474.5459G} {474, 5459}

\bibitem[\protect\citeauthoryear{{Habouzit}, {Volonteri}, {Latif}, {Dubois}  \&
  {Peirani}}{{Habouzit} et~al.}{2016}]{Habouzit16}
{Habouzit} M.,  {Volonteri} M.,  {Latif} M.,  {Dubois} Y.,   {Peirani} S.,
  2016, \mn@doi [\mnras] {10.1093/mnras/stw1924}, \href
  {https://ui.adsabs.harvard.edu/abs/2016MNRAS.463..529H} {463, 529}

\bibitem[\protect\citeauthoryear{{Holoien}, {Marshall}  \&
  {Wechsler}}{{Holoien} et~al.}{2017}]{Holoien17}
{Holoien} T. W.~S.,  {Marshall} P.~J.,   {Wechsler} R.~H.,  2017, \mn@doi [\aj]
  {10.3847/1538-3881/aa68a1}, \href
  {https://ui.adsabs.harvard.edu/abs/2017AJ....153..249H} {153, 249}

\bibitem[\protect\citeauthoryear{{Inayoshi}, {Haiman}  \&
  {Ostriker}}{{Inayoshi} et~al.}{2016}]{Inayoshi16}
{Inayoshi} K.,  {Haiman} Z.,   {Ostriker} J.~P.,  2016, \mn@doi [\mnras]
  {10.1093/mnras/stw836}, \href
  {https://ui.adsabs.harvard.edu/abs/2016MNRAS.459.3738I} {459, 3738}

\bibitem[\protect\citeauthoryear{{Inayoshi}, {Visbal}  \& {Haiman}}{{Inayoshi}
  et~al.}{2020}]{Inayoshi20}
{Inayoshi} K.,  {Visbal} E.,   {Haiman} Z.,  2020, \mn@doi [\araa]
  {10.1146/annurev-astro-120419-014455}, \href
  {https://ui.adsabs.harvard.edu/abs/2020ARA&A..58...27I} {58, 27}

\bibitem[\protect\citeauthoryear{{Jiang} et~al.,}{{Jiang}
  et~al.}{2006}]{Jiang06}
{Jiang} L.,  et~al., 2006, \mn@doi [\aj] {10.1086/503745}, \href
  {https://ui.adsabs.harvard.edu/abs/2006AJ....131.2788J} {131, 2788}

\bibitem[\protect\citeauthoryear{{Jiang} et~al.,}{{Jiang}
  et~al.}{2016}]{Jiang16}
{Jiang} L.,  et~al., 2016, \mn@doi [\apj] {10.3847/1538-4357/833/2/222}, \href
  {https://ui.adsabs.harvard.edu/abs/2016ApJ...833..222J} {833, 222}

\bibitem[\protect\citeauthoryear{{Kelly}, {Bechtold}  \&
  {Siemiginowska}}{{Kelly} et~al.}{2009}]{Kelly09}
{Kelly} B.~C.,  {Bechtold} J.,   {Siemiginowska} A.,  2009, \mn@doi [\apj]
  {10.1088/0004-637X/698/1/895}, \href
  {https://ui.adsabs.harvard.edu/abs/2009ApJ...698..895K} {698, 895}

\bibitem[\protect\citeauthoryear{{Kuhn}, {Elvis}, {Bechtold}  \&
  {Elston}}{{Kuhn} et~al.}{2001}]{Khun01}
{Kuhn} O.,  {Elvis} M.,  {Bechtold} J.,   {Elston} R.,  2001, \mn@doi [\apjs]
  {10.1086/322535}, \href
  {https://ui.adsabs.harvard.edu/abs/2001ApJS..136..225K} {136, 225}

\bibitem[\protect\citeauthoryear{{Laureijs} et~al.,}{{Laureijs}
  et~al.}{2011}]{Laureijs11}
{Laureijs} R.,  et~al., 2011, arXiv e-prints, \href
  {https://ui.adsabs.harvard.edu/abs/2011arXiv1110.3193L} {p. arXiv:1110.3193}

\bibitem[\protect\citeauthoryear{{Madau} \& {Rees}}{{Madau} \&
  {Rees}}{2001}]{MadauRees01}
{Madau} P.,  {Rees} M.~J.,  2001, \mn@doi [\apjl] {10.1086/319848}, \href
  {https://ui.adsabs.harvard.edu/abs/2001ApJ...551L..27M} {551, L27}

\bibitem[\protect\citeauthoryear{{Mainzer} et~al.,}{{Mainzer}
  et~al.}{2011}]{Mainzer11}
{Mainzer} A.,  et~al., 2011, \mn@doi [\apj] {10.1088/0004-637X/743/2/156},
  \href {https://ui.adsabs.harvard.edu/abs/2011ApJ...743..156M} {743, 156}

\bibitem[\protect\citeauthoryear{{Matsuoka} et~al.,}{{Matsuoka}
  et~al.}{2016}]{Matsuoka16}
{Matsuoka} Y.,  et~al., 2016, \mn@doi [\apj] {10.3847/0004-637X/828/1/26},
  \href {https://ui.adsabs.harvard.edu/abs/2016ApJ...828...26M} {828, 26}

\bibitem[\protect\citeauthoryear{{Matsuoka} et~al.,}{{Matsuoka}
  et~al.}{2018a}]{Matsuoka18a}
{Matsuoka} Y.,  et~al., 2018a, \mn@doi [\pasj] {10.1093/pasj/psx046}, \href
  {https://ui.adsabs.harvard.edu/abs/2018PASJ...70S..35M} {70, S35}

\bibitem[\protect\citeauthoryear{{Matsuoka} et~al.,}{{Matsuoka}
  et~al.}{2018b}]{Matsuoka18b}
{Matsuoka} Y.,  et~al., 2018b, \mn@doi [\apjs] {10.3847/1538-4365/aac724},
  \href {https://ui.adsabs.harvard.edu/abs/2018ApJS..237....5M} {237, 5}

\bibitem[\protect\citeauthoryear{{Matsuoka} et~al.,}{{Matsuoka}
  et~al.}{2018c}]{Matsuoka18c}
{Matsuoka} Y.,  et~al., 2018c, \mn@doi [\apj] {10.3847/1538-4357/aaee7a}, \href
  {https://ui.adsabs.harvard.edu/abs/2018ApJ...869..150M} {869, 150}

\bibitem[\protect\citeauthoryear{{Matsuoka} et~al.,}{{Matsuoka}
  et~al.}{2019a}]{Matsuoka19b}
{Matsuoka} Y.,  et~al., 2019a, \mn@doi [\apjl] {10.3847/2041-8213/ab0216},
  \href {https://ui.adsabs.harvard.edu/abs/2019ApJ...872L...2M} {872, L2}

\bibitem[\protect\citeauthoryear{{Matsuoka} et~al.,}{{Matsuoka}
  et~al.}{2019b}]{Matsuoka19a}
{Matsuoka} Y.,  et~al., 2019b, \mn@doi [\apj] {10.3847/1538-4357/ab3c60}, \href
  {https://ui.adsabs.harvard.edu/abs/2019ApJ...883..183M} {883, 183}

\bibitem[\protect\citeauthoryear{{McGreer}, {Mesinger}  \& {Fan}}{{McGreer}
  et~al.}{2011}]{Mcgreer11}
{McGreer} I.~D.,  {Mesinger} A.,   {Fan} X.,  2011, \mn@doi [\mnras]
  {10.1111/j.1365-2966.2011.18935.x}, \href
  {https://ui.adsabs.harvard.edu/abs/2011MNRAS.415.3237M} {415, 3237}

\bibitem[\protect\citeauthoryear{{McGreer} et~al.,}{{McGreer}
  et~al.}{2013}]{McGreer13}
{McGreer} I.~D.,  et~al., 2013, \mn@doi [\apj] {10.1088/0004-637X/768/2/105},
  \href {https://ui.adsabs.harvard.edu/abs/2013ApJ...768..105M} {768, 105}

\bibitem[\protect\citeauthoryear{{McGreer}, {Mesinger}  \&
  {D'Odorico}}{{McGreer} et~al.}{2015}]{Mcgreer15}
{McGreer} I.~D.,  {Mesinger} A.,   {D'Odorico} V.,  2015, \mn@doi [\mnras]
  {10.1093/mnras/stu2449}, \href
  {https://ui.adsabs.harvard.edu/abs/2015MNRAS.447..499M} {447, 499}

\bibitem[\protect\citeauthoryear{{Mortlock} et~al.,}{{Mortlock}
  et~al.}{2011}]{Mortlock11}
{Mortlock} D.~J.,  et~al., 2011, \mn@doi [\nat] {10.1038/nature10159}, \href
  {https://ui.adsabs.harvard.edu/abs/2011Natur.474..616M} {474, 616}

\bibitem[\protect\citeauthoryear{{Mortlock}, {Patel}, {Warren}, {Hewett},
  {Venemans}, {McMahon}  \& {Simpson}}{{Mortlock} et~al.}{2012}]{Mortlock12}
{Mortlock} D.~J.,  {Patel} M.,  {Warren} S.~J.,  {Hewett} P.~C.,  {Venemans}
  B.~P.,  {McMahon} R.~G.,   {Simpson} C.,  2012, \mn@doi [\mnras]
  {10.1111/j.1365-2966.2011.19710.x}, \href
  {https://ui.adsabs.harvard.edu/abs/2012MNRAS.419..390M} {419, 390}

\bibitem[\protect\citeauthoryear{Pedregosa et~al.,}{Pedregosa
  et~al.}{2011}]{scikit-learn}
Pedregosa F.,  et~al., 2011, Journal of Machine Learning Research, 12, 2825

\bibitem[\protect\citeauthoryear{{Reed} et~al.,}{{Reed} et~al.}{2015}]{Reed15}
{Reed} S.~L.,  et~al., 2015, \mn@doi [\mnras] {10.1093/mnras/stv2031}, \href
  {https://ui.adsabs.harvard.edu/abs/2015MNRAS.454.3952R} {454, 3952}

\bibitem[\protect\citeauthoryear{{Reed} et~al.,}{{Reed} et~al.}{2017}]{Reed17}
{Reed} S.~L.,  et~al., 2017, \mn@doi [\mnras] {10.1093/mnras/stx728}, \href
  {https://ui.adsabs.harvard.edu/abs/2017MNRAS.468.4702R} {468, 4702}

\bibitem[\protect\citeauthoryear{{Richards} et~al.,}{{Richards}
  et~al.}{2006}]{Richards06}
{Richards} G.~T.,  et~al., 2006, \mn@doi [\aj] {10.1086/503559}, \href
  {https://ui.adsabs.harvard.edu/abs/2006AJ....131.2766R} {131, 2766}

\bibitem[\protect\citeauthoryear{{Richards} et~al.,}{{Richards}
  et~al.}{2011}]{Richards11}
{Richards} G.~T.,  et~al., 2011, \mn@doi [\aj] {10.1088/0004-6256/141/5/167},
  \href {https://ui.adsabs.harvard.edu/abs/2011AJ....141..167R} {141, 167}

\bibitem[\protect\citeauthoryear{{Ross} \& {Cross}}{{Ross} \&
  {Cross}}{2020}]{RossCross20}
{Ross} N.~P.,  {Cross} N. J.~G.,  2020, \mn@doi [\mnras]
  {10.1093/mnras/staa544}, \href
  {https://ui.adsabs.harvard.edu/abs/2020MNRAS.494..789R} {494, 789}

\bibitem[\protect\citeauthoryear{{Schauer}, {Regan}, {Glover}  \&
  {Klessen}}{{Schauer} et~al.}{2017}]{Schauer17}
{Schauer} A. T.~P.,  {Regan} J.,  {Glover} S. C.~O.,   {Klessen} R.~S.,  2017,
  \mn@doi [\mnras] {10.1093/mnras/stx1915}, \href
  {https://ui.adsabs.harvard.edu/abs/2017MNRAS.471.4878S} {471, 4878}

\bibitem[\protect\citeauthoryear{{Schindler}, {Fan}, {McGreer}, {Yang}, {Wu},
  {Jiang}  \& {Green}}{{Schindler} et~al.}{2017}]{Schindler17}
{Schindler} J.-T.,  {Fan} X.,  {McGreer} I.~D.,  {Yang} Q.,  {Wu} J.,  {Jiang}
  L.,   {Green} R.,  2017, \mn@doi [\apj] {10.3847/1538-4357/aa9929}, \href
  {https://ui.adsabs.harvard.edu/abs/2017ApJ...851...13S} {851, 13}

\bibitem[\protect\citeauthoryear{{Schindler} et~al.,}{{Schindler}
  et~al.}{2018}]{Schindler18}
{Schindler} J.-T.,  et~al., 2018, \mn@doi [\apj] {10.3847/1538-4357/aad2dd},
  \href {https://ui.adsabs.harvard.edu/abs/2018ApJ...863..144S} {863, 144}

\bibitem[\protect\citeauthoryear{{Schindler} et~al.,}{{Schindler}
  et~al.}{2019}]{Schindler19}
{Schindler} J.-T.,  et~al., 2019, \mn@doi [\apj] {10.3847/1538-4357/aaf86c},
  \href {https://ui.adsabs.harvard.edu/abs/2019ApJ...871..258S} {871, 258}

\bibitem[\protect\citeauthoryear{{Schlafly}, {Meisner}  \& {Green}}{{Schlafly}
  et~al.}{2019}]{Schlafly19}
{Schlafly} E.~F.,  {Meisner} A.~M.,   {Green} G.~M.,  2019, \mn@doi [\apjs]
  {10.3847/1538-4365/aafbea}, \href
  {https://ui.adsabs.harvard.edu/abs/2019ApJS..240...30S} {240, 30}

\bibitem[\protect\citeauthoryear{{Schmidt}, {Marshall}, {Rix}, {Jester},
  {Hennawi}  \& {Dobler}}{{Schmidt} et~al.}{2010}]{Schmidt10}
{Schmidt} K.~B.,  {Marshall} P.~J.,  {Rix} H.-W.,  {Jester} S.,  {Hennawi}
  J.~F.,   {Dobler} G.,  2010, \mn@doi [\apj] {10.1088/0004-637X/714/2/1194},
  \href {https://ui.adsabs.harvard.edu/abs/2010ApJ...714.1194S} {714, 1194}

\bibitem[\protect\citeauthoryear{{Tanaka} \& {Haiman}}{{Tanaka} \&
  {Haiman}}{2009}]{TanakaHaiman09}
{Tanaka} T.,  {Haiman} Z.,  2009, \mn@doi [\apj]
  {10.1088/0004-637X/696/2/1798}, \href
  {https://ui.adsabs.harvard.edu/abs/2009ApJ...696.1798T} {696, 1798}

\bibitem[\protect\citeauthoryear{{Trakhtenbrot}, {Volonteri}  \&
  {Natarajan}}{{Trakhtenbrot} et~al.}{2017}]{Trakhtenbrot17}
{Trakhtenbrot} B.,  {Volonteri} M.,   {Natarajan} P.,  2017, \mn@doi [\apjl]
  {10.3847/2041-8213/836/1/L1}, \href
  {https://ui.adsabs.harvard.edu/abs/2017ApJ...836L...1T} {836, L1}

\bibitem[\protect\citeauthoryear{{Vanden Berk} et~al.,}{{Vanden Berk}
  et~al.}{2004}]{VandenBerk04}
{Vanden Berk} D.~E.,  et~al., 2004, \mn@doi [\apj] {10.1086/380563}, \href
  {https://ui.adsabs.harvard.edu/abs/2004ApJ...601..692V} {601, 692}

\bibitem[\protect\citeauthoryear{{Venemans} et~al.,}{{Venemans}
  et~al.}{2013}]{Venemans13}
{Venemans} B.~P.,  et~al., 2013, \mn@doi [\apj] {10.1088/0004-637X/779/1/24},
  \href {https://ui.adsabs.harvard.edu/abs/2013ApJ...779...24V} {779, 24}

\bibitem[\protect\citeauthoryear{{Venemans} et~al.,}{{Venemans}
  et~al.}{2015}]{Venemans15}
{Venemans} B.~P.,  et~al., 2015, \mn@doi [\mnras] {10.1093/mnras/stv1774},
  \href {https://ui.adsabs.harvard.edu/abs/2015MNRAS.453.2259V} {453, 2259}

\bibitem[\protect\citeauthoryear{{Volonteri}}{{Volonteri}}{2012}]{Volonteri12}
{Volonteri} M.,  2012, \mn@doi [Science] {10.1126/science.1220843}, \href
  {https://ui.adsabs.harvard.edu/abs/2012Sci...337..544V} {337, 544}

\bibitem[\protect\citeauthoryear{{Volonteri} \& {Begelman}}{{Volonteri} \&
  {Begelman}}{2010}]{Volonteri10}
{Volonteri} M.,  {Begelman} M.~C.,  2010, \mn@doi [\mnras]
  {10.1111/j.1365-2966.2010.17359.x}, \href
  {https://ui.adsabs.harvard.edu/abs/2010MNRAS.409.1022V} {409, 1022}

\bibitem[\protect\citeauthoryear{{Wang} et~al.,}{{Wang} et~al.}{2017}]{Wang17}
{Wang} F.,  et~al., 2017, \mn@doi [\apj] {10.3847/1538-4357/aa689f}, \href
  {https://ui.adsabs.harvard.edu/abs/2017ApJ...839...27W} {839, 27}

\bibitem[\protect\citeauthoryear{{Wang} et~al.,}{{Wang} et~al.}{2018}]{Wang18}
{Wang} F.,  et~al., 2018, \mn@doi [\apjl] {10.3847/2041-8213/aaf1d2}, \href
  {https://ui.adsabs.harvard.edu/abs/2018ApJ...869L...9W} {869, L9}

\bibitem[\protect\citeauthoryear{{Wang} et~al.,}{{Wang} et~al.}{2019}]{Wang19}
{Wang} F.,  et~al., 2019, \mn@doi [\apj] {10.3847/1538-4357/ab2be5}, \href
  {https://ui.adsabs.harvard.edu/abs/2019ApJ...884...30W} {884, 30}

\bibitem[\protect\citeauthoryear{{Wang} et~al.,}{{Wang} et~al.}{2020}]{Wang20}
{Wang} F.,  et~al., 2020, \mn@doi [\apj] {10.3847/1538-4357/ab8c45}, \href
  {https://ui.adsabs.harvard.edu/abs/2020ApJ...896...23W} {896, 23}

\bibitem[\protect\citeauthoryear{{Wang} et~al.,}{{Wang} et~al.}{2021}]{Wang21}
{Wang} F.,  et~al., 2021, \mn@doi [\apjl] {10.3847/2041-8213/abd8c6}, \href
  {https://ui.adsabs.harvard.edu/abs/2021ApJ...907L...1W} {907, L1}

\bibitem[\protect\citeauthoryear{{Wenzl} et~al.,}{{Wenzl}
  et~al.}{2021}]{Wenzl21}
{Wenzl} L.,  et~al., 2021, \mn@doi [\aj] {10.3847/1538-3881/ac0254}, \href
  {https://ui.adsabs.harvard.edu/abs/2021AJ....162...72W} {162, 72}

\bibitem[\protect\citeauthoryear{{Willott} et~al.,}{{Willott}
  et~al.}{2009}]{Willott09}
{Willott} C.~J.,  et~al., 2009, \mn@doi [\aj] {10.1088/0004-6256/137/3/3541},
  \href {https://ui.adsabs.harvard.edu/abs/2009AJ....137.3541W} {137, 3541}

\bibitem[\protect\citeauthoryear{{Worseck} \& {Prochaska}}{{Worseck} \&
  {Prochaska}}{2011}]{Wor_Pro11}
{Worseck} G.,  {Prochaska} J.~X.,  2011, \mn@doi [\apj]
  {10.1088/0004-637X/728/1/23}, \href
  {https://ui.adsabs.harvard.edu/abs/2011ApJ...728...23W} {728, 23}

\bibitem[\protect\citeauthoryear{{Wright} et~al.,}{{Wright}
  et~al.}{2010}]{Wright10}
{Wright} E.~L.,  et~al., 2010, \mn@doi [\aj] {10.1088/0004-6256/140/6/1868},
  \href {https://ui.adsabs.harvard.edu/abs/2010AJ....140.1868W} {140, 1868}

\bibitem[\protect\citeauthoryear{{Wu} et~al.,}{{Wu} et~al.}{2015}]{Wu15}
{Wu} X.-B.,  et~al., 2015, \mn@doi [\nat] {10.1038/nature14241}, \href
  {https://ui.adsabs.harvard.edu/abs/2015Natur.518..512W} {518, 512}

\bibitem[\protect\citeauthoryear{{Yang} et~al.,}{{Yang} et~al.}{2016}]{Yang16}
{Yang} J.,  et~al., 2016, \mn@doi [\apj] {10.3847/0004-637X/829/1/33}, \href
  {https://ui.adsabs.harvard.edu/abs/2016ApJ...829...33Y} {829, 33}

\bibitem[\protect\citeauthoryear{{Yang} et~al.,}{{Yang} et~al.}{2019}]{Yang19}
{Yang} J.,  et~al., 2019, \mn@doi [\aj] {10.3847/1538-3881/ab1be1}, \href
  {https://ui.adsabs.harvard.edu/abs/2019AJ....157..236Y} {157, 236}

\bibitem[\protect\citeauthoryear{{Yang} et~al.,}{{Yang}
  et~al.}{2020a}]{Yang20b}
{Yang} J.,  et~al., 2020a, \mn@doi [\apjl] {10.3847/2041-8213/ab9c26}, \href
  {https://ui.adsabs.harvard.edu/abs/2020ApJ...897L..14Y} {897, L14}

\bibitem[\protect\citeauthoryear{{Yang} et~al.,}{{Yang}
  et~al.}{2020b}]{Yang20a}
{Yang} J.,  et~al., 2020b, \mn@doi [\apj] {10.3847/1538-4357/abbc1b}, \href
  {https://ui.adsabs.harvard.edu/abs/2020ApJ...904...26Y} {904, 26}

\bibitem[\protect\citeauthoryear{{Yip} et~al.,}{{Yip} et~al.}{2004}]{Yip04}
{Yip} C.~W.,  et~al., 2004, \mn@doi [\aj] {10.1086/425626}, \href
  {https://ui.adsabs.harvard.edu/abs/2004AJ....128.2603Y} {128, 2603}

\bibitem[\protect\citeauthoryear{{Zou} et~al.,}{{Zou} et~al.}{2019}]{Zou19}
{Zou} H.,  et~al., 2019, \mn@doi [\apjs] {10.3847/1538-4365/ab48e8}, \href
  {https://ui.adsabs.harvard.edu/abs/2019ApJS..245....4Z} {245, 4}

\makeatother
\end{thebibliography}
\bibliographystyle{mnras}

%\bsp	% typesetting comment
\label{lastpage}
\end{document}